\documentclass[aps, rmp, amsmath, floats, floatfix, twocolumn, nofootinbib, showpacs]{revtex4-1}
\usepackage{graphicx}
\usepackage{color}

\newcommand{\be}{\begin{eqnarray}}
\newcommand{\ee}{\end{eqnarray}}

\def\apjl{Astrophys. J.}
\def\apjs{Astrophys. J. Suppl.}
\def\jcap{JCAP}
\def\jhep{JHEP}
\def\mnras{MNRAS}
\def\nar{New Astronomy Reviews}
\def\ssr{Space Science Reviews}
\def\aap{Astron. Astrophys.}
\def\araa{Ann. Rev. Astron. Astrophys.}
\def\physrep{Physics Reports}
\def\apss{Astrophysics and Space Science}
\def\procspie{Proc. SPIE}
\def\pasj{Publ. Astron. Soc. Japan}
\def\physrep{Physics Reports}

\begin{document}

\title{Testing black hole candidates with electromagnetic radiation}

\author{Cosimo Bambi}
\email{bambi@fudan.edu.cn}

\affiliation{Center for Field Theory and Particle Physics and Department of Physics, Fudan University, 200433 Shanghai, China}

\affiliation{Theoretical Astrophysics, Eberhard-Karls Universit\"at T\"ubingen, 72076 T\"ubingen, Germany}

\date{\today}

\begin{abstract}
Astrophysical black hole candidates are thought to be the Kerr black holes of general relativity, but there is not yet direct observational evidence that the spacetime geometry around these objects is described by the Kerr solution. The study of the properties of the electromagnetic radiation emitted by gas or stars orbiting these objects can potentially test the Kerr black hole hypothesis. This paper reviews the state of the art of this research field, describing the possible approaches to test the Kerr metric with current and future observational facilities and discussing current constraints.
\end{abstract}


\maketitle

\tableofcontents


\section{Introduction \label{s1-intro}}

\subsection{Motivations}

Einstein's theory of general relativity was proposed in 1915~\citep{einstein1916}. The first experimental confirmation was attained in 1919, from the measurement of light bending in the vicinity of the surface of the Sun by Eddington~\citep{eddington1919}. Thanks to this observation, the theory immediately became very popular. However, systematic tests of general relativity started much later. Experiments in the Solar System started in the 1960s. Tests using the observations of binary pulsars began in the 1970s. In the past 50 years, a large number of experiments have confirmed the predictions of general relativity in weak gravitational fields~\citep{will}. The focus of current experiments has now shifted to testing the theory in other regimes.

Tests of general relativity at large scales are mainly motivated by cosmological observations. While the dark matter problem is more likely due to some weakly interacting massive particles beyond the Standard Model of particle physics rather than a breakdown of Newton's law of gravitation~\citep{darkmatter}, see in particular the Bullet Cluster~\citep{bullet}, the problem of dark energy is completely open~\citep{darkenergy}. At the moment, the accelerating expansion rate of the Universe may be explained by a small positive {\it ad hoc} cosmological constant. However, it is possible that the actual explanation is either a breakdown of general relativity at large scales or the existence of some new field with peculiar properties. In the past 20 years, testing general relativity at cosmological scales has been a very active research field~\citep{test-gr-c}.

More recently, there has been increasing interest in testing general relativity in strong gravitational fields, which is the regime where the theory is more likely to encounter deviations from its predictions\footnote{There is no single definition of ``strong field'' in general relativity. Even binary pulsar tests are sometimes classified as tests in strong gravitational fields. If we consider the strength of the Newtonian potential (when the metric is close to Minkowski), binary pulsar measurements test weak gravitational fields, because $M/r \sim 10^{-5}$, where $M$ is the mass of the system and $r$ is the distance between the pulsar and the companion. In this context, strong gravitational fields have $M/r \sim 0.1$-1. A more coordinate-independent measure is given by some curvature invariant, like the Kretschmann scalar $R_{\mu\nu\rho\sigma} R^{\mu\nu\rho\sigma}$. However, this introduces a length scale. If we identify the Planck length $L_{\rm Pl} = G_{\rm N}^{1/2} \sim 10^{-33}$~cm as the natural scale in general relativity, no astronomical observation would probably ever reach strong fields.}. One of the most fascinating predictions of general relativity is the existence of Kerr black holes (BHs). The so-called BH candidates are astrophysical compact objects that can be naturally interpreted as the Kerr BHs of general relativity or they could be something else in the light of new physics.

A direct observational confirmation of the nature of these objects could be seen as an important test of general relativity in the strong gravity regime~\citep{review1,review2}. Deviations from the Kerr metric may be expected from classical extensions of general relativity~\citep{c1,berti15} as well as from macroscopic quantum gravity effects~\citep{q1,q2,q3}. Indeed, there are some arguments suggesting that the critical scale at which classical predictions may break down is not at the Planck one, as expected in a collision between two particles, but at the gravitational radius of the system~\citep{q1,q2,q3}.

There are two possible approaches to test BH candidates: electromagnetic and gravitational wave observations. The aim of this paper is to review current attempts with electromagnetic radiation. Generally speaking, this goal is very challenging, because it is difficult to have a reliable and sophisticated astrophysical model to test fundamental physics. Note that such a research field is young and still a developing one. While tests of general relativity in the Solar System and with binary pulsars started about 60 years ago, studies to directly test the nature of BH candidates are recent and became a hot topic only in the past 10~years. As a not yet fully mature research field, in order to make progress it is common to employ some approaches lacking the necessary scientific rigor and that are sometimes criticized.

The results of current electromagnetic observations can be summarized as follows. BH candidates cannot be astrophysical bodies with a surface made of ordinary matter~\citep{eea1,eeb2}, nor can be compact stars made of exotic non-interacting particles~\citep{exotic1}. Even if the constraints are currently weak, the quadrupole moment of these objects seems to be in agreement at the level of 30\% with that which is expected for a Kerr BH~\citep{valtonen1}. The thermal spectrum of the accretion disk around the stellar-mass BH candidates in GRS~1915+105 and Cygnus~X-1 looks like that expected for a very fast-rotating Kerr BH, and this can constrain at least some kinds of deviations from the Kerr geometry~\citep{cfm-cc,lingyao}. Similar constraints can be obtained for supermassive BH candidates under some reasonable assumptions about the evolution of the spin parameter~\citep{spin-c4}. These results are reviewed in the next sections. However, there is much work to do to test the nature of BH candidates, and new and more powerful observational facilities are necessary if we want to achieve this goal.

Gravitational wave tests have been reviewed, e.g., in~\citet{rev-gw-13}. Gravitational wave constraints on alternative theories of gravity can be obtained from the observed decay of the orbital period in binary pulsars, see e.g.~\citet{ny-sah-10}. The best system for gravitational wave tests may be an extreme mass ratio inspiral (commonly called EMRI), in which a stellar-mass compact object slowly falls onto a supermassive BH candidate~\citep{tests-gw0,tests-gw1,tests-gw2}. Accurate measurements of BH quasi-normal modes may also do the job~\citep{berti06,gossan}. The detection of gravitational waves by LIGO in September 2015 has opened a new window~\citep{v3-ligo}. Gravitational wave detectors now promise precision tests of the strong gravity regime in 5-10~years.

The content of the paper is as follows: an introduction section devoted to briefly review the motivations to test the Kerr metric, a discussion of some important properties of the Kerr metric, and a discussion of basic astronomical observations of astrophysical BH candidates. In Section~\ref{s2-tests}, I discuss the general approach to test the Kerr metric with electromagnetic radiation, its limitations, and some important phenomena that may show up in non-Kerr metrics. In Section~\ref{s3-xray}, I review the possibility of testing the Kerr metric with X-ray observations. This is currently the only available electromagnetic approach to probe the strong gravity field close to these objects. Section~\ref{s4-sgra} is reserved to discuss possible measurements of SgrA$^*$, the supermassive BH candidate at the center of our Galaxy, which can be considered as a special case. In Section~\ref{s5-other}, I briefly introduce a few more approaches to test the Kerr metric. Section~\ref{s-new} discusses the differences between tests with electromagnetic and gravitational wave observations. Summary and conclusions are reported in Section~\ref{s6-sc}.

Throughout the paper, I employ the convention of a metric with signature $(-+++)$ and units in which $G_{\rm N} = c = 1$, unless stated otherwise. In these units, the size of a stellar-mass BH is
\be
M &=& 14.77 \left(\frac{M}{10 \, M_\odot}\right) \, {\rm km} \nonumber\\
&=& 49.23 \left(\frac{M}{10 \, M_\odot}\right) \, \mu{\rm s} \, . 
\ee
In the case of supermassive BHs with mass $M = 10^6$~$M_\odot$ ($10^9$~$M_\odot$), we have $M \approx 1.5 \cdot 10^6$~km ($1.5 \cdot10^9$~km) and $\approx 5$~s (1.4~hr).

\subsection{Kerr black holes \label{ss-kerr}}

Technically, a BH in an asymptotically flat spacetime ${\mathcal M}$ is defined as the set of events that do not belong to the causal past of future null infinity $J^-({\mathcal I}^+)$, i.e. 
\be
{\mathcal B} = {\mathcal M} - J^-({\mathcal I}^+) \neq \emptyset \, .
\ee
The {\it event horizon} is the boundary of the BH region. See, e.g., \citet{MTW73}, \citet{v3-wald} or \citet{poisson} for the details. In other words, all future-directed curves (either time-like or null) starting from the region ${\mathcal B}$ fail to reach null infinity ${\mathcal I}^+$. A BH is thus an actual one-way membrane: if something crosses the event horizon it can no longer send any signal to the exterior region.

While the event horizon is a global property of an entire spacetime, the apparent horizon is a local property and is slicing-dependent. The apparent horizon can be defined after introducing the concept of a trapped surface. A trapped surface is a 2-dimensional surface ${\mathcal S}$ with the property that the expansion of ingoing and outgoing congruences of null geodesics orthogonal to ${\mathcal S}$ is negative everywhere on ${\mathcal S}$. An {\it apparent horizon} is the 2-dimensional intersection of the 3-dimensional boundary of the region of the spacetime that contains a trapped surface with  a space-like hyper-surface. Physically speaking, outward-pointing light rays behind an apparent horizon actually move inward and therefore they cannot cross the apparent horizon. Under certain assumptions, the existence of an apparent horizon implies that the slice contains an event horizon; the converse may not be true~\citep{v3-wi}. More details can be found, e.g., in \citet{MTW73}, \citet{v3-wald} or \citet{poisson}.

Note that there are scenarios beyond classical general relativity in which a collapsing object does not generate a BH with a central spacetime singularity. For an observer in the asymptotic flat region, the collapse may generate an apparent horizon for a finite time; the latter may be interpreted as an event horizon if the observational timescale is shorter than the lifetime of the apparent horizon~\citep{frolov81,bmm,v3-bmm2}.

In general relativity, the simplest BH solution is the Schwarzschild metric that describes an uncharged and non-rotating BH in vacuum and has only one parameter, the BH mass $M$. If the BH additionally has a non-vanishing electric charge $Q$, we have the Reissner-Nordstr\"om solution. If the BH has a non-vanishing spin angular momentum $J$, we have the Kerr solution. The Kerr-Newman metric describes a rotating BH with a non-vanishing electric charge.

The fact that BHs are described by a small number of parameters ($M$, $J$, $Q$) is explained by the so-called ``no-hair'' theorems\footnote{Note that there are a number of assumptions concerning these theorems. Specifically, the spacetime must be stationary, asymptotically flat, and have 4 dimensions; the only stress-energy tensor is due to the electromagnetic field; the exterior region must be regular (no naked singularities)~\citep{nh3}.}, which were pioneered in~\citet{nh1} and \citet{nh2} and the final version is still a work in progress~\citep{nh3}. The name no-hair indicates that BHs have no features (hairs), in the sense that they are completely specified by a small number of parameters. In the context of tests of general relativity and of the Kerr metric, the ``uniqueness'' of the Kerr-Newman solution is also relevant: under the same assumptions of the no-hair theorems, there are only Kerr-Newman BHs.

``Hairy'' BHs generically arise when gravity couples to non-Abelian gauge fields~\citep{volkov1,volkov2} or when scalar fields non-minimally couple to gravity, e.g. a dilation field in Einstein-dilaton-Gauss-Bonnet gravity~\citep{v3-EdGB}. An important example of the violation of the no-hair theorem was presented in~\cite{v2-hr0}, in which hairy BHs are possible by combining rotation with a harmonic time-dependence in the scalar field. Another interesting example is Chern-Simons dynamical gravity. In this framework, non-rotating BHs are described by the Schwarzschild solution, but rotating BHs are different from those of Kerr~\citep{chern-simons}. A review on hairy BHs is \cite{v2-hr}. A review on a number of BHs in alternative theories of gravity is~\cite{berti15}.

If BHs were merely theoretical solutions of the Einstein equations, they would not be so interesting. They become interesting because we have compelling observational evidence of their existence in the Universe. BHs can be created by the gravitational collapse of matter. The simplest example for which there is an analytic solution is the collapse of a spherically symmetric cloud of dust, the so-called Oppenheimer-Snyder model~\citep{op-sn}. This example shows how the final product of collapse is a Schwarzschild BH with a central singularity. Scenarios closer to reality require numerical calculations~\citep{luca1,luca2}. For a general review on gravitational collapse, see e.g.~\citet{daniele}. A BH can be created by the gravitational collapse of a very heavy star, after the latter has exhausted all its nuclear fuel and in the case the degenerate neutron pressure cannot balance the gravitational force. The process should be so common that from stellar population arguments we expect that today there are about $10^8$-$10^9$ BHs in our Galaxy formed from core-collapse~\citep{v2-remnant}.

It is worth stressing that the stationary Kerr solution of general relativity should well be capable of describing the spacetime around astrophysical BHs formed from gravitational collapse. In general relativity, initial deviations from the Kerr metric are quickly radiated away through the emission of gravitational waves~\citep{price}. For macroscopic BHs, the equilibrium electric charge is completely irrelevant in the spacetime geometry~\citep{petrov}. The presence of an accretion disk is normally negligible, because the disk is extended and has a low density~\citep{v2-bcp}. Moreover, the disk mass is many orders of magnitude smaller than that of the BH, so its impact on the measurement of the metric would be extremely small in any case~\citep{disk}.

\subsection{Basic properties of the Kerr metric}

In Boyer-Lindquist coordinates, the line element of the Kerr metric reads~\citep{v3-kerr,v3-chandra}
\be
\hspace{-0.5cm}
ds^2 &=& - \left(1 - \frac{2 M r}{\Sigma}\right) dt^2 
- \frac{4 M a r \sin^2\theta}{\Sigma} dt d\phi 
+ \frac{\Sigma}{\Delta} dr^2 \nonumber\\
&& + \Sigma d\theta^2 
+ \left(r^2 + a^2 + \frac{2 M a^2 r \sin^2\theta}{\Sigma} \right) 
\sin^2\theta d\phi^2 \, ,
\ee
where $a = J/M$, $\Sigma = r^2 + a^2 \cos^2 \theta$, and $\Delta = r^2 - 2 M r + a^2$. The dimensionless spin parameter\footnote{Throughout this review article, I use $a_*$ to indicate the dimensionless spin parameter of the spacetime, namely $a_* = J/M^2$. In the Kerr metric and in the non-Kerr metrics discussed in this paper, $a/M$ coincide with $J/M^2$, but this is not universally true when we consider deviations from the Kerr solution. $J/M^2$ is a physically measurable quantity. In the general case, $a/M$ is a parameter of a metric to be indirectly inferred, and affecting more than just the asymptotical mass-current dipole moment of the spacetime.} is $a_* = a/M = J/M^2$.

In the Kerr metric in Boyer-Lindquist coordinates, the event horizon is defined by the larger root of $g^{rr} = 0$~\citep{MTW73,poisson}. The solution of $g^{rr} = 0$, which is equivalent to $\Delta = 0$, is
\be\label{eq-horizon}
r_\pm = M \pm \sqrt{M^2 - a^2} \, .
\ee
$r_+$ is the radius of the event horizon and ranges from $2M$, when $a_*=0$, to $M$, when $|a_*|=1$. $r_-$ is the radius of the inner horizon that turns out to be a Cauchy horizon. Metric perturbations are unstable for $r < r_-$~\citep{massinfl1,massinfl2}. This means that the interior solution of the Kerr metric breaks down for $r < r_-$. Deviations from the Kerr geometry should thus be expected already at $r \approx r_-$, which coincides with $r_+$ in the case of an extremal Kerr BH with $|a_*| = 1$, and not just at a Planck length distance from the central singularity. The exterior solution is regular (no spacetime singularities and closed time-like curves) only for $|a_*| \le 1$. When $|a_*| > 1$, there is no horizon and the spacetime describes a naked singularity, see Subsection~\ref{naked}.

The properties of the orbits around a BH are an important tool to connect possible observational effects with the geometry of the spacetime~\citep{bpt72}. Time-like circular orbits in the equatorial plane play a special role, because accretion disk models usually assume that the disk is in the equatorial plane and parallel to the BH spin (see Subsection~\ref{ss-cfm} and reference therein for more details).

Circular orbits in the equatorial plane exist only for radii larger than a critical radius, called the radius of the {\it photon orbit}~\citep{bpt72}
\be
r_\gamma = 2 M \left\{1 + \cos\left[ \frac{2}{3} \arccos \left(\mp \frac{a}{M}\right)\right]\right\} \, ,
\ee 
where here and in what follows the upper sign refers to corotating orbits (orbital angular momentum parallel to the BH spin), while the lower sign is for counterrotating orbits (orbital angular momentum antiparallel to the BH spin). No circular orbits exist for $r < r_\gamma$. Massive particles can be in the photon orbit in the limit in which they have infinite energy.

The {\it marginally bound circular orbit} separates unbound circular orbits ($r < r_{\rm mb}$) from bound circular orbits ($r > r_{\rm mb}$). The specific energy of a test-particle is $E > 1$ in unbound orbits (the particle has the energy to escape to infinity) and $E < 1$ in bound orbits. In the Kerr metric in Boyer-Lindquist coordinates, the radius of the marginally bound circular orbit is~\citep{bpt72}
\be
r_{\rm mb} = 2 M \mp a + 2 \sqrt{M \left(M \mp a\right)} \, .
\ee

\begin{figure}[t]
\begin{center}
\hspace{-1.428cm}
\includegraphics[type=pdf,ext=.pdf,read=.pdf,width=10cm]{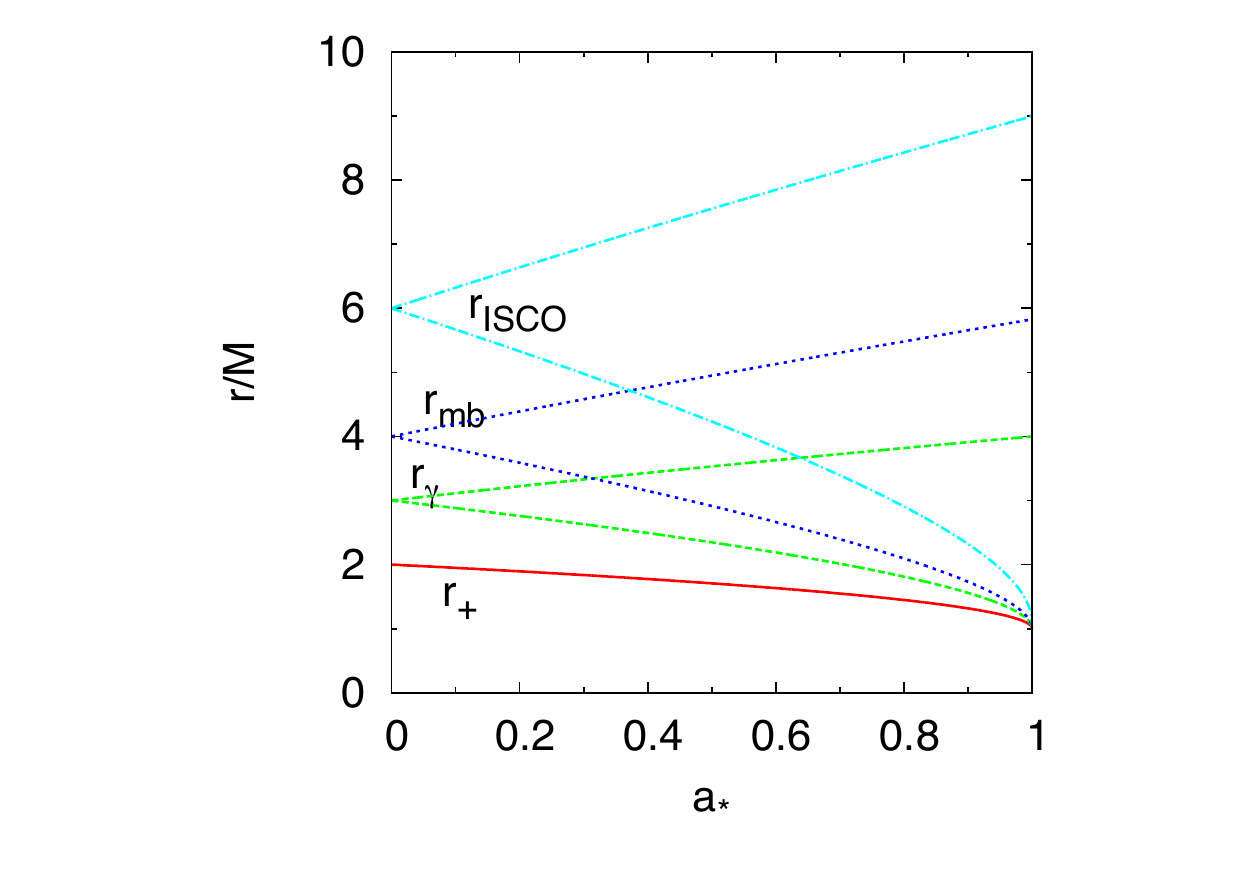}
\end{center}
\vspace{-0.5cm}
\caption{Radial coordinates of the event horizon $r_+$, the photon circular orbit $r_\gamma$, the marginally bound circular orbit $r_{\rm mb}$, and the ISCO $r_{\rm ISCO}$ in the Kerr metric as functions of the spin parameter $a_*$. For every radius, the upper curve refers to the counterrotating orbits, and the lower curve to the corotating orbits. }
\label{f-orbit}
\end{figure}

Finally, we address the topic of stable orbits. The critical radius separating unstable and stable circular orbits is the radius of the marginally stable circular orbit, $r_{\rm ms}$, more often called the radius of the {\it innermost stable circular orbit} (ISCO), $r_{\rm ISCO}$. In the Kerr metric in Boyer-Lindquist coordinates, we have~\citep{bpt72}
\be
r_{\rm ISCO} &=& \left[ 3 + Z_2 \mp 
\sqrt{\left(3 - Z_1\right)\left(3 + Z_1 + 2 Z_2\right)} \right] M \, , \nonumber\\
Z_1 &=& 1 + \left(1 - a_*^2\right)^{1/3} \left[ \left(1 + a_*\right)^{1/3} 
+ \left(1 - a_*\right)^{1/3}\right] \, , \nonumber\\
Z_2 &=& \sqrt{3 a^2_* + Z_1^2} \, .
\ee
In the Kerr metric, equatorial circular orbits are always vertically stable, while they are radially unstable for $r < r_{\rm ISCO}$.

Fig.~\ref{f-orbit} shows the radial coordinates of $r_+$, $r_\gamma$, $r_{\rm mb}$, and $r_{\rm ISCO}$ as functions of the BH spin parameter $a_*$. For every radius, the upper curve refers to the counterrotating orbits, and the lower curve to the corotating orbits. More details on the properties of the orbits in the Kerr metric can be found in~\citet{bpt72}. As briefly reviewed in Subsection~\ref{prop-nk}, in non-Kerr backgrounds there may be substantial differences, which may lead to specific observational signatures.

\subsection{Black hole candidates}

Astronomical observations have discovered a number of ``BH candidates''~\citep{narayan1}. In this article, I adopt quite a conservative attitude and I use this term to indicate very compact and massive objects, whose properties match those of BHs in general relativity. A BH is currently the least exotic hypothesis for most of these objects. However, many others adopt a less conservative perspective and call them BHs when there is a dynamical measurement of their mass (without testing the Kerr-ness of the metric or the possible existence of some kind of horizon), bestowing the term BH candidate to those objects for which there is no dynamical measurement of their mass but share common features observed in sources with a BH.

BH candidates are grouped into two classes: stellar-mass and supermassive BH candidates. There is probably a third class of objects, intermediate-mass BH candidates~\citep{r-imbh-cmc}, filling the mass gap between the stellar-mass and the supermassive ones, but the measurement of these objects is more difficult. They will not be considered in the following discussion, because they are currently unsuitable sources to test the Kerr metric with electromagnetic radiation.

\begin{figure*}
\vspace{-2.8cm}
\begin{center}
\includegraphics[type=pdf,ext=.pdf,read=.pdf,width=16cm]{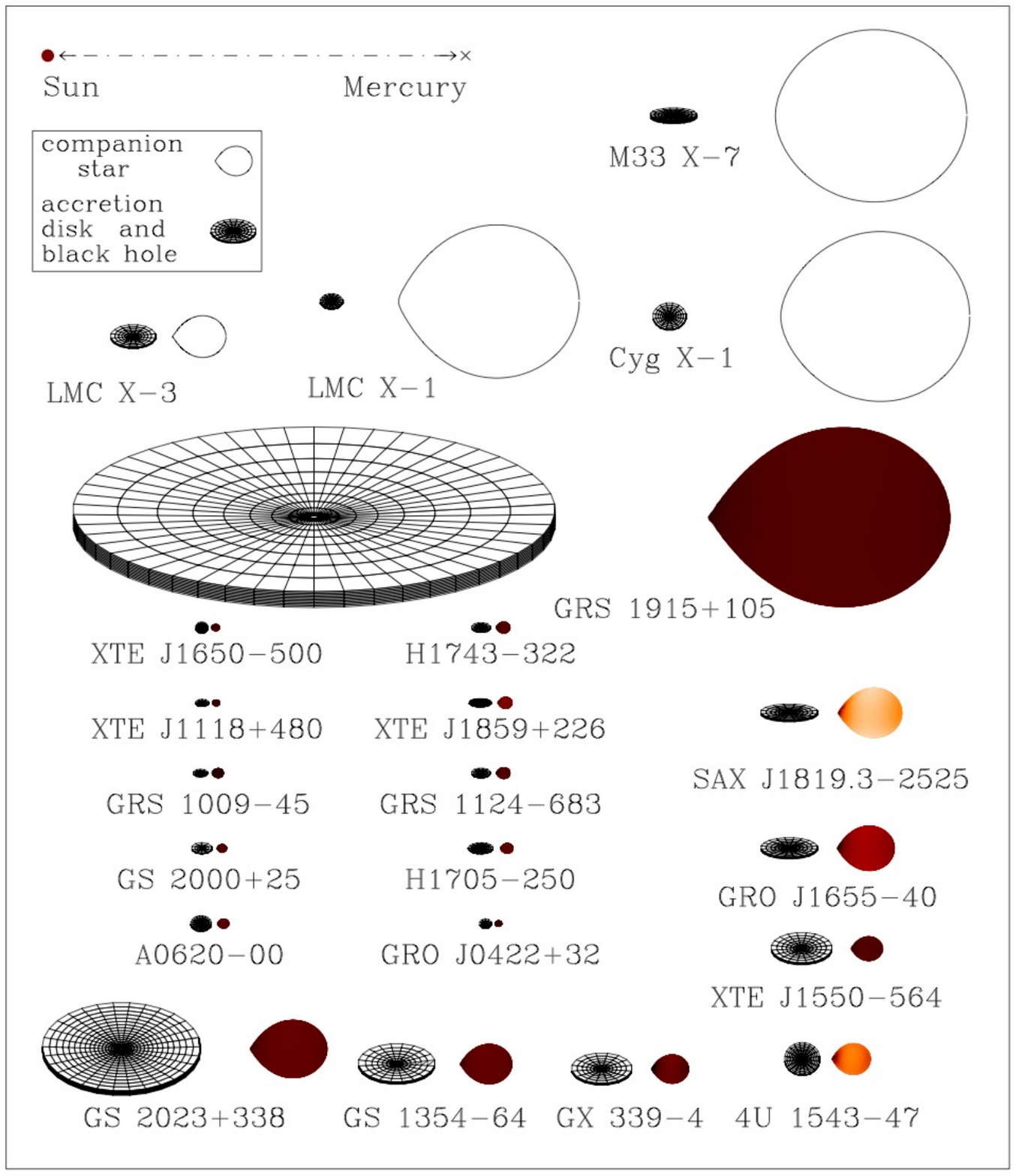}
\end{center}
\vspace{-3.2cm}
\caption{Sketches of 22 binaries with a stellar-mass BH candidate confirmed by dynamical measurements. For every system, the BH accretion disk is on the left and the companion star is on the right. The orientation of the disks indicates the inclination angles $i$ of the binaries. The distorted shapes of the stellar companions are due to the gravitational fields of the BH candidates. The size of the latter should be about 50~km, compared with the distance of the Sun to Mercury of about 50~million km and the radius of the Sun of 0.7~million km (top left corner). Figure courtesy of Jerome Orosz.}
\label{f-bhb}
\end{figure*}

\subsubsection{Stellar-mass black hole candidates}

In general relativity, the maximum mass for a compact star made of neutrons, mesons, or quarks for plausible matter equations of state is about 3~$M_\odot$~\citep{rr74,kb96,v3-lattimer}\footnote{In alternative theories of gravity, this bound may somewhat change~\citep{ns-atg,dandon}.}. Within standard physics, if an object is compact and exceeds 3~$M_\odot$ it can only be a BH and is classified as a BH candidate. In this context, a reliable mass measurement is thus crucial. BH candidates with a mass ranging from 3 to about 100~$M_\odot$ can be called stellar-mass BH candidates.

Almost all the currently known stellar-mass BH candidates are in X-ray binaries. Their masses range from about 5 to about 20~$M_\odot$~\citep{mcrem06}. We can infer that they are compact from their short timescale variability. Their mass can instead be measured by studying the orbital motion of the companion star, typically with optical observations. These systems can be studied within the framework of Newtonian mechanics, because the companion star is always far from the BH candidate\footnote{Systems in which the companion star is very close to the BH candidate may not exist, because the strong gravitational field around the compact object would disrupt an ordinary star.}. The key-quantity is the mass function~\citep{v3-casares}
\be\label{eq-fffff}
f (M) = \frac{K_c^3 T}{2 \pi G_{\rm N}} = \frac{M^3 \sin^3 i}{\left(M + M_c\right)^2} \, ,
\ee
where $K_c = v_c \sin i$ is the maximum line-of-sight Doppler velocity of the companion star, $v_c$ is the velocity of the companion star, $i$ is the inclination angle of the orbital plane with respect to the line of sight of the observer, $T$ is the orbital period, and $M$ and $M_c$ are the mass of the compact object and of the stellar companion, respectively. Here I reintroduced $G_{\rm N}$ because this formula is usually presented in this form. If we can get an independent estimate of $i$ and $M_c$ and we can measure $K_c$ and $T$, it is possible to determine the mass of the compact object $M$. Note that $T$ and $K_c$, and therefore the measurement of the mass function, can be obtained from light curves and spectroscopy. Moreover, from the right-hand side in Eq.~(\ref{eq-fffff}), we can see that $f(M) < M$; that is, from the estimate of the mass function we can directly infer a lower bound on the mass of the dark object. When the mass $M$ of a compact object exceeds 3~$M_\odot$, the object is classified as a BH candidate\footnote{Massive stars exceeding the bound 3~$M_\odot$ are not compact objects and they can be identified by optical observations.}.

Today we have 24 ``confirmed'' BH candidates in X-ray binaries, where the term confirmed is used to indicate that there are dynamical measurements of their masses and they exceed the 3~$M_\odot$ bound. Actually these masses are typically higher than 5~$M_\odot$, and in some cases the same mass function $f(M)$ exceeds 5~$M_\odot$, which means the mass of the compact object is larger than 5~$M_\odot$ independent of the estimate of the mass of the companion star $M_c$ and the viewing angle of the orbit $i$ [because $M > f(M)$]. Fig.~\ref{f-bhb} is a schematic diagram of 22 binaries with a dynamically confirmed BH candidate. In addition to the 24 confirmed BH candidates, we currently know another $\sim$40~BH candidates in X-ray binaries that are without a dynamical measurement of their masses. They are nevertheless classified as BH candidates since they have features of BH candidates, but it is possible that some of them are actually neutron stars.

The presently known stellar-mass BH candidates in binaries are mainly in our Galaxy, and a minor number are in nearby galaxies. Note that we expect a population of about $10^8$-$10^9$ stellar-mass BHs in our Galaxy~\citep{v2-remnant}. So we presumably know only a very small fraction of the stellar-mass BHs in our Galaxy and this is because the detection of these objects is only possible under some special conditions related to the properties of the binary system. A list of confirmed and unconfirmed stellar-mass BH candidates in binaries with their properties can be found in~\citet{ozel}.

X-ray binaries can be grouped into 2 classes: low-mass X-ray binaries and high-mass X-ray binaries. Low-mass X-ray binaries are systems in which the stellar companion is not more than a few Solar masses ($\lesssim 3$~$M_\odot$) and the mass transfer occurs by Roche lobe overflow~\citep{v3-sav}. These systems are typically ``transient'' X-ray sources because the mass transfer is not continuous. For instance, a similar system may be bright for some months and then be in a quiescent state for decades. High-mass X-ray binaries are systems in which the stellar companion is massive ($\gtrsim 10$~$M_\odot$) and the mass transfer from the companion star to the BH is due to the wind from the former. These systems are ``persistent'' X-ray sources. Among the 24 confirmed BH candidates in X-ray binaries, 6 are persistent sources (Cyg~X-1, LMC~X-1, LMC~X-3, M33~X-7, NGC~300-1, and IC~10~X-1) and the other 18 sources are transient.

In addition to BHs in binary systems with ordinary stars, we expect also the existence of isolated BHs and BHs in binary systems with neutron stars or other BHs. Attempts to identify isolated BHs are based on microlensing techniques~\citep{v3-micro1,v3-micro2}, but these observations are difficult and there are currently only some weak candidates. BHs in binary systems with neutron stars or other BHs can potentially be discovered with gravitational waves in the next years. The LIGO/Virgo collaboration recently announced the detection of gravitational waves from a binary system in which two BHs of about 30~$M_\odot$ merged into a BH of $\sim 60$~$M_\odot$~\citep{v3-ligo}. The detection of another binary BH merger was reported in~\citet{v4-gw}. Gravitational wave detectors now promise to discover tens of stellar-mass BH candidates in the next few years. For the time being, we do not know any BH candidate in a binary in which the stellar companion is a pulsar, but a similar system is not expected to be too rare and there are efforts to find BH binaries with pulsar companions with radio observations.

\subsubsection{Supermassive black hole candidates}

Supermassive BH candidates have a mass in the range $M \sim 10^5 - 10^{10}$~$M_\odot$~\citep{kormendy}. They are harbored in the center of galaxies, and it is supposed that any normal galaxy has a supermassive BH at its center, namely any galaxy that is not too small\footnote{Exceptions might be possible. For example, the galaxy A2261-BCG has a very large mass but it might not have any supermassive BH at its center~\citep{a2261}.}. Small galaxies usually do not have a supermassive BH. Once again, these objects are classified as BH candidates because, from the estimate of their mass and volume, a BH is the least exotic hypothesis for most of these objects. In general, their mass can be dynamically measured by the motion of the orbiting gas. The estimate of the volume follows from the timescale variability of these sources.

The strongest evidence for the existence of a supermassive BH comes from the center of our Galaxy. Much work in this direction was done by Genzel and his group~\citep{v2-genzel1,v2-genzel2,v2-genzel3,v2-genzel4}. From the study of the Newtonian motion of individual stars, we can infer that the mass of the compact object is about $4\cdot 10^6$~$M_\odot$. The upper bound on the size of the BH candidate comes from the minimum distance approached by one of these stars, which is less than 45~AU and corresponds to 600~$M$ for a $4\cdot 10^6$~$M_\odot$ BH~\citep{ghez}. With such a large mass in a relatively small volume, such an object cannot be a cluster of non-luminous compact bodies such as neutron stars. This is because the lifetime of the cluster due to evaporation and physical collisions would be shorter than the age of this system~\citep{maoz}. The most natural interpretation is that it is a BH and, according to general relativity, the spacetime geometry around an astrophysical BH should be well approximated by the Kerr metric.

\subsubsection{Black hole horizon}

The key-feature that makes an object a BH is the existence of an event horizon. Interestingly, there are some observations that show that BH candidates have no ``normal'' surface but instead have something that may indeed be an event horizon. In particular:
\begin{enumerate}

\item Most X-ray binaries are transient sources: therefore they spend a long period in a quiescent state with a very low mass accretion rate and luminosity. This is true for X-ray binaries containing either neutron stars or BH candidates. However, it turns out that BH candidates can be extremely underluminous in comparison to neutron stars in the quiescent state. This should indeed be expected, because the thermal energy locked in the gas can be completely lost when the gas crosses the horizon, while in the case of the neutron star the gas hits the surface of the compact object and can subsequently release energy~\citep{eeb2}. More interestingly, during the quiescent period, neutron star binaries show a thermal blackbody-like component in the X-ray band, which is interpreted as the emission from the neutron star surface. No similar component is found in binaries with a BH candidate~\citep{eeb1}. The fact that there has been no detection of any thermal component may be interpreted as evidence for the absence of a normal surface and therefore the existence of a horizon.

\item In the case of SgrA$^*$, the supermassive BH candidate at the center of the Galaxy, millimeter and infrared observations strongly constrain any possible thermal blackbody component emitted from the surface of this object~\citep{eea1,grava-eh}. Again, the non-detection of a thermal component is consistent with the fact that the object has no surface but instead possesses an event horizon.

\item In the case of a neutron star, the gas of accretion accumulates on the neutron star surface and eventually develops a thermonuclear instability. The result is a thermonuclear explosion called a Type~I X-ray burst. The phenomenon seems to be well understood and theoretical models agree well with observations. These X-ray bursts are observed from sources that are supposed to be neutron stars, and they have never been observed from BH candidates~\citep{tournear}. This should indeed be the case if BH candidates have an event horizon, because no gas can accumulate on their surface~\citep{eec1,yifei}.

\end{enumerate}

Bearing in mind the definition of a BH, see Subsection~\ref{ss-kerr}, it is fundamentally impossible to test the existence of an event horizon. These observations may instead suggest that BH candidates have an apparent horizon. See also~\citet{hhh1} and \citet{hhh2} on this point. In particular, there are theoretical frameworks in which a BH never forms, instead, only an apparent horizon can be created~\citep{frolov81,bmm,v3-bmm2}. Another issue is that the non-detection of a thermal component does not necessarily indicate the absence of a surface. There are scenarios, such as those in~\citet{q1,q2}, in which a ``classical'' BH is actually a Bose-Einstein condensate of gravitons. If this were the case, one should not expect any blackbody-like component from the BH surface.


\section{Testing the Kerr metric \label{s2-tests}}

\subsection{Non-Kerr metrics}

Tests of the Schwarzschild metric in the weak field limit are commonly and conveniently discussed within the PPN (Parametrized Post-Newtonian) formalism, see e.g.~\citet{will}. We write the most general line element for a static and spherically symmetric spacetime in terms of an expansion in $M/r$, namely
\be\label{eq-ppn}
ds^2 &=& - \left(1 - \frac{2 M}{r} + \beta\frac{2 M^2}{r^2} + . . .  \right) dt^2 
\nonumber\\ &&
+ \left(1 + \gamma \frac{2 M}{r} + . . .  \right) \left(dx^2 + dy^2 + dz^2 \right) \, ,
\ee
where $\beta$ and $\gamma$ are two coefficients that parametrize our ignorance. In the isotropic coordinates of Eq.~(\ref{eq-ppn}), the Schwarzschild solution has $\beta = \gamma = 1$~\citep{v3-eddington}\footnote{The PPN approach is traditionally formulated in isotropic coordinates and the line element of the most general static and spherically symmetric distribution of matter is given by Eq.~(\ref{eq-ppn}). When arranged in the more familiar Schwarzschild coordinates, the line element becomes
\be
ds^2 &=& - \left(1 - \frac{2 M}{r} + \left(\beta - \gamma\right)\frac{2 M^2}{r^2} + . . .  \right) dt^2 \nonumber\\ 
&&+ \left(1 + \gamma \frac{2 M}{r} + . . .  \right) dr^2 + r^2 d\theta^2 + r^2 \sin^2\theta d\phi^2 \, .
\ee}. In Solar System experiments, one assumes that $\beta$ and $\gamma$ are free parameters to be determined by observations. Current observational data provide the following constraints~\citep{v3-ppn1,v3-ppn2}
\be\label{eq-ppn2}
|\beta - 1| < 2.3 \cdot 10^{-4} \, , \quad
|\gamma - 1| < 2.3 \cdot 10^{-5} \, ,
\ee
confirming the validity of the Schwarzschild solution in the weak field limit within the precision of current observations. For the details on how the constraints in Eq.~(\ref{eq-ppn2}) are obtained, see the original papers \citet{v3-ppn2} and \citet{v3-ppn1}.

We can attempt to use the same reasoning to test the Kerr metric around BH candidates. We first consider a metric more general than the Kerr solution. Deviations from Kerr are parametrized by a number of ``deformation parameters'', which are unknown constants to be determined by observations. If the latter require vanishing deformation parameters, the Kerr BH hypothesis is satisfied. If observations require that at least one deformation parameter is non-vanishing, BH candidates may not be the Kerr BHs of general relativity.

In the Kerr case, it is problematic to find a very general solution counterpart of that in Eq.~(\ref{eq-ppn}). The point is that we want to test the strong gravitational field near BH candidates and therefore we cannot use an expansion in $M/r$. Moreover, the metric is now stationary and axisymmetric rather than static and spherically symmetric. In the ideal case, we want to have a sufficiently general metric that can reduce to any non-Kerr BH in any (known and unknown) alternative theory of gravity for a specific choice of the values of its deformation parameters. At the moment, such a suitable metric with a countable number of degrees of freedom does not exist: nevertheless there are some proposals in the literature.

A very popular choice is the Johannsen-Psaltis (JP) metric~\citep{jp-m}. It is an {\it ad hoc} metric and is not a solution of a specific alternative theory of gravity. In Boyer-Lindquist coordinates, the line element reads
\be\label{eq-jp}
ds^2 &=& - \left(1 - \frac{2 M r}{\Sigma}\right)\left(1 + h\right) dt^2 \nonumber\\ &&
 - \frac{4 M a r \sin^2\theta}{\Sigma}\left(1 + h\right) dt d\phi \nonumber\\ &&
+ \frac{\Sigma \left(1 + h\right)}{\Delta + h a^2 \sin^2\theta} dr^2
+ \Sigma d\theta^2 \nonumber\\
&& + \Bigg[r^2 + a^2 + \frac{2 a^2 M r \sin^2\theta}{\Sigma} \nonumber\\ && \;\;\;\;\;\;
+ \frac{a^2 \left(\Sigma + 2 M r\right) \sin^2\theta}{\Sigma} h \Bigg] 
\sin^2\theta d\phi^2 \, ,
\ee
where $h$ is 
\be\label{eq-jp-hhh}
h &=& \sum_{k=0}^{+\infty} \left(\epsilon_{2k}
+ \epsilon_{2k+1} \frac{M r}{\Sigma}\right)\left(\frac{M^2}{\Sigma}
\right)^k \, .
\ee
The metric has an infinite set of deformation parameters $\{\epsilon_k\}$ and it reduces to the Kerr solution when all the deformation parameters vanish. However, $\epsilon_0$ must vanish in order to recover the correct Newtonian limit, while $\epsilon_1$ and $\epsilon_2$ are already strongly constrained by experiments in the Solar System through Eq.~(\ref{eq-ppn2})~\citep{cpr-m}. The simplest non-trivial metric is thus that with $\epsilon_3$ free and with all the other deformation parameters set to zero\footnote{The JP metric is sometimes criticized because of its derivation. It is obtained by applying the Newman-Janis prescription, although it is not guaranteed that such an algorithm works outside general relativity~\citep{ny-nja}. Moreover, the transformation to eliminate some off-diagonal terms is not correct, see the discussion in~\citet{azreg}.}.

In~\cite{jp-m}, $h$ is introduced as an arbitrary function of $r$ describing deviations from the Schwarzschild solution. The rotating JP metric is obtained by applying the Newman-Janis prescription~\citep{v2-nj-alg,v2-nj-alg2}, and $h$ becomes a function of both $r$ and $\theta$. If we believe that deviations from the Kerr geometry must be small, which is consistent with the fact that the known non-Kerr BH solutions in alternative theories of gravity are close to the Kerr one, $h$ is always a small quantity. In this case, all the deformation parameters $\{\epsilon_k\}$ are small and it is possible to consider only the unconstrained leading order term, neglecting all the others. Alternatively, one may adopt a more phenomenological approach, on the basis that whatever is not forbidden may be allowed, and consider the JP metric with one or a few deformation parameters, which are not necessarily small, and simply try to constrain that specific choice of the metric. In the following sections, I adopt this more relaxed point of view, and the deformation parameters will be allowed to have any value permitted by observations without the restrictions of being small quantities.

The weak field tests discussed in Subsections~\ref{v2-weak} and \ref{s5-weak} can potentially constrain $\epsilon_3$, but can unlikely test higher order terms. In particular, if we believe in the interpretation of~\cite{valtonen1}, $\epsilon_3$ would already be constrained to a small value. Strong field tests, particularly in those sources in which the inner edge of the disk is very close to the compact object, can instead constrain even high order deformation parameters, because at the inner edge of the disk $M/r$ can be close to 1 and there is not much difference between lower and higher order terms. In the case of strong field tests, with a phenomenological metric like that in Eq.~(\ref{eq-jp}), it is common to consider just one deformation parameter without the restriction of being small.

\newpage

An extension of the JP metric is the Cardoso-Pani-Rico (CPR) parametrization~\citep{cpr-m}. The line element is
\begin{widetext}
\be\label{eq-cpr}
ds^2 &=& - \left(1 - \frac{2 M r}{\Sigma}\right)\left(1 + h^t\right) dt^2
- 2 a \sin^2\theta \left[\sqrt{\left(1 + h^t\right)\left(1 + h^r\right)} 
- \left(1 - \frac{2 M r}{\Sigma}\right)\left(1 + h^t\right)\right] dt d\phi \nonumber\\
&& + \frac{\Sigma \left(1 + h^r\right)}{\Delta + h^r a^2 \sin^2\theta} dr^2
+ \Sigma d\theta^2 
+ \sin^2\theta \left\{\Sigma + a^2 \sin^2\theta \left[ 2 \sqrt{\left(1 + h^t\right)
\left(1 + h^r\right)} - \left(1 - \frac{2 M r}{\Sigma}\right)
\left(1 + h^t\right)\right]\right\} d\phi^2 \, , \quad
\ee
\end{widetext}
where 
\be
h^t &=& \sum_{k=0}^{+\infty} \left(\epsilon_{2k}^t 
+ \epsilon_{2k+1}^t \frac{M r}{\Sigma}\right)\left(\frac{M^2}{\Sigma}
\right)^k\, , \\
h^r &=& \sum_{k=0}^{+\infty} \left(\epsilon_{2k}^r
+ \epsilon_{2k+1}^r \frac{M r}{\Sigma}\right)\left(\frac{M^2}{\Sigma}
\right)^k \, .
\ee
Now there are two infinite sets of deformation parameters, $\{\epsilon_k^t\}$ and $\{\epsilon_k^r\}$. The JP metric is recovered when $\epsilon_k^t = \epsilon_k^r$ for all $k$, and the Kerr metric when $\epsilon_k^t = \epsilon_k^r = 0$ for all $k$.

There are also other proposals in the literature. For instance, the quasi-Kerr metric of~\citet{tests-gw1}, which is based on a multipole moment expansion and the mass-quadrupole moment of the object is
\be
Q = Q_{\rm Kerr} - \epsilon M^3 \, ,
\ee
where $Q_{\rm Kerr} = - J^2/M$ is the mass-quadrupole in the Kerr case and $\epsilon$ is the deformation parameter. Another framework is that of the bumpy BHs~\citep{bumpy1,bumpy2}.

An appealing parametrization, which has not been explored very far, is represented by the family of regular BH metrics of~\citet{reg-j1,reg-j2}, which is a generalization of the family of metrics proposed in~\citet{vigeland-y-s}. Here ``regular'' is to indicate that these spacetimes have no naked singularities or closed time-like curves outside of the event horizon and the equations of motion are separable (as in the Kerr metric, but it is not true in general).

Recently, several others proposed new parametrizations to test the Kerr metric. \citet{zhidenko} and \citet{v3-krz} suggested a parametrization that seems to fairly well reproduce some known non-Kerr BH solutions in alternative theories of gravity with a small number of deformation parameters. \citet{v3-nan} introduced a parametrization suitable for the calculations of the electromagnetic spectrum from thin accretion disks. \citet{v3-masume} discussed a test-metric in which the free parameters are all equal to 1 to recover the Kerr metric and are different from 1 in the case of deviations from it.

Tests of the Kerr metric may also employ the Manko-Novikov metric~\citep{manko1}, which was not originally proposed to test BH candidates and it is an exact solution of the vacuum Einstein equations with arbitrary mass-multipole moments\footnote{A particular choice of the mass-multipole of order $n$ fixes the value of the current-multipole moment of the same order, which is completely determined by the former for given mass-monopole and current-dipole moments~\citep{manko1}.}. Here the no-hair theorem does not apply because the exterior spacetime is not regular, namely there are closed time-like curves and naked singularities. There are also some variations, such as the Manko-Mielke-Sanabria-G{\'o}mez solution~\citep{manko2}, which can be extended to include fast-rotating objects with $J/M^2 > 1$~\citep{spin-c2}.

In this subsection we have discussed some model-independent parametrizations. An alternative strategy is to consider a specific non-Kerr BH solution from a known alternative theory of gravity rather than a general ansatz like those in~(\ref{eq-jp}) and (\ref{eq-cpr}). Such an approach is clearly much less general. We simply compare two background metrics: the Kerr solution against the metric of the alternative theory of gravity of interest. However, there is another issue. Usually we know non-rotating solutions in alternative theories of gravity, and in a small number of cases we know the rotating solution in the slow-rotation approximation. Rotating non-Kerr BH solutions in alternative theories of gravity are difficult to find. The spin plays a crucial role in the properties of the radiation emitted close to BH candidates and therefore, without the rotating solution, it is not possible to test the Kerr metric, because we are not able to distinguish the effects of the spin from those due to possible deviations from the Kerr geometry.

\subsection{Basic properties of non-Kerr metrics \label{prop-nk}}

General statements or properties that hold in the case of the Kerr metric may not be true in other backgrounds. This may lead to new phenomena with specific astrophysical implications and observational signatures.

The uniqueness of the Kerr solution in general relativity is valid under some specific assumptions and it can be violated in the presence of exotic fields~\citep{v2-hr}. An alternative theory of gravity may potentially have several kinds of BHs, which may be created by gravitational collapse from different initial conditions. In a similar context, it is possible that astrophysical BH candidates are not all of the same type and therefore the possible confirmation that a specific object is a Kerr BH does not necessarily imply that all BH candidates are Kerr BHs.

\subsubsection{Horizon}

An event horizon is a null surface in spacetime. If we introduce a scalar function $f$ such that at the event horizon $f=0$, the normal to the event horizon is $n^\nu = \partial^\mu f$ and is a null vector. The condition for the surface $f=0$ to be null is thus~\citep{pat5,pat3}
\be\label{eq-v3-hh}
g^{\mu\nu} \left(\partial_\mu f\right) \left(\partial_\nu f\right) = 0 \, .
\ee
In general, one can find the event horizon by integrating null geodesics backward in time, see~\citet{pat5} and \citet{pat3} for details. In the case of a stationary and axisymmetric spacetime, the procedure can significantly simplify. In a coordinate system adapted to the two Killing isometries (stationarity and axisymmetry), and such that $f$ is also compatible with the Killing isometries, Eq.~(\ref{eq-v3-hh}) reduces to 
\be\label{eq-v3-hh2}
g^{rr} \left(\partial_r f\right)^2 + 2 g^{r\theta} \left(\partial_r f\right) 
\left(\partial_\theta f\right) + g^{\theta\theta} \left(\partial_\theta f\right)^2 = 0
\ee
in spherical-like coordinates $\left(t,r,\theta,\phi\right)$. The surface must be closed and non-singular (namely geodesically complete) in order to be an event horizon and not just a null surface.

If we assume that there is a unique horizon radius for any angle $\theta$ (Strahlk\"orper assumption), we can write $f$ as $f = r - H(\theta)$, where $H(\theta)$ is a function of $\theta$ only and the event horizon is $r_H = H(\theta)$; see~\citet{pat3} and \citet{pat5} for details and the limitations of the Strahlk\"orper assumption. The problem is thus reduced to finding the solution of the differential equation
\be\label{eq-h-rt-j}
g^{rr} + 2 g^{r\theta} \left(\frac{dH}{d\theta}\right) 
+ g^{\theta\theta} \left(\frac{dH}{d\theta}\right)^2 = 0 \, .
\ee
The event horizon equation $g^{rr}=0$ valid in the Kerr spacetime in Boyer-Lindquist coordinates only holds when the surfaces $r = {\it const.}$, which must be closed, have a well-defined causal structure, in the sense that the surfaces $r = {\it const.}$ are null, space-like, or time-like~\citep{pat2}. Such a condition clearly depends on the coordinate system, so the event horizon equation $g^{rr}=0$ can only be valid with certain coordinates.

A {\it Killing horizon} is a null hyper-surface on which there is a null Killing vector field. In a stationary and axisymmetric spacetime and employing a coordinate system adapted to the two Killing isometries, the Killing horizon is given by the largest root of 
\be\label{eq-kill}
g_{tt} g_{\phi\phi} - g_{t\phi}^2 = 0 \, .
\ee
In general relativity, the Hawking rigidity theorem shows that the event and the Killing horizons coincide~\citep{haw-r}, so Eq.~(\ref{eq-h-rt-j}) and Eq.~(\ref{eq-kill}) provide the same result. In alternative theories of gravity, this is not guaranteed~\citep{pat2}.

In general relativity, the event horizon must have $S^2 \times \bf{R}$ topology, and even this property is regulated by certain theorems~\citep{haw-r,jacobson}. For instance, toroidal horizons can form, but they can only exist for a short time, in agreement with these theorems~\citep{teuk}. If we want to test the Kerr metric and general relativity, we cannot exclude scenarios with BHs with a topologically non-trivial event horizon~\citep{pat1,pat2}.

\subsubsection{Particle orbits}

In a generic stationary and axisymmetric spacetime, there are three constants of motion, namely the mass $m$, the energy $E$, and the axial component of the angular momentum $L_z$. The Kerr solution is a Petrov type~D spacetime and therefore there is a fourth constant $\mathcal{Q}$ called the Carter constant~\citep{carter68}\footnote{In the Schwarzschild limit ($a_*=0$), the Carter constant reduces to $\mathcal{Q} = L^2 - L_z^2$, where $L$ is the total angular momentum. In the Kerr metric, there is not a direct physical interpretation of $\mathcal{Q}$.}. If the spacetime has a forth constant of motion, it is always possible to choose a coordinate system in which the equations of motion are separable. In the Kerr metric, this is the case for the Boyer-Lindquist coordinates and the calculations of the photon trajectories from the observer to the region around the BH can be reduced to the calculations of some elliptic integrals, which is computationally more efficient. In generic spacetimes with a Carter-like constant, these calculations are already more complicated and, in the simplest case, the elliptic integrals of the Kerr metric become hyper-elliptic integrals~\citep{v2-hyper}. If the spacetime has no Carter-like constant, it is necessary to solve a system of coupled second order differential equations.

In accretion disk models and astrophysical measurements, the ISCO radius plays an important role. For instance, in the Novikov-Thorne model~\citep{nt1}, which is the standard set-up for thin accretion disks, the inner edge of the disk is at the ISCO radius and it is the crucial parameter to measuring the spin or possible deviations from the Kerr solution. In the Kerr metric, equatorial circular orbits are always vertically stable, whereas they are only radially stable for $r > r_{\rm ISCO}$. In non-Kerr metrics, the picture is more complicated. The ISCO may be either radially or vertically marginally stable~\citep{cfm-c1,bb2011}, and it is also possible to have disconnected stable regions among which equatorial orbits are unstable~\citep{glg}. Moreover, in the Kerr metric the usual picture is that a particle\footnote{It is worth pointing out that, in this discussion and in what follows, ``particle'' is used to refer to a parcel of gas.} of the accretion disk reaches the ISCO and then quickly plunges onto the BH, without emitting much radiation after leaving the ISCO. In non-Kerr metrics, it is possible that the particle plunges to a region inside the ISCO, but it cannot be immediately swallowed by the BH~\citep{bb2011}. On the contrary, there may be an accumulation of gas between the ISCO and the compact object, and the gas has to lose additional energy and angular momentum before plunging to the central body. There are also spacetimes in which there is no ISCO, namely the orbits are always stable. For instance, this is the case of some metrics describing the spacetime around certain exotic non-interacting matter~\citep{no-isco1,exotic1,v2-mpcc}.

In the Kerr metric, the ISCO radius is located at the minimum of the energy of equatorial circular orbits\footnote{This can be seen as follows. The specific energy of a test-particle in an equatorial circular orbit is given, for instance, in Eq.~(2.12) in \citet{bpt72}. Its derivative with respect to the radial coordinate is
\be
\frac{dE}{dr} = \frac{r^2 - 6 M r + 8 a M^{1/2} r^{1/2} - 3 a^2}{2 r^{7/4} \left( r^{3/2} - 3 M r^{1/2} + 2 a M^{1/2} \right)^{3/2}} \, .
\ee 
The minimum can be found from $dE/dr = 0$, which is the same equation as that for the orbital stability, see e.g. \citet{bpt72} or \citet{v3-chandra}.}. At larger radii, the specific energy monotonically increases to approach 1 at infinity. From the ISCO radius to smaller radii, the specific energy monotonically increases to diverge to infinity at the photon orbit. At the radius of the marginally bound circular orbit, which is located between the photon radius and the ISCO radius, the specific energy of a test-particle is $E=1$. When the ISCO is marginally vertically unstable, the energy of equatorial circular orbits may be a monotonic function without minimum (the energy decreases as the radial coordinate decreases). In this case, there is no marginally bound circular orbit. Accretion disk structures around similar BHs are qualitatively different~\citep{z1}.

\begin{figure*}
\begin{center}
\includegraphics[type=pdf,ext=.pdf,read=.pdf,width=8.5cm]{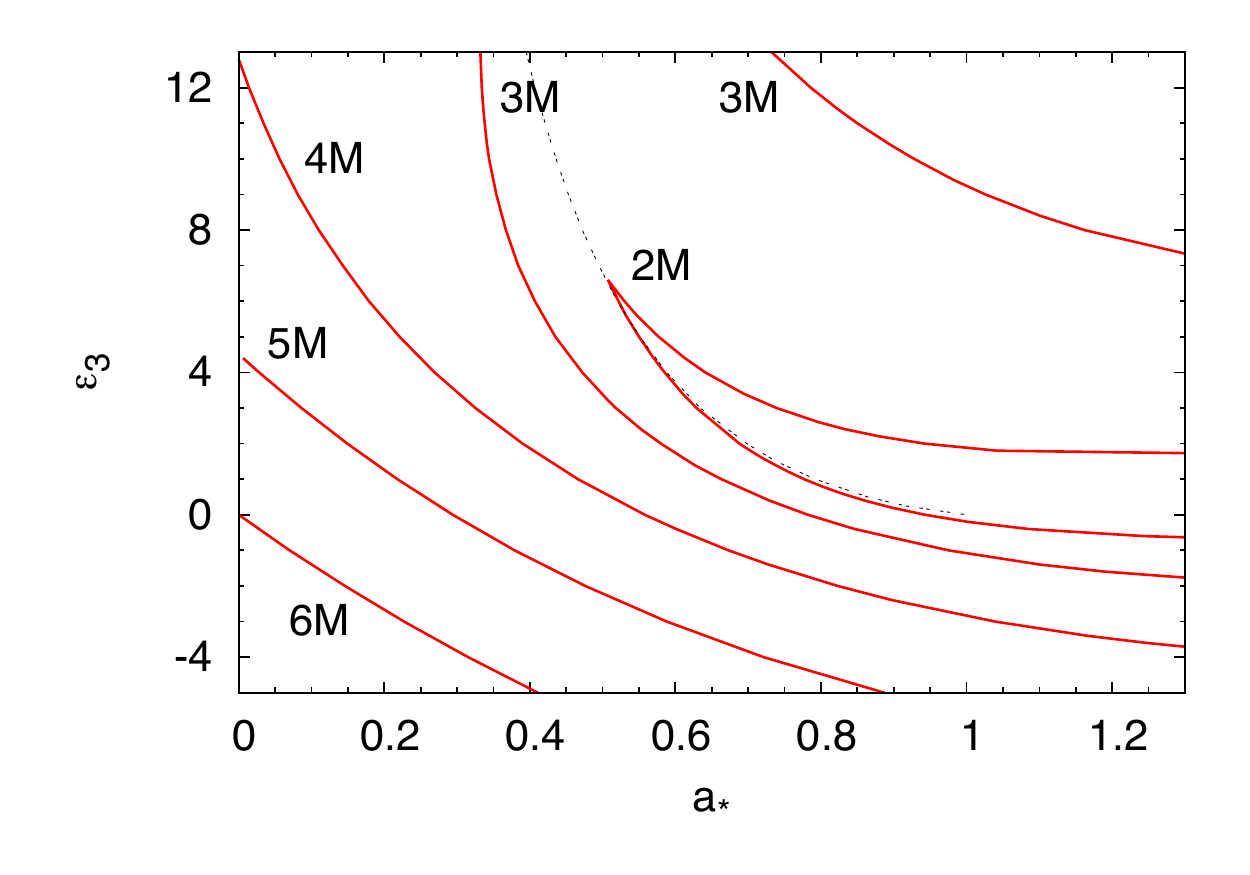}
\includegraphics[type=pdf,ext=.pdf,read=.pdf,width=8.5cm]{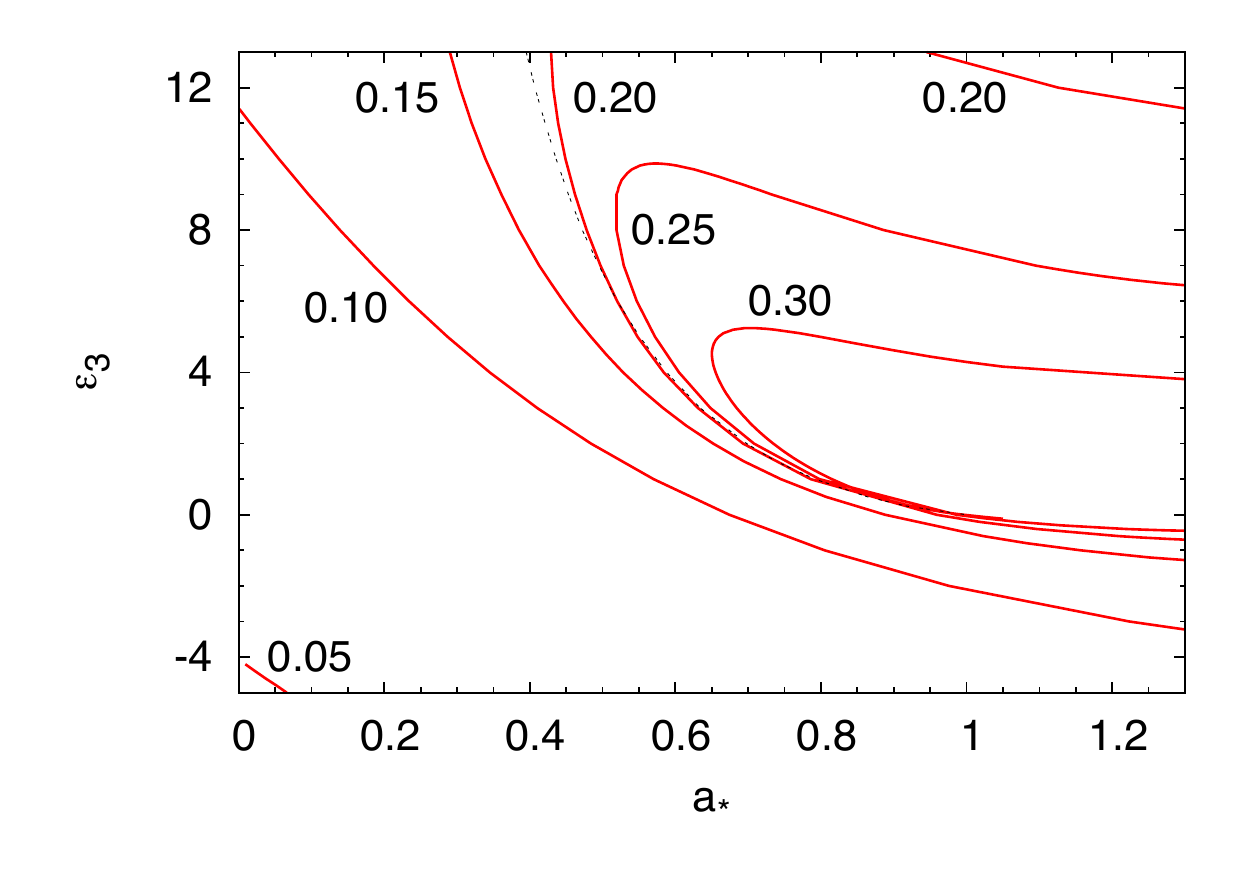}
\end{center}
\vspace{-0.5cm}
\caption{Contour levels of the ISCO radius $r_{\rm ISCO}$ (left panel) and of the Novikov-Thorne radiative efficiency $\eta_{\rm NT} = 1 - E_{\rm ISCO}$ (right panel) in the JP metric with non-vanishing deformation parameter $\epsilon_3$. The black dotted line separates spacetimes with a regular event horizon (on the left of the line) from those with topologically non-trivial event horizons and naked singularities (on the right). See the text for more details. }
\label{f-isco}
\end{figure*}

Since astrophysical observations are often sensitive to the position of the ISCO, it is useful to have an idea of the correlation between the spin and the deformation parameter in the determination of the ISCO radius. For instance, \citet{iron-jp} showed how iron lines in the JP metric with different $a_*$ and $\epsilon_3$ but the same ISCO radius are very similar, as well as orbital fundamental frequencies in the JP metric are correlated to the position of the ISCO radius. The left panel of Fig.~\ref{f-isco} shows the contour levels for $r_{\rm ISCO} = 2\,M$, $3\,M$, $4\,M$, $5\,M$, and $6\,M$ on the plane spin parameter $a_*$ and JP deformation parameter $\epsilon_3$. As already mentioned, throughout this article I do not impose that the deformation parameter must be a small quantity.

For very large and positive $\epsilon_3$, the ISCO is at a larger radius because of the strong instability along the vertical direction. The contour levels of the ISCO can give a simple idea of which spacetimes may look similar in astrophysical observations. The right panel in Fig.~\ref{f-isco} shows the Novikov-Thorne radiative efficiency $\eta_{\rm NT} = 1 - E_{\rm ISCO}$, where $E_{\rm ISCO}$ is the specific energy of a test-particle at the ISCO radius. The contour levels are for $\eta_{\rm NT} = 0.05$, 0.10, 0.15, 0.20, 0.25, and 0.30. $\eta_{\rm NT}$ is indeed a more appropriate quantity than the ISCO radius. First, $\eta_{\rm NT}$ is independent of the coordinate system (the radial coordinate of the ISCO has no physical meaning) and it can be directly measured. Second, it is indeed a better estimator to figure out which spacetimes may look similar in astrophysical observations (see the two panels in Fig.~\ref{f-isco} and compare the shape of their contour levels with, e.g., the shape of the constraints in Figs.~\ref{f-cfm2} and \ref{f-cfm2bis}).

Finally, note that non-Kerr metrics often have some pathological features for some choices of the deformation parameters.  Naked singularities, regions with closed time-like curves, etc, are possible. Some caution has to be taken. However, even pathological spacetimes can be used to test the Kerr metric if we assume that the spacetime solution is only valid outside of some interior region. For example, the interior would be different because of matter source terms. In the case of a compact object made of exotic matter, the vacuum solution would hold up to the surface of the object, while at smaller radii the metric would be described by an interior solution. Pathological features like naked singularities may also be removed by unknown quantum gravity effects~\citep{v2-gh}. A study of the pathologies in these spacetimes is reported in~\citet{pat2}.

\subsection{Violation of the Kerr bound $|a_*| \le 1$ \label{naked}}

A fundamental limit for a Kerr BH is the bound $|a_*| \le 1$. As seen from Eq.~(\ref{eq-horizon}), for $|a_*| > 1$ there is no horizon, and the Kerr metric describes the spacetime of a naked singularity. According to the cosmic censorship conjecture, naked singularities cannot be created by gravitational collapse~\citep{wccc}, even if we know some counterexamples in which a naked singularity exists for an infinitesimal time~\citep{daniele}. The question is thus if BH candidates may be objects with $|a_*| > 1$\footnote{We remind the reader that here, as well as in the remainder of the article, $a_* = J/M^2$ is the dimensionless spin parameter of the compact object, not merely a parameter of the metric.}.

First, an object with $|a_*| \sim 1$ is a fast-rotating body only in the case where it is very compact. For instance, the spin parameter of Earth is about $10^3$ and there is no violation of any principle because the vacuum solution holds up to that of the Earth surface. The Earth's radius is $r_{\rm Earth} \approx 6,400$~km, which is much larger than its gravitational radius $r_g = M \approx 4.4$~mm.

Second, the Kerr metric with $|a_*| > 1$ is not a viable astrophysical scenario. For instance, \citet{super3} found that in general relativity stellar models with $|a_*| > 1$ do not collapse without losing angular momentum. If we consider an existing Kerr BH and we try to overspin it up to $|a_*| > 1$, we fail~\citep{super2}. The same negative result is found by considering the collision of two BHs at the speed of light in full non-linear general relativity~\citep{v2-scp}. None of these studies proves that it is impossible to create a Kerr spacetime with $|a_*| > 1$, but at least they suggest that naked Kerr singularities may not be physical merely because they may not be created in an astrophysical process. Assuming it is somehow possible to create a similar object, \citet{super1} showed that the spacetime would be unstable independently of the boundary conditions imposed at the excision radius $r$, which means that possible unknown quantum gravity corrections at the singularity cannot change these conclusions. Very compact objects described by the Kerr solution with $|a_*| > 1$ are thus unlikely astrophysical viable BH candidates.

Third, if the measurement of the spin parameter of a BH candidate gave a value larger than 1 assuming the Kerr metric, the measurement would be presumably wrong, but it would be a clear indication of new physics. As previously discussed, the Kerr metric with $|a_*| > 1$ is not a viable astrophysical scenario. The metric around the BH candidate should thus be different from that of the Kerr metric. Since spin measurements strongly depend on the exact background metric, the measurement would be wrong and it is possible that the actual value of the spin parameter is instead smaller than 1.

Last, note that $|a_*| = 1$ is a critical value only in the Kerr metric. In the case of non-Kerr BHs, the critical bound is typically different, and it may be either larger or smaller than 1, depending on the spacetime geometry. Moreover, there are examples in which one can create a non-Kerr compact object with $|a_*| > 1$ \citep{spin-c1,spin-c2} or in which one can overspin a BH up and ``destroy''\footnote{An event horizon and a BH cannot be destroyed by definition. Here we start from a stationary spacetime describing a BH and we introduce some small particles to destroy the BH of the stationary metric. The small particles thus completely change the causal structure of the spacetime.} its horizon~\citep{spin-cz}.

\subsection{Important remarks \label{ss-v3-ir}}

In order to test the Kerr metric with electromagnetic radiation, we need to study the properties of the radiation emitted by the gas in the accretion disk or by stars orbiting the BH candidate. Assuming we are in a {\it metric theory of gravity}~\citep{will}, namely that test-particles follow the geodesics of the spacetime, the metric provides all the answers. The spectrum of the BH candidate depends on the motion of the gas in the accretion flow or of orbiting stars and by the propagation of the photons from the point of emission to the distant observer. We thus note the following:
\begin{enumerate}

\item With this approach, we test the Kerr metric, just like with the PPN formalism we can test the Schwarzschild solution. We do not directly test the Einstein equations. For instance, we cannot distinguish a Kerr BH of general relativity from a Kerr BH in an alternative metric theory of gravity, because the motion of particles and photons is the same. Actually the Kerr metric is quite a common solution to many gravity theories~\citep{kpsaltis}. However, the observation of a non-Kerr BH could rule out the Einstein equations because, within general relativity, astrophysical BHs should be well described by the Kerr solution.

\item If we want to directly test the Einstein equations, we should use gravitational (rather than electromagnetic) waves, because their emission is governed by the Einstein equations~\citep{kenrico}.

\item Even if we assume the Kerr metric, analyze the data, and obtain a very good fit, it is not enough to claim that the object is a Kerr BH. This is made clearer in the next sections. There is typically a degeneracy among the parameters of the model, and in particular between the spin and possible deviations from the Kerr background. For this reason we adopt a PPN-like approach.

\end{enumerate}


\section{X-ray observations \label{s3-xray}}

The X-ray radiation in the spectrum of BH candidates is thought to be generated in the vicinity of these objects. Some features depend on the strong gravitational field near the BH candidate and therefore, if properly understood, they may be used to test the Kerr BH hypothesis.

There is some difference between stellar-mass and supermassive BH candidates, because of the different masses and environments. Generally speaking, BH candidates are observed in different {\it spectral states}, which correspond to different geometries and emission mechanisms of their accretion flow~\citep{v2-tb10}. In the case of stellar-mass BH candidates, a source can change its spectral state on a timescale of weeks or months. For supermassive BH candidates, the timescales are too long and the source can only be observed in its current spectral state. Note that the spectral state classification is still a work in progress, some spectral states and their physical interpretation are not yet well understood, and different authors may use a different nomenclature. The key-point in any measurement is to have the correct astrophysical model.

\begin{figure}
\begin{center}
\includegraphics[type=pdf,ext=.pdf,read=.pdf,width=8.5cm]{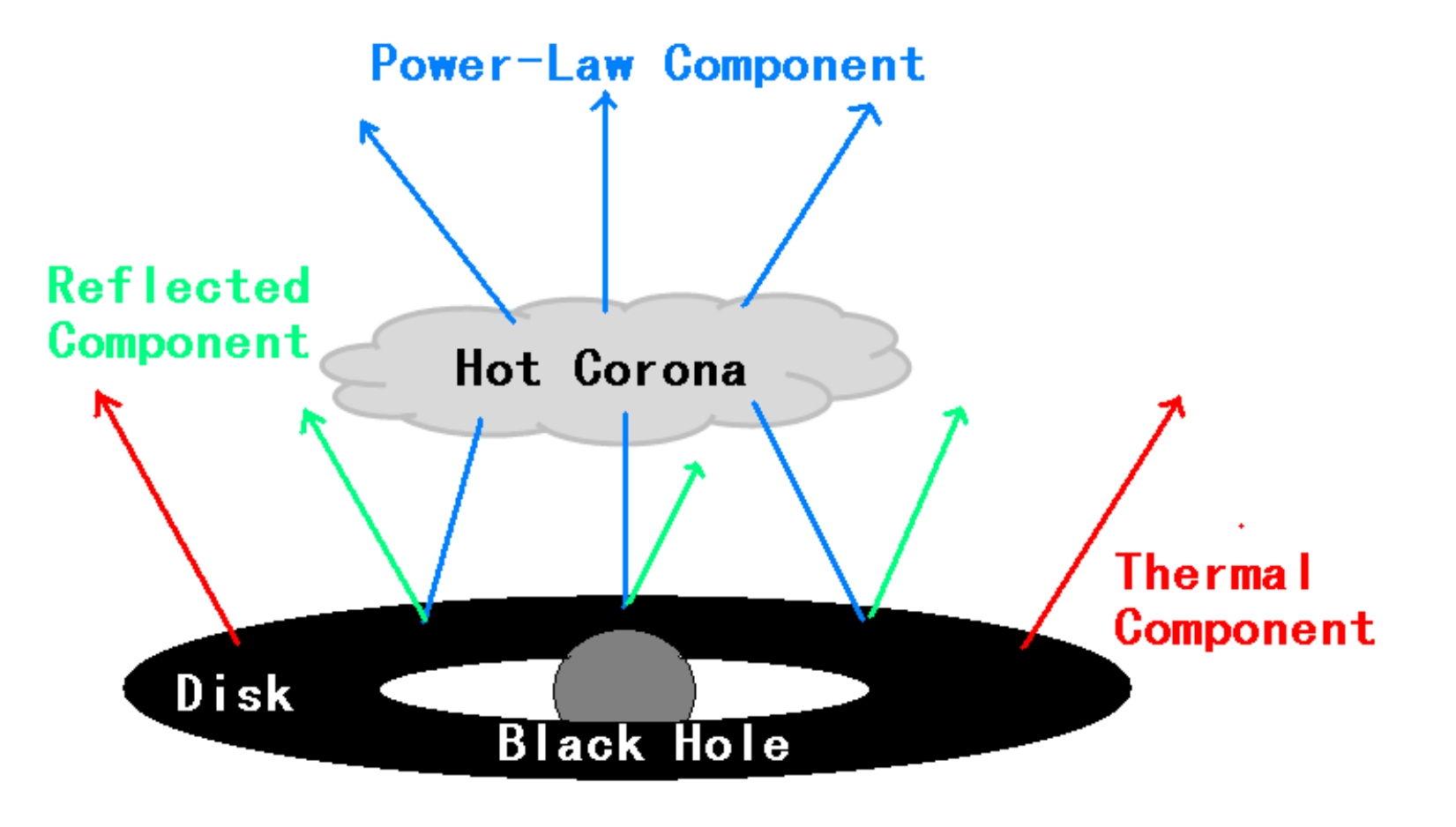}
\end{center}
\vspace{-0.5cm}
\caption{Corona-disk model with lamppost geometry. The BH is surrounded by a geometrically thin and optically thick accretion disk. The accretion disk radiates like a blackbody locally and as a multi-color blackbody when integrated radially. The hot electron cloud called corona acts as an X-ray source and it is located just above the BH. The power-law component represents the direct radiation from the hot corona. The latter illuminates also the disk, producing a reflection component and some emission lines, the most prominent of which is usually the iron K$\alpha$ line. Figures courtesy of Jiachen Jiang. }
\label{f-diskcorona}
\end{figure}

Within the {\it corona-disk model} with {\it lamppost geometry}, the set-up is shown in Fig.~\ref{f-diskcorona}~\citep{matt1,matt2}. The accretion disk is geometrically thin and optically thick. It radiates like a blackbody locally and as a multi-color blackbody when integrated radially. The temperature of the disk depends on the BH mass and the mass accretion rate [see e.g. \citet{v3-cqg}], and the disk's thermal spectrum is in the X-ray band only for stellar-mass BH candidates. The ``corona'' is a hotter, usually optically-thin, electron cloud which enshrouds the central disk and acts as a source of X-rays, due to inverse Compton scattering of the thermal photons from the accretion disk off the electrons in the corona. The corona is often approximated as a point source located on the axis of the accretion disk and just above the BH. This arrangement is often referred to as a lamppost geometry. The lamppost geometry requires a plasma of electrons very close to the BH and such a set-up may be realized in the case of the base of a jet. However, other geometries are possible, and an example is the family of ``sandwich models''~\citep{sandm}. The direct radiation from the hot corona produces a power-law component in the X-ray spectrum. The corona illuminates also the disk, producing a reflected component and some emission lines, the most prominent of which is usually the iron K$\alpha$ line.

X-ray techniques to test the Kerr metric are discussed in the next subsections: continuum-fitting method (\ref{ss-cfm}), iron K$\alpha$ line with time-integrated (\ref{ss-iii}) and time-resolved data (\ref{ss-iii2}), quasi-periodic oscillations (QPOs, \ref{ss-qpos}), and polarization measurements (\ref{ss-polar}).

Currently, the two leading techniques to probe the spacetime geometry around BH candidates with X-ray measurements are the study of the thermal spectrum of thin disks (continuum-fitting method)~\citep{cfm1,cfm2,cfm-rev} and the analysis of the iron K$\alpha$ line~\citep{iron1,iron2,BR06}. Both techniques have been developed to measure the spin parameter of BH candidates under the assumption of the Kerr background, but they can be naturally extended to test the Kerr metric. Current spin measurements of stellar-mass BH candidates reported in the literature under the assumption of the Kerr background are summarized in Tab.~\ref{tab1}. Iron line spin measurements of supermassive BH candidates under the assumption of the Kerr background are reported in Tab.~\ref{tab2}.

QPOs are seen as peaks in the X-ray power density spectra of BH candidates and they may be used to measure the properties of the metric around these objects. For the time being, we do not know the exact mechanism responsible for these oscillations, and therefore QPOs cannot yet be used to test fundamental physics. Different models provide different results. However, QPOs are a promising tool for the future, because the value of their central frequency can be measured with high precision.

The observation of the X-ray polarization of the thermal spectrum of thin accretion disks is another potential technique to test the Kerr metric. There are currently no space missions with polarization detectors, but there are a few proposals that may be operative within 10~years: the Chinese-European mission eXTP~\citep{v4-extp}, the European project XIPE~\citep{xipe}, or the two NASA missions IXPE~\citep{ixpe} and PRAXYS\footnote{http://asd.gsfc.nasa.gov/praxys/}.

\begin{table*}
\centering
\begin{tabular}{ccccccc}
\hline \hline
BH Binary & \hspace{0.1cm} & $a_*$ (CF) & \hspace{0.1cm} & $a_*$ (iron) & \hspace{0.1cm} & Principal References \\
\hline \hline
GRS~1915-105 && $> 0.98$ && $0.98 \pm 0.01$ && \citet{1915,1915b} \\
Cyg~X-1 && $> 0.98$ && $0.97^{+0.014}_{-0.02}$ && \citet{cyg1,cyg2,cyg3} \\
LMC~X-1 && $0.92 \pm 0.06$ && $0.97^{+0.02}_{-0.25}$ && \citet{lmcx1,lmcx1b} \\
GX~339-4 && $< 0.9$ && $0.95^{+0.03}_{-0.05}$ && \citet{gx339,gx339b} \\
MAXI~J1836-194 && --- && $0.88 \pm 0.03$ && \citet{maxi} \\
M33~X-7 && $0.84 \pm 0.05$ && --- && \citet{liu08,liu10} \\
4U~1543-47 && $0.80 \pm 0.10^\star$ && --- && \citet{sh06} \\
Swift~J1753.5 && --- && $0.76^{+0.11}_{-0.15}$ && \citet{swift} \\
XTE~J1650-500 && --- && $0.84 \sim 0.98$ && \citet{1650} \\
IC~10~X-1 &&  $\gtrsim0.7$  && --- && \cite{v2-newspin1} \\
GRO~J1655-40 && $0.70 \pm 0.10^\star$ && $> 0.9$ && \citet{sh06,swift} \\
GS~1124-683 && $0.63^{+0.16}_{-0.19}$ && --- && \cite{v2-newspin2} \\
XTE~J1752-223 && --- && $0.52 \pm 0.11$ && \citet{1752} \\
XTE~J1652-453 && --- && $< 0.5$ && \citet{1652} \\
XTE~J1550-564 && $0.34 \pm 0.28$ && $0.55^{+0.15}_{-0.22}$ && \citet{xte} \\
LMC~X-3 && $0.25 \pm 0.15$ && --- && \citet{lmcx3} \\
H1743-322 && $0.2 \pm 0.3$ && --- && \citet{h1743} \\
A0620-00 &&  $0.12 \pm 0.19$ && --- && \citet{62} \\
XMMU~J004243.6 && $< -0.2$ && --- && \citet{m31} \\
\hline \hline
\end{tabular}
\caption{Summary of the continuum-fitting (CF) and iron line measurements of the spin parameter of stellar-mass BH candidates under the assumption of the Kerr background. See the references in the last column for more details. $^\star$These sources were studied in an early work on the continuum-fitting method, within a more simple model, so the published 1-$\sigma$ error estimates are doubled following~\citet{cfm-rev}. \label{tab1}}
\end{table*}

\begin{table*}
\centering
\begin{tabular}{ccccccc}
\hline \hline
AGN & \hspace{0.1cm} & $a_*$ (iron) & \hspace{0.1cm} & $L_{\rm Bol}/L_{\rm Edd}$ & \hspace{0.1cm} & Principal References \\
\hline \hline
IRAS~13224-3809 && $> 0.995$ && $0.71$ && \citet{W13} \\
Mrk~110         && $> 0.99$  && $0.16 \pm 0.04$ && \citet{W13} \\
NGC~4051        && $> 0.99$  && $0.03$ && \citet{P11} \\
MCG-6-30-15     && $> 0.98$  && $0.40 \pm 0.13$ && \citet{BR06,Miniutti2007} \\
1H~0707-495     && $> 0.98$  && $\sim 1$ && \citet{Calle-Perez2010,W13}; \\
                              &&                   &&                 &&  \citet{Zoghbi2010} \\
NGC~3783        && $> 0.98$  && $0.06 \pm 0.01$ && \citet{B11,P11} \\
RBS~1124        && $> 0.98$  && $0.15$ && \citet{W13} \\
NGC~1365        && $0.97^{+0.01}_{-0.04}$ && $0.06^{+0.06}_{-0.04}$ && \citet{Risaliti2009b,Risaliti2013,Brenneman2013} \\
Swift~J0501.9-3239 && $> 0.96$  && --- && \citet{W13} \\
Ark~564         && $0.96^{+0.01}_{-0.06}$ && $> 0.11$ && \citet{W13} \\
3C~120          && $> 0.95$  && $0.31 \pm 0.20$ && \citet{Lohfink2013} \\
Ark~120         && $0.94 \pm 0.01$ && $0.04 \pm 0.01$ && \citet{P12,Nardini2011}; \\
                        &&                                &&                               && \citet{W13} \\
Ton~S180        && $0.91^{+0.02}_{-0.09}$ && $2.1^{+3.2}_{-1.6}$ && \citet{W13} \\
1H~0419-577     && $> 0.88$ && $1.3 \pm 0.4$ && \citet{W13} \\
Mrk~509         && $0.86^{+0.02}_{-0.01}$ && --- && \citet{W13} \\
IRAS~00521-7054 && $> 0.84$ && --- && \citet{Tan2012} \\
3C~382          && $0.75^{+0.07}_{-0.04}$ && --- && \citet{W13} \\
Mrk~335         && $0.70^{+0.12}_{-0.01}$ && $0.25 \pm 0.07$ && \citet{P12,W13} \\
Mrk~79          && $0.7 \pm 0.1$ && $0.05 \pm 0.01$ && \citet{Gallo2005,Gallo2011} \\
Mrk~359         && $0.7^{+0.3}_{-0.5}$ && $0.25$ && \citet{W13} \\
NGC~7469 && $0.69 \pm 0.09$ && --- && \citet{P12} \\
Swift~J2127.4+5654 && $0.6 \pm 0.2$ && $0.18 \pm 0.03$ && \citet{Miniutti2009a,P12} \\ 
Mrk~1018        && $0.6^{+0.4}_{-0.7}$ && $0.01$ && \citet{W13} \\
Mrk~841         && $> 0.56$  && $0.44$ && \citet{W13} \\
Fairall~9       && $0.52^{+0.19}_{-0.15}$ && $0.05 \pm 0.01$ && \citet{P12,L12}; \\
                       &&                                            &&                               && \citet{Schmoll2009,W13} \\
\hline \hline
\end{tabular}
\caption{Summary of the iron line measurements of the spin parameter of supermassive BH candidates under the assumption of the Kerr background. See the references in the last column and \citet{brenneman-review13} for more details. \label{tab2}}
\end{table*}

\subsection{Continuum-fitting method \label{ss-cfm}}

The continuum-fitting method consists of the analysis of the thermal spectrum of geometrically thin and optically thick accretion disks~\citep{cfm1,cfm2,cfm-rev}. The standard theoretical framework is the Novikov-Thorne model~\citep{nt1,nt2}. The plane of the disk is assumed to be perpendicular to the BH spin\footnote{If the BH formed from the supernova explosion of a heavy star in a binary, its spin would more likely already be aligned with the orbital angular momentum vector of the system (at least in the case of a symmetric explosion without strong shock and kick)~\citep{fragos10k}. If this is not the case, the inner part of the accretion disk can still be expected to be on the plane perpendicular to the BH spin as a consequence of the Bardeen-Petterson effect~\citep{alignment1,kumar85,alignment2}.}. The particles of the gas move on nearly geodesic circular orbits, and the inner edge of the disk is at the ISCO radius. The latter assumption plays a crucial role and it is confirmed by some observations that show that the inner edge of the disk does not change appreciably over several years when the source is in the thermal state. The most compelling evidence comes from LMC~X-3. The analysis of many spectra collected during eight X-ray missions and spanning 26~years shows that the radius of the inner edge of the disk is quite constant~\citep{r_in}. The most natural interpretation is that the inner edge is associated with some intrinsic property of the geometry of the spacetime, namely the radius of the ISCO, and it is not affected by variable phenomena like the accretion process.

An accretion disk meets these conditions when the source is in the so-called high/soft state. Here ``high'' refers to the accretion luminosity, which must be higher than 5\% of the Eddington limit. ``Soft'' refers to the fact that the soft X-ray component, corresponding to the thermal spectrum of the disk, is the dominant one. This is true only for stellar-mass BH candidates, because the temperature of the disk depends on the BH mass and the mass accretion rate. As it turns out, the thermal spectrum of a thin disk is in the X-ray range for stellar-mass BH candidates and in the UV/optical bands for supermassive BH candidates. In the latter case, extinction and dust absorption limit the ability to make accurate measurements. The continuum-fitting method is thus normally applied to stellar-mass BH candidates\footnote{The continuum-fitting method was applied to supermassive BH candidates in some very special cases in~\citet{4-czerny2011} and \citet{4-done2013}.}. Thin disks are present when the accretion luminosity is 5-30\% of the Eddington limit~\citep{1915}. For the validity of the method, see e.g.~\citet{cfm-rev} and references therein.

\begin{figure*}
\begin{center}
\includegraphics[type=pdf,ext=.pdf,read=.pdf,width=8.5cm]{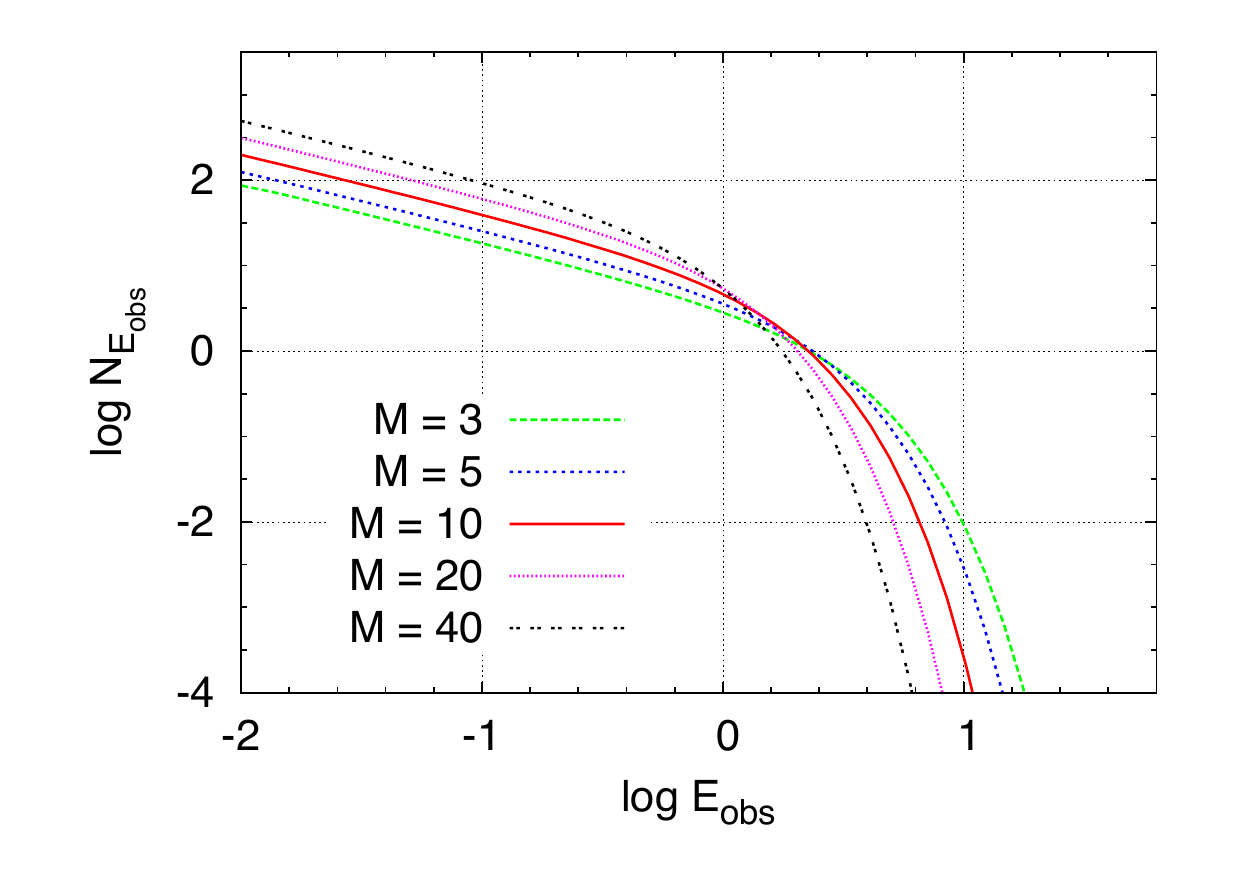}
\includegraphics[type=pdf,ext=.pdf,read=.pdf,width=8.5cm]{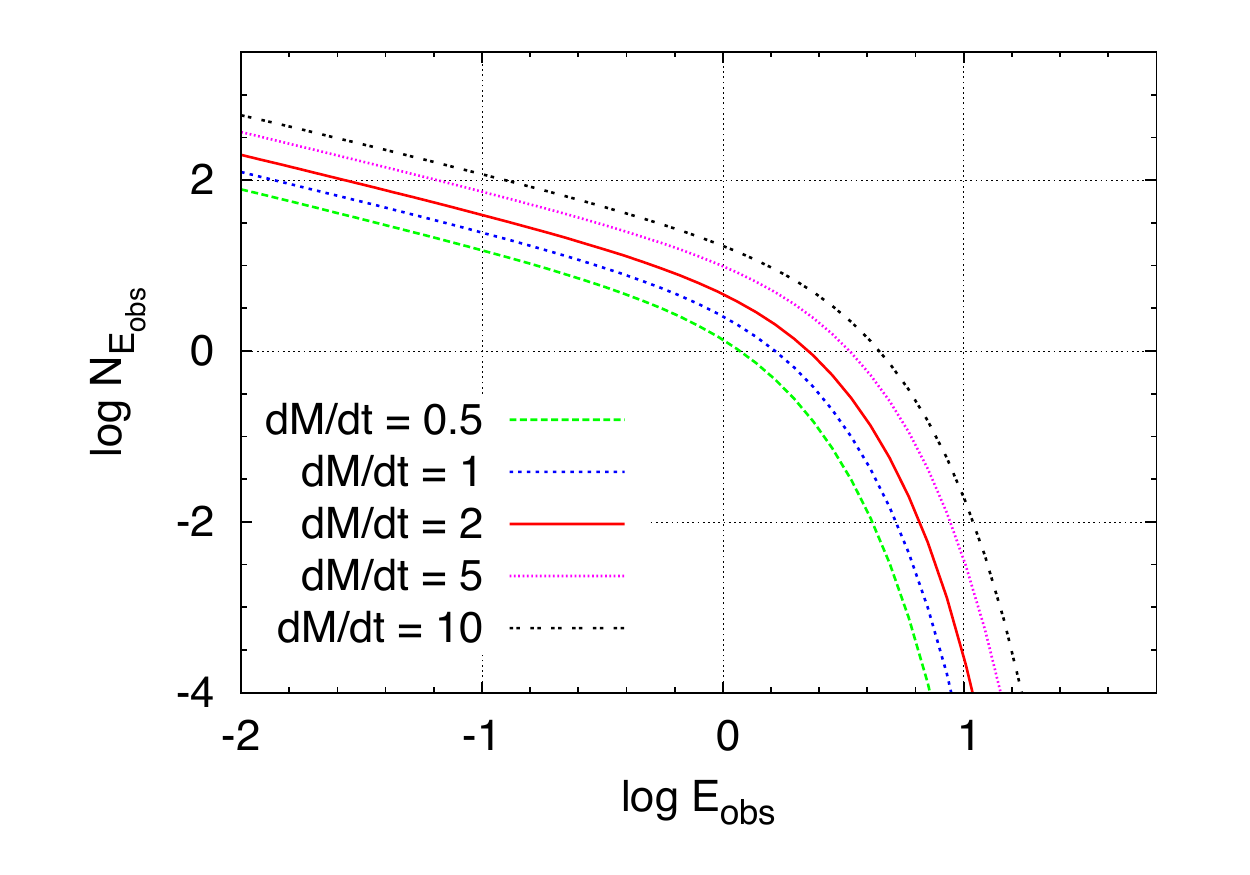} \\
\vspace{0.2cm}
\includegraphics[type=pdf,ext=.pdf,read=.pdf,width=8.5cm]{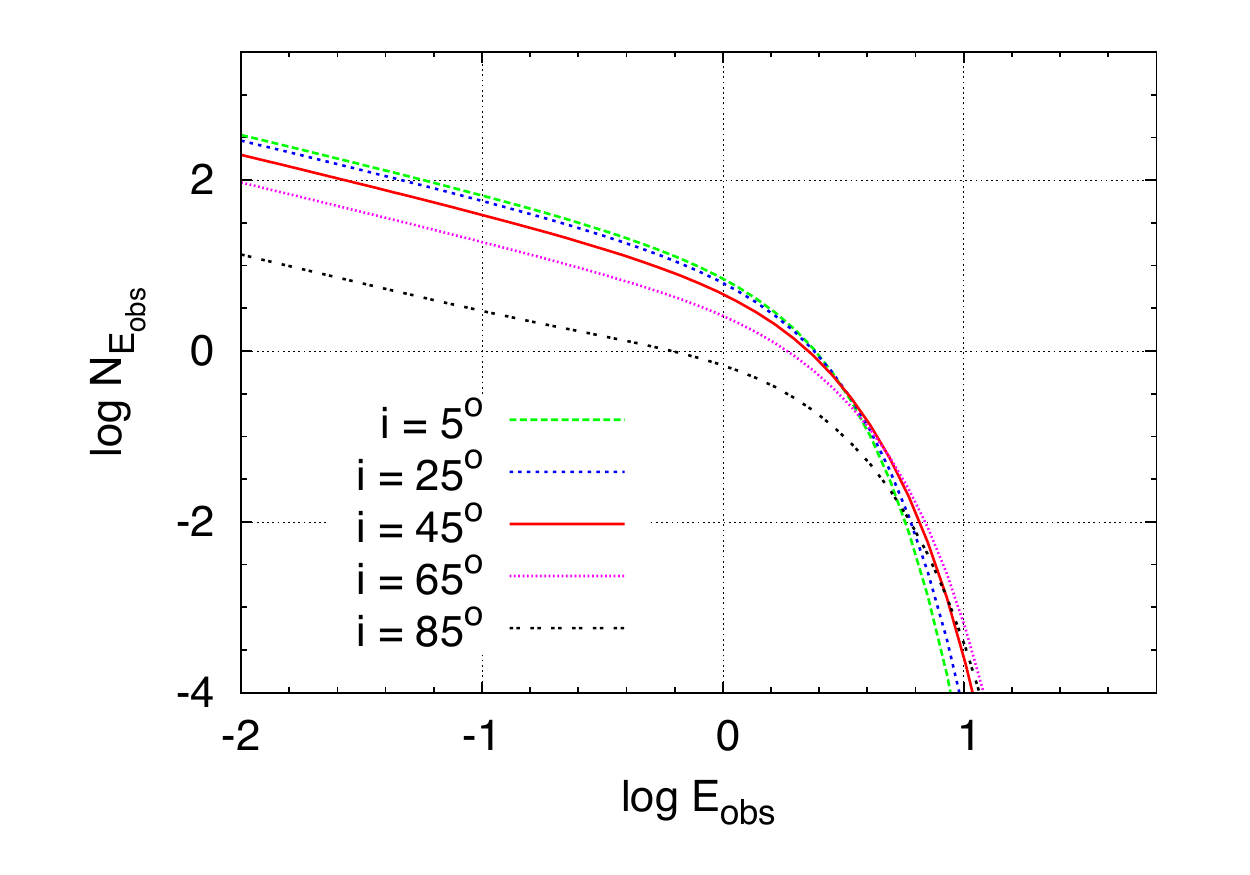}
\includegraphics[type=pdf,ext=.pdf,read=.pdf,width=8.5cm]{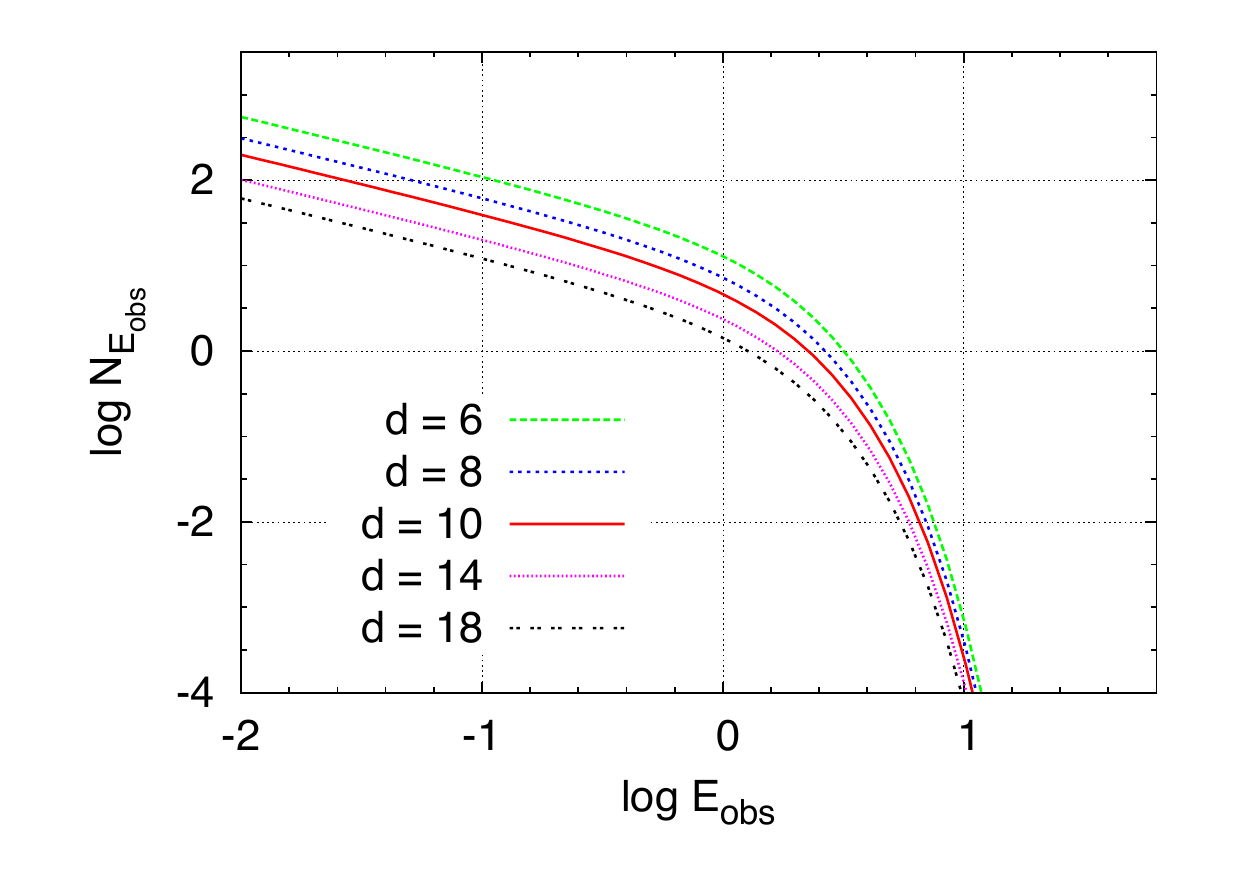} \\
\vspace{0.2cm}
\includegraphics[type=pdf,ext=.pdf,read=.pdf,width=8.5cm]{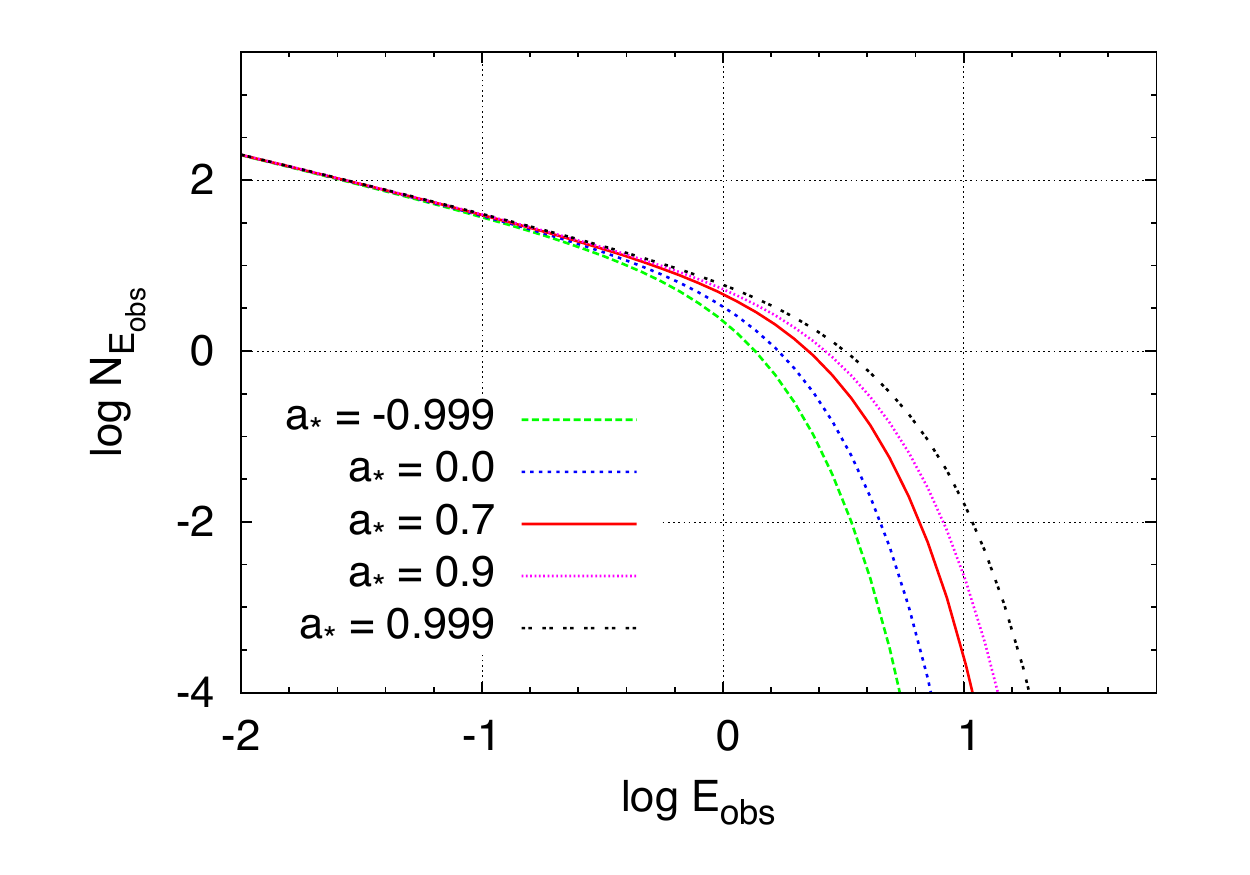}
\includegraphics[type=pdf,ext=.pdf,read=.pdf,width=8.5cm]{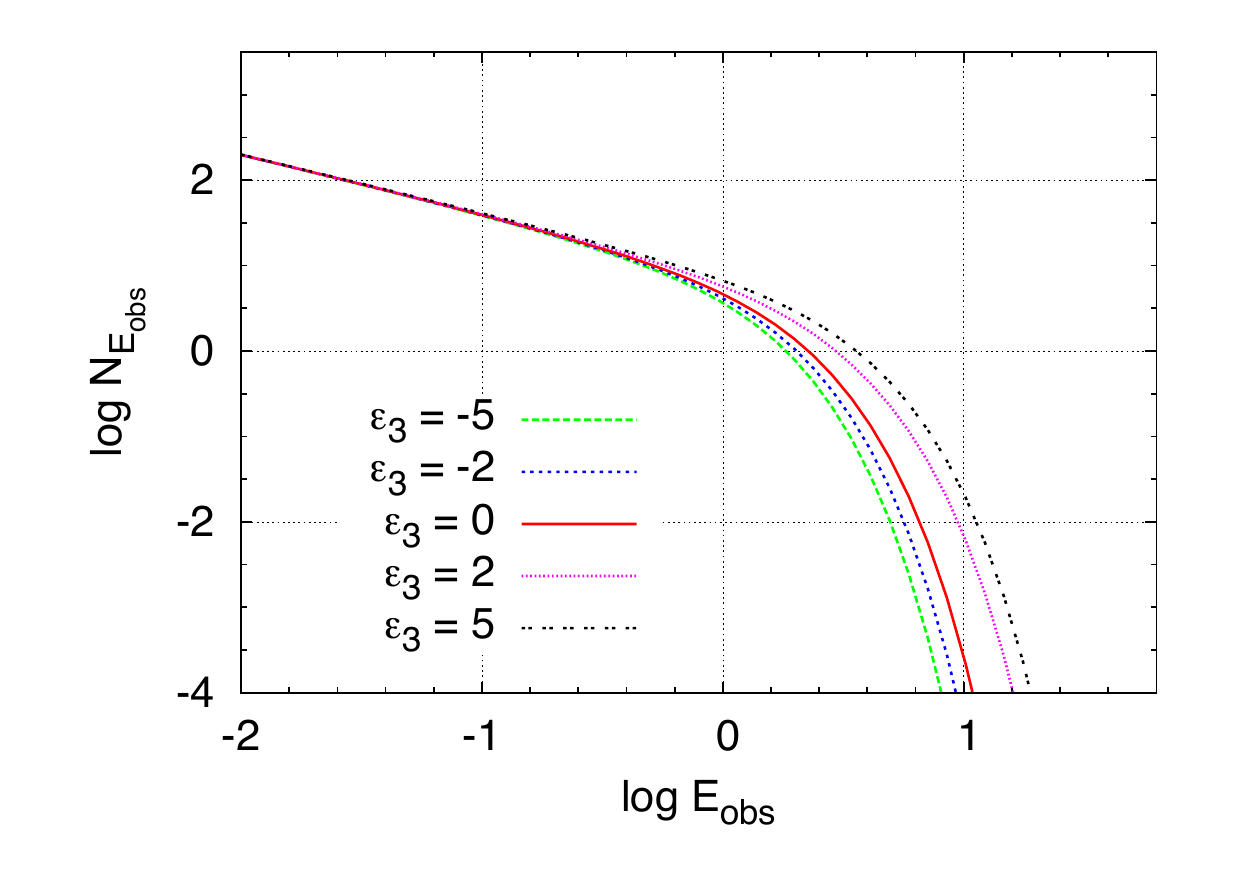} \\
\end{center}
\vspace{-0.2cm}
\caption{Impact of the model parameters on the thermal spectrum of a thin disk: mass $M$ (top left panel), mass accretion rate $\dot{M}$ (top right panel), viewing angle $i$ (central left panel), distance $d$ (central right panel), spin parameter $a_*$ (bottom left panel), and JP deformation parameter $\epsilon_3$ (bottom right panel). When not shown, the values of the parameters are: $M = 10$~$M_\odot$, $\dot{M} = 2 \cdot 10^{18}$~g~s$^{-1}$, $d = 10$~kpc, $i = 45^\circ$, $a_*=0.7$, and $\epsilon_3 = 0$. $M$ in units of $M_\odot$, $\dot{M}$ in units of $10^{18}$~g~s$^{-1}$, $d$ in kpc, flux density $N_{E_{\rm obs}}$ in photons~keV$^{-1}$~cm$^{-2}$~s$^{-1}$, and photon energy $E_{\rm obs}$ in keV. }
\label{f-cfm1}
\end{figure*}

In the Kerr background, the thermal spectrum of a thin disk only depends on 5 parameters: the BH mass $M$, the mass accretion rate $\dot{M}$, the inclination angle of the disk with respect to the line of sight of the observer $i$, the distance $d$ of the source, and the spin parameter of the BH $a_*$. The impact of the model parameters on the shape of the spectrum is shown in Fig.~\ref{f-cfm1}. If it is possible to have independent measurements of $M$, $i$, and $d$, one can fit the thermal component of the spectrum and determine the BH spin $a_*$ and the mass accretion rate $\dot{M}$. Current spin measurements with the continuum-fitting method are reported in Tab.~\ref{tab1}. Some of these objects are not dynamically confirmed BH candidates, so their mass is not estimated from the motion of the companion star.

The technique can be naturally extended to non-Kerr backgrounds. Preliminary studies considered some specific non-Kerr objects and did not take into account all the relativistic effects~\citep{cfm-t,harko1,harko2,harko3,harko4,harko5,cfm-c1}. A ray-tracing code that includes all the relativistic effects and can compute the thermal spectrum of a thin disk in a generic stationary and axisymmetric spacetime was presented in~\citet{cfm-c2}.

The impact of a possible non-vanishing deformation parameter on the thermal spectrum of a thin disk is shown in the bottom right panel in Fig.~\ref{f-cfm1}. Even without a quantitative analysis, it is clear that the effect of the spin and of the deformation parameter is very similar. The point is that the cut-off of the spectrum is determined by the position of the inner edge of the disk, which is assumed to be at the ISCO radius, or, more precisely, by the radiative efficiency in the Novikov-Thorne model~\citep{lingyao}. The radiative efficiency in the Novikov-Thorne model is
\be
\eta_{\rm NT} = 1 - E_{\rm ISCO} \, ,
\ee
where $E_{\rm ISCO}$ is the specific energy of a test-particle at the ISCO radius. In the Kerr metric, $\eta_{\rm NT}$ ranges from 0.057 for a non-rotating BH ($a_* = 0$) to about 0.42 for a maximally rotating BH and a corotating disk ($a_* = 1$). In the case of retrograde disks, $\eta_{\rm NT}$ decreases as the spin parameter increases until about 0.038 for $a_* = 1$. In the Kerr metric, there is a one-to-one correspondence between $\eta_{\rm NT}$ and $a_*$, and therefore the measurement of the former can provide an estimate of the BH spin parameter. If we relax the Kerr BH hypothesis and we allow for a non-vanishing deformation parameter, the same value of the radiative efficiency can be obtained for different combinations of the spin parameter and the deformation parameter. The result is that there is a degeneracy and it is impossible to measure both the spin and the deformation parameter. In general, we can measure only a combination of them.

\begin{figure*}
\begin{center}
\includegraphics[type=pdf,ext=.pdf,read=.pdf,width=8.5cm]{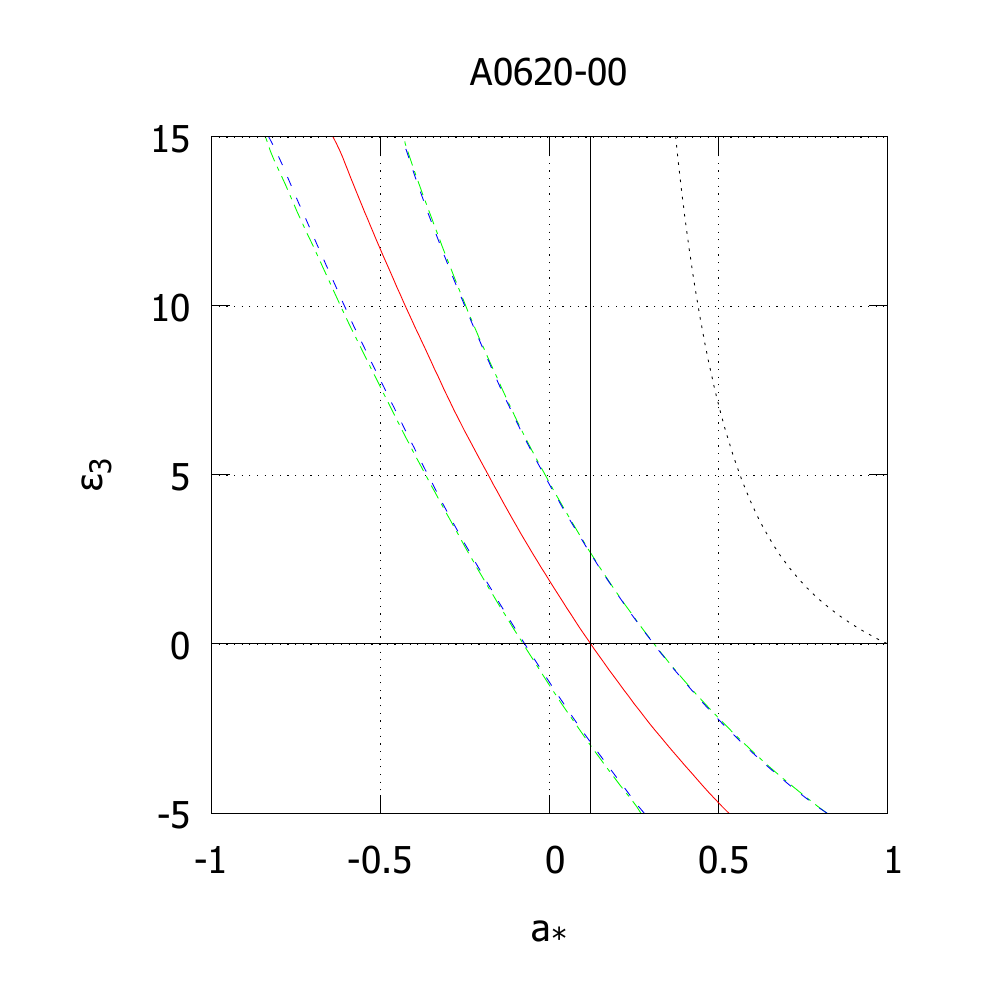}
\includegraphics[type=pdf,ext=.pdf,read=.pdf,width=8.5cm]{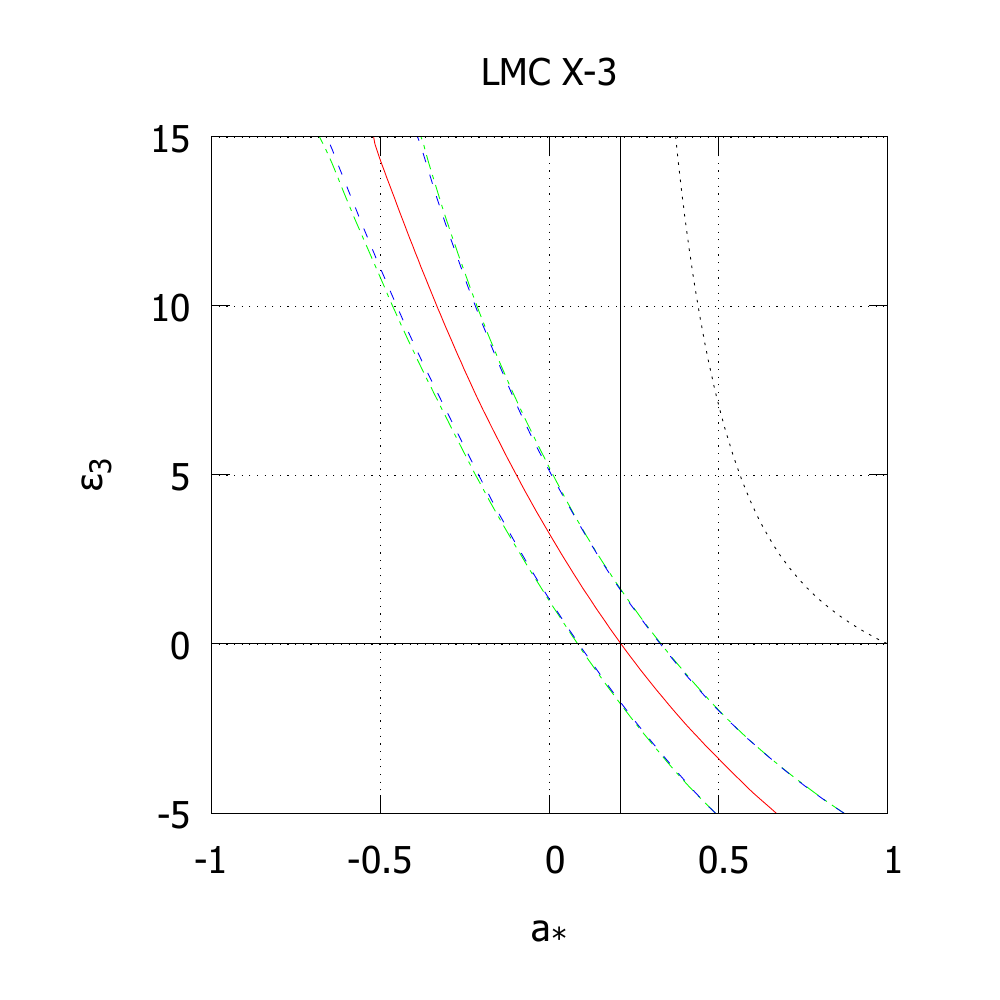} \\
\includegraphics[type=pdf,ext=.pdf,read=.pdf,width=8.5cm]{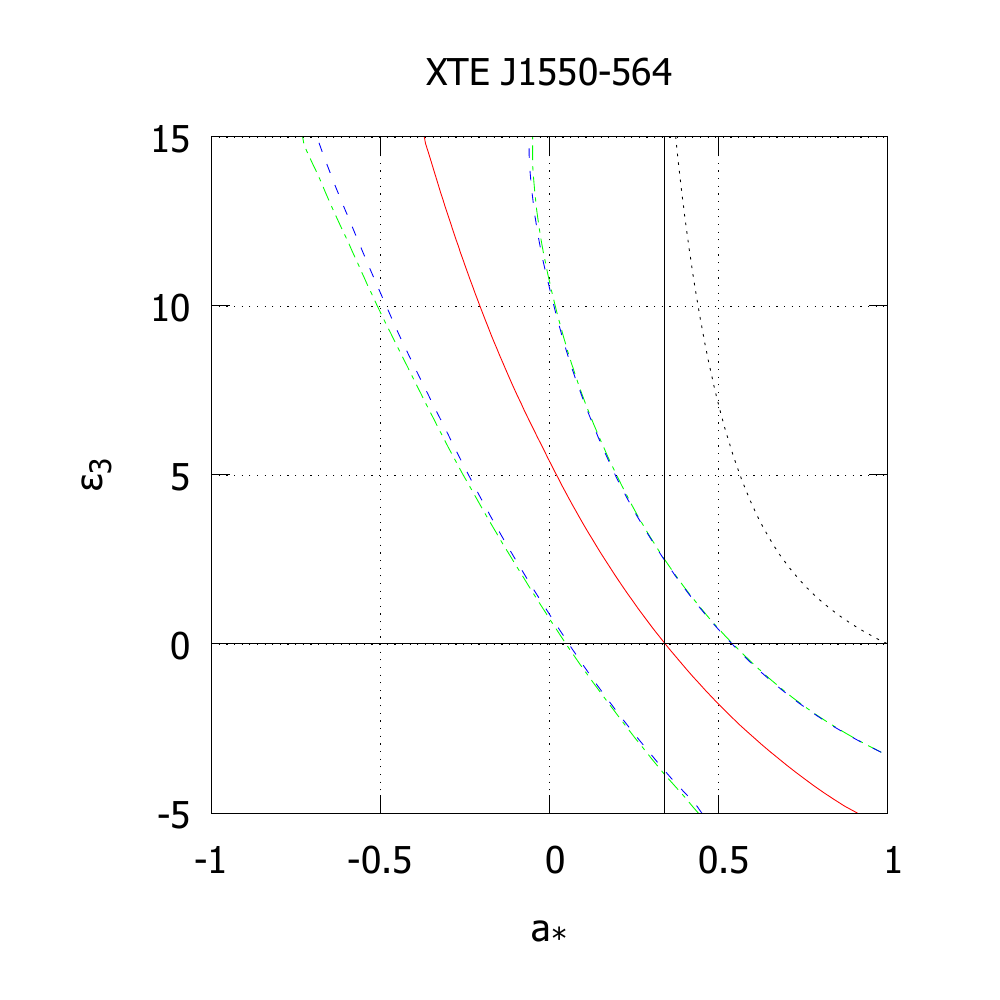}
\includegraphics[type=pdf,ext=.pdf,read=.pdf,width=8.5cm]{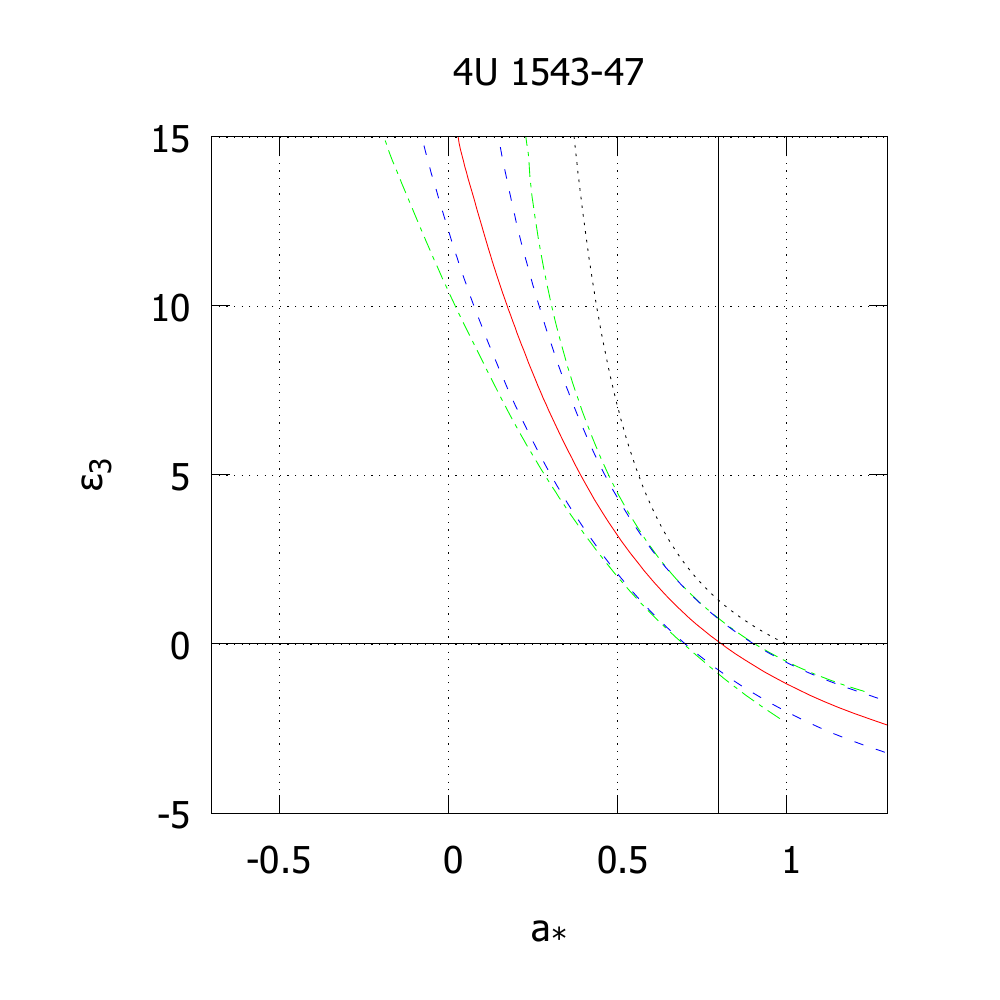}
\end{center}
\vspace{-0.5cm}
\caption{Continuum-fitting constraints on the JP parameter $\epsilon_3$ from A0620-00 (top left panel), LMC~X-3 (top right panel), XTE~J1550-564 (bottom left panel), and 4U~1543-47 (bottom right panel). The spacetimes along the red solid lines cannot be distinguished and represent the central values of the measurement. In every panel, the dash-dotted green lines are the boundaries of the allowed region (1-$\sigma$ error) along the red line inferred within the analysis of \citet{lingyao} from current X-ray measurements. The blue dashed lines are the same boundaries inferred if the best fit were exactly for $\epsilon_3 = 0$. The dotted black curve on the right of each panel separates spacetimes with a regular exterior (on the left of the curve) from those with naked singularities (on the right of the curve). From~\citet{lingyao}. See \citet{lingyao} and the text for more details. }
\label{f-cfm2}
\end{figure*}

\begin{figure*}
\begin{center}
\includegraphics[type=pdf,ext=.pdf,read=.pdf,width=8.5cm]{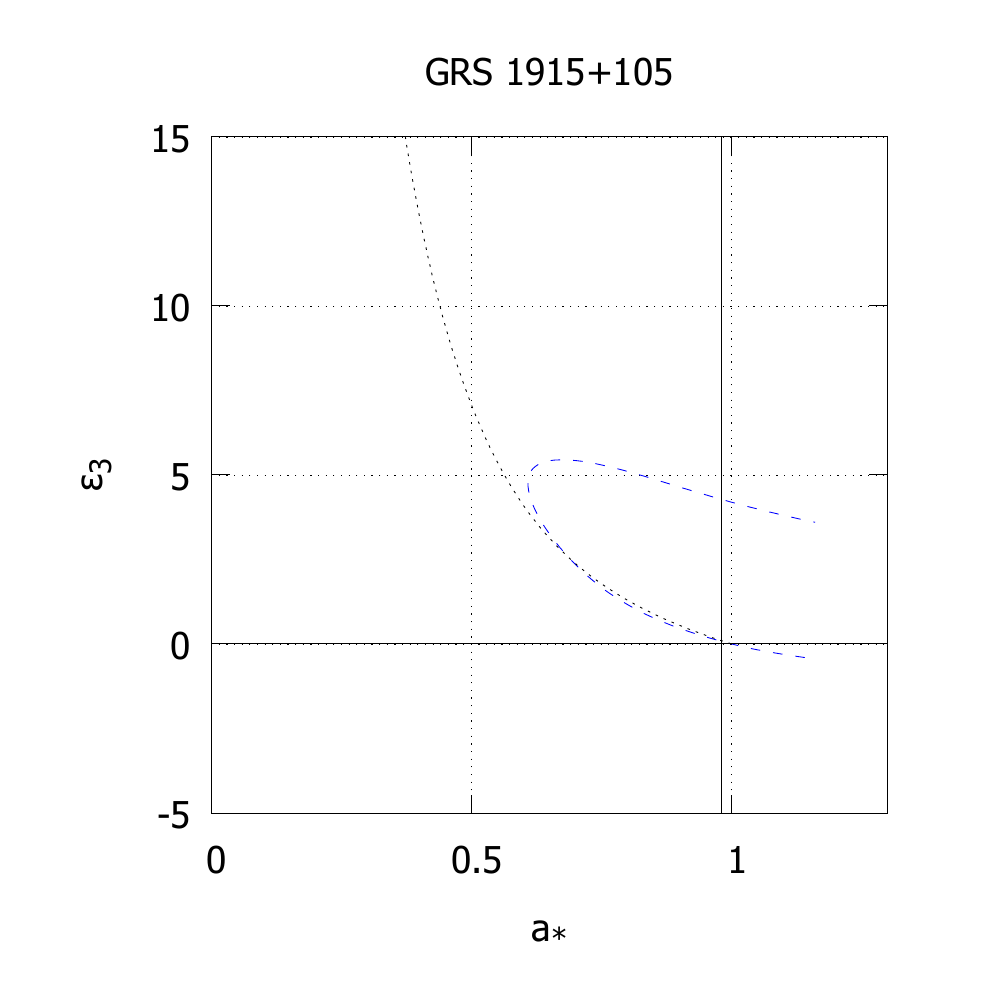}
\put(-93,102){Allowed Region}
\put(-160,145){Excluded Region}
\includegraphics[type=pdf,ext=.pdf,read=.pdf,width=8.5cm]{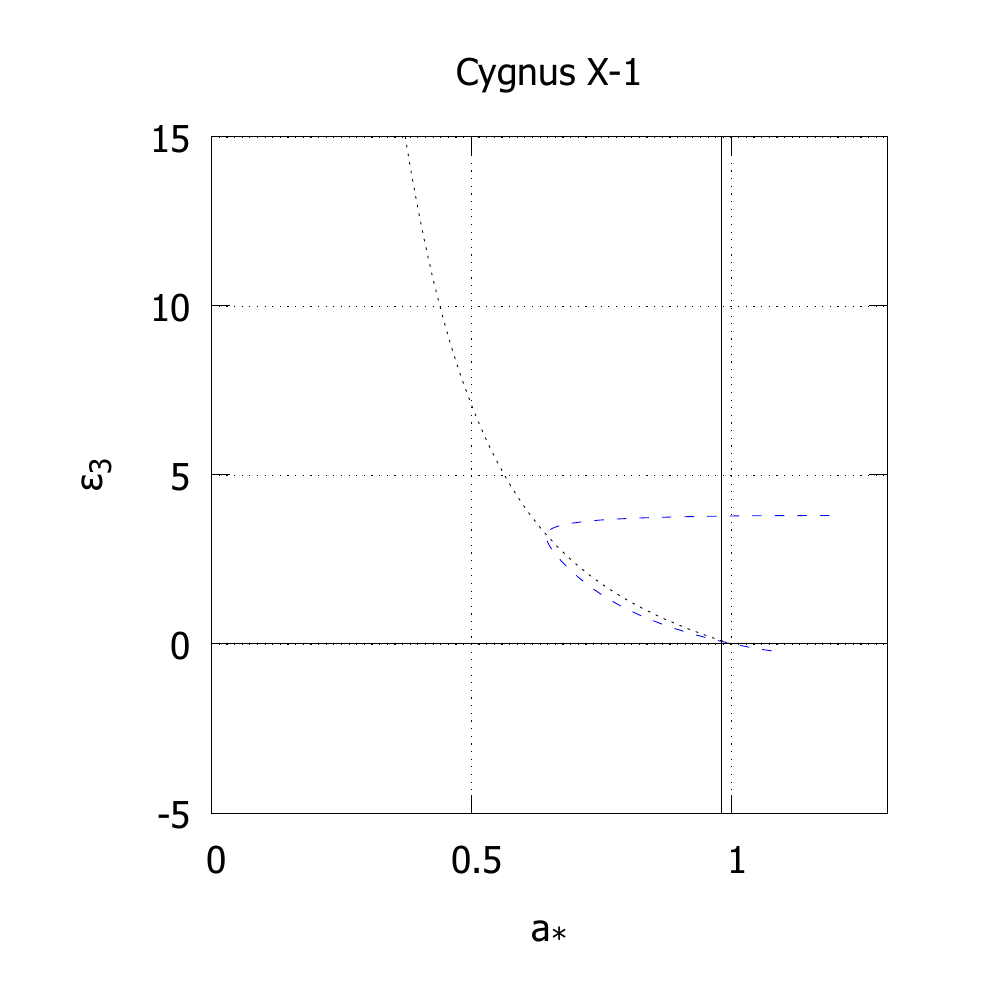}
\put(-93,102){Allowed Region}
\put(-160,145){Excluded Region}
\end{center}
\vspace{-0.5cm}
\caption{Continuum-fitting constraints on the JP parameter $\epsilon_3$ from GRS~1915+105 (left panel) and Cygnus~X-1 (right panel). The allowed regions are those inside the dashed blue lines. The dotted black curve separates spacetime with a regular exterior (on the left of the curve) from those with naked singularities (on the right of the curve). From~\citet{lingyao}. See \citet{lingyao} and the text for more details. }
\label{f-cfm2bis}
\end{figure*}

In~\citet{lingyao}, we reconsidered the spin measurements with the continuum-fitting method reported in the literature and under the assumption of the Kerr background. We obtained the constraints on the spin parameter -- deformation parameter plane within the JP background. Some examples are shown in Figs.~\ref{f-cfm2} and \ref{f-cfm2bis}. In the case of A0620-00, LMC~X-3, XTE~J1550-564, and 4U~1543-47, there is a specific measurement in the Kerr metric, and this can be translated into an allowed region on the spin parameter -- deformation parameter plane. These constraints are obtained with the following method that can be justified {\it a posteriori} [more details can be found in the original paper~\citet{lingyao}]. Instead of working on some observational data, we adopt the theoretical spectrum of a disk around a Kerr black hole with the parameter values reported in the literature ($M$, $i$, and $d$ inferred without the assumption of the Kerr metric, and $a_*$ and $\dot{M}$ obtained with the continuum-fitting method and assuming the Kerr metric). Such a spectrum is then compared with the theoretical spectra computed in spacetimes in which $a_*$, $\dot{M}$, and $\epsilon_3$ vary. Employing a $\chi^2$ analysis, the red solid lines in Fig.~\ref{f-cfm2} are the best fits for a fixed $\epsilon_3$. The blue dashed lines are the boundaries of the allowed regions at 1-$\sigma$. However, in this way we are assuming that these sources have vanishing $\epsilon_3$. For a fixed $\epsilon_3$, the 1-$\sigma$ error on the spin is given by the dash-dotted green lines, which should better approximate the actual 1-$\sigma$ error if we analyzed the real data. This approach can be justified {\it a posteriori} because the constraints provided by the dash-dotted green lines and by the dashed blue lines are very similar considering the spin uncertainty. There is a quasi-degeneracy in the theoretical prediction of the spectra and therefore a spin measurement inferred in the Kerr background can be translated into an allowed region on the spin parameter -- deformation parameter plane.

Fig.~\ref{f-cfm2bis} shows the constraints on $\epsilon_3$ from GRS~1915+105 and Cygnus~X-1. In both cases, the Kerr measurement is $a_* > 0.98$, and therefore in the JP metric we have an allowed region in which the spectra look more like a Kerr BH with $a_* > 0.98$ and an excluded region in which the spectra are more like that of a Kerr BH with $a_* < 0.98$. The difference between the two allowed regions is only due to the different inclination angle, $i = 66^\circ$ for GRS~1915+105 and $i = 27.1^\circ$ for Cygnus~X-1. Note the similarity of the shapes of the allowed regions in Figs.~\ref{f-cfm2} and \ref{f-cfm2bis} with the contour levels of $\eta_{\rm NT}$ in the right panel in Fig.~\ref{f-isco}.

\begin{figure}
\begin{center}
\includegraphics[type=pdf,ext=.pdf,read=.pdf,width=8.5cm]{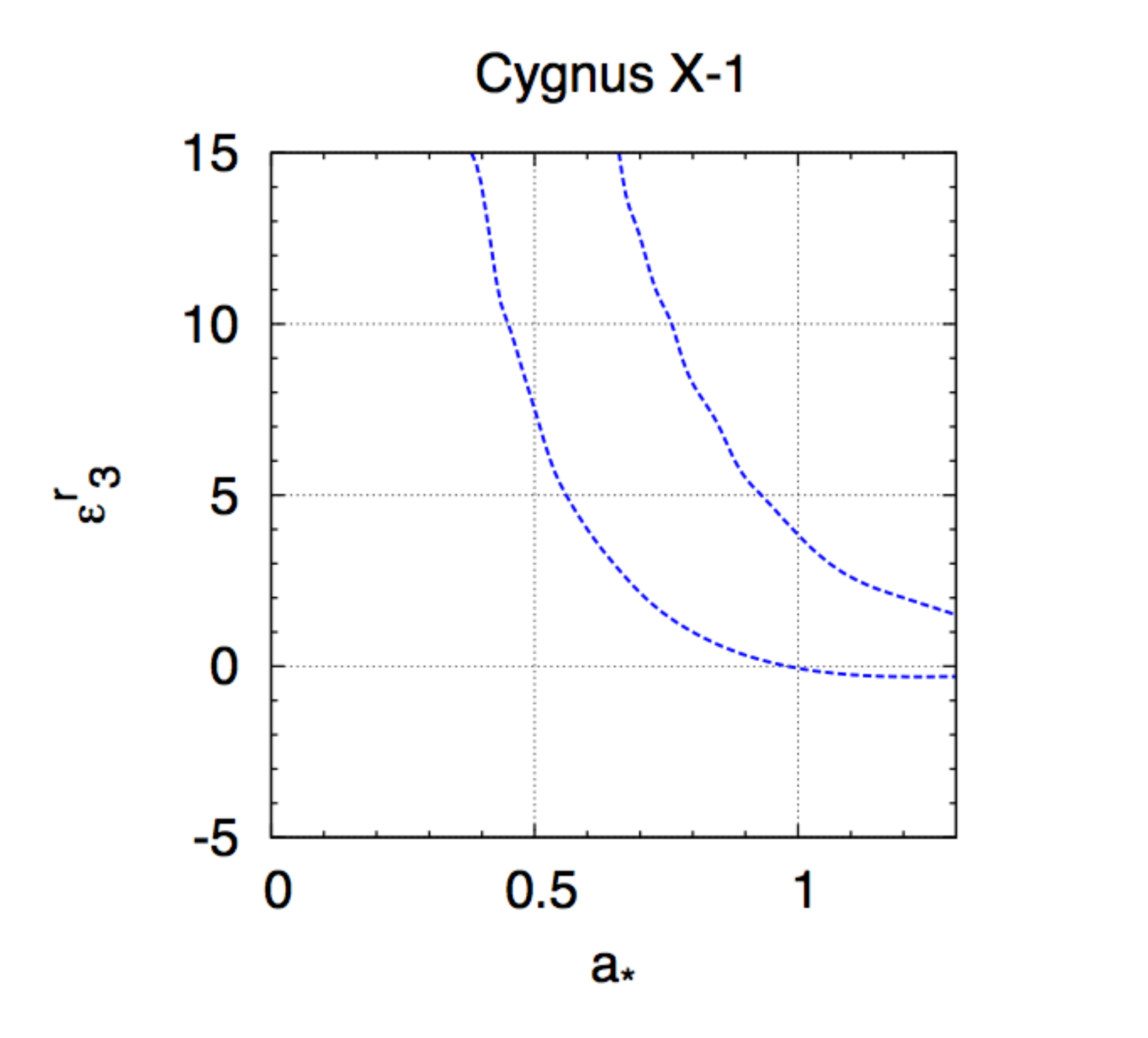}
\end{center}
\vspace{-0.5cm}
\caption{Continuum-fitting constraints on the CPR parameter $\epsilon_3^r$ from Cygnus~X-1. The allowed region is between the two dashed blue lines. See the text and the original paper \citet{cfm-cc2} for more details.}
\label{f-cfm3}
\end{figure}

If we have a BH candidate that looks like a very fast-rotating Kerr BH, similar to that in Cygnus~X-1 in the example in Fig.~\ref{f-cfm2bis}, it is sometimes possible to constrain the deformation parameter. Another example with a different background metric was shown in~\citet{cfm-cc}. The reason is that, in general, if one considers very large deviations from Kerr, in both directions in the deformation parameter, the ISCO radius increases and therefore the Novikov-Thorne radiative efficiency decreases. The result is that very deformed objects cannot mimic a fast-rotating Kerr BH. However, this is not a general statement, and some deformations may be extremely large. An example of the latter case is the CPR deformation parameter $\epsilon^r_3$, which cannot be constrained with Cygnus~X-1~\citep{cfm-cc2}, see Fig.~\ref{f-cfm3}.

In principle, one could find a source with a thermal spectrum harder than that which is expected for a Kerr BH with $a_* = 1$. This would essentially correspond to a spacetime in which the Novikov-Thorne radiative efficiency exceeds the Kerr BH bound $\eta_{\rm NT} = 0.42$ and could be an indication of deviations from the Kerr geometry. For the time being, there are no observations of this kind and therefore all the data are consistent with the Kerr metric.

The continuum-fitting method is probably the most robust technique among those available today. Its assumptions have been tested in a number of studies; for instance, the observed temporal constancy of the accretion disk's inner radius in the thermal state supports the assumption that the inner edge is at the ISCO~\citep{r_in}. However, the method also has some weak points. It cannot be applied to AGN, which represent the majority of the BH candidates, because their disk temperature is in the UV/optical bands. The measurements of $M$, $i$, and $d$ from optical observations are sometimes difficult and may be affected by systematic effects. Corrections for non-blackbody effects are usually taken into account by introducing the color factor, which is obtained from disk atmosphere models and there is not a unanimous consensus on their reliability~\citep{v2-davis1,v2-davis2}.

Assuming the systematics are under control, the thermal spectrum of a thin disk has a very simple shape, and it cannot provide much information on the spacetime geometry around the BH candidate. If we assume the Kerr metric, we can determine the spin parameter. If we have just one possible non-vanishing deformation parameter, we meet a degeneracy and, in general, we cannot constrain the spin and possible deviations from the Kerr solution at the same time. If we have a source that looks like a very fast-rotating Kerr BH, we can constrain some deformation parameters (e.g. the JP deformation parameters $\epsilon_k$), but other deviations from the Kerr geometry cannot be constrained (e.g. the CPR deformation parameters $\epsilon^r_k$). The reason is that the spectrum is simply a multi-color blackbody spectrum without additional features. Different parameters of the model have a quite similar impact on the shape of the spectrum and therefore there is a strong parameter degeneracy. The best that we can do is to combine the continuum-fitting measurements with other observations to break the parameter degeneracy.

\begin{figure}
\begin{center}
\includegraphics[type=pdf,ext=.pdf,read=.pdf,width=8.5cm]{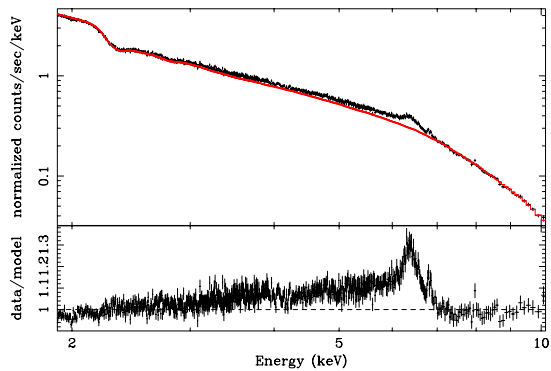}
\end{center}
\vspace{-0.5cm}
\caption{Photon count (in black) and fit without the iron line (in red) of MGC-6-30-15. From~\citet{BR06}. \copyright AAS. Reproduced with permission. }
\label{f-iron0}
\end{figure}

\begin{figure*}
\begin{center}
\vspace{-0.3cm}
\includegraphics[type=pdf,ext=.pdf,read=.pdf,width=8.5cm]{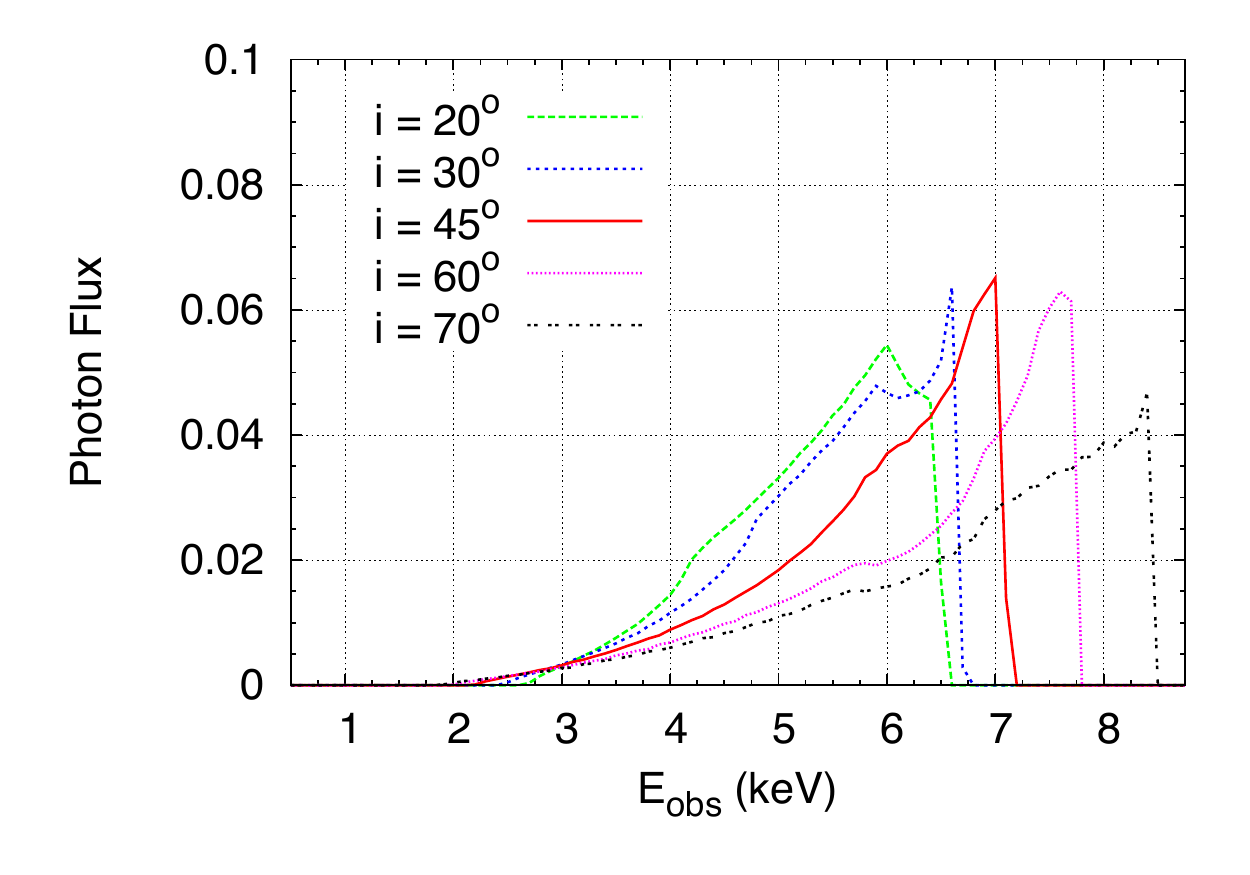}
\includegraphics[type=pdf,ext=.pdf,read=.pdf,width=8.5cm]{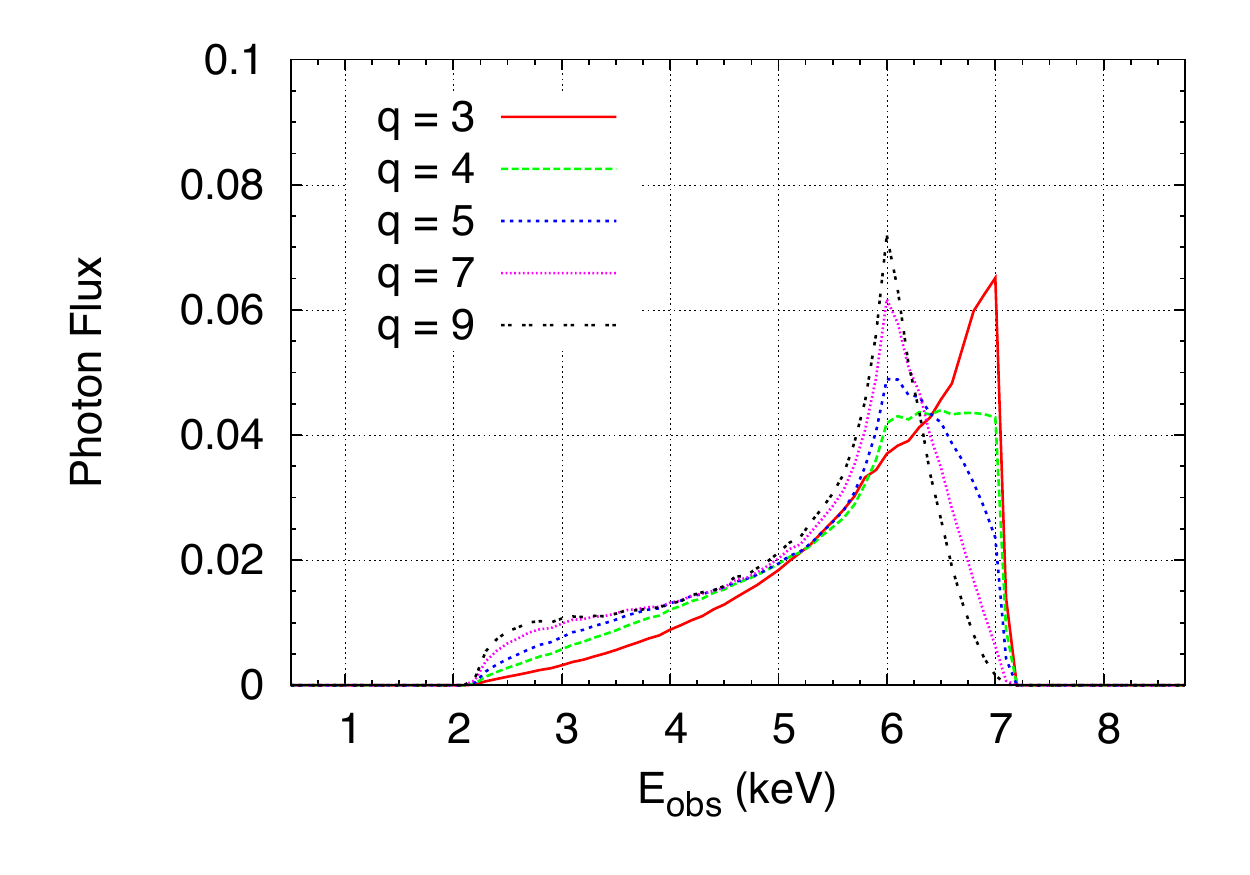} \\
\vspace{0.3cm}
\includegraphics[type=pdf,ext=.pdf,read=.pdf,width=8.5cm]{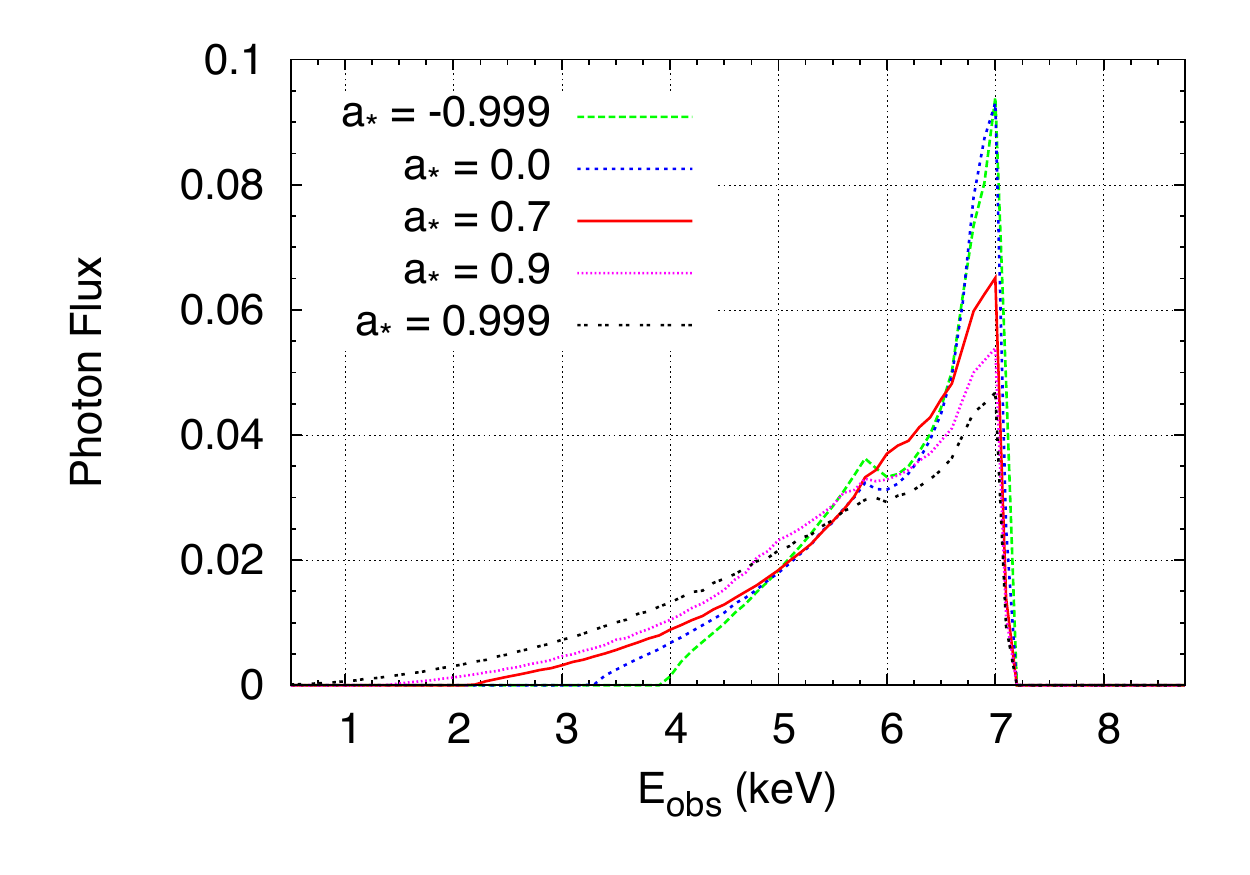}
\includegraphics[type=pdf,ext=.pdf,read=.pdf,width=8.5cm]{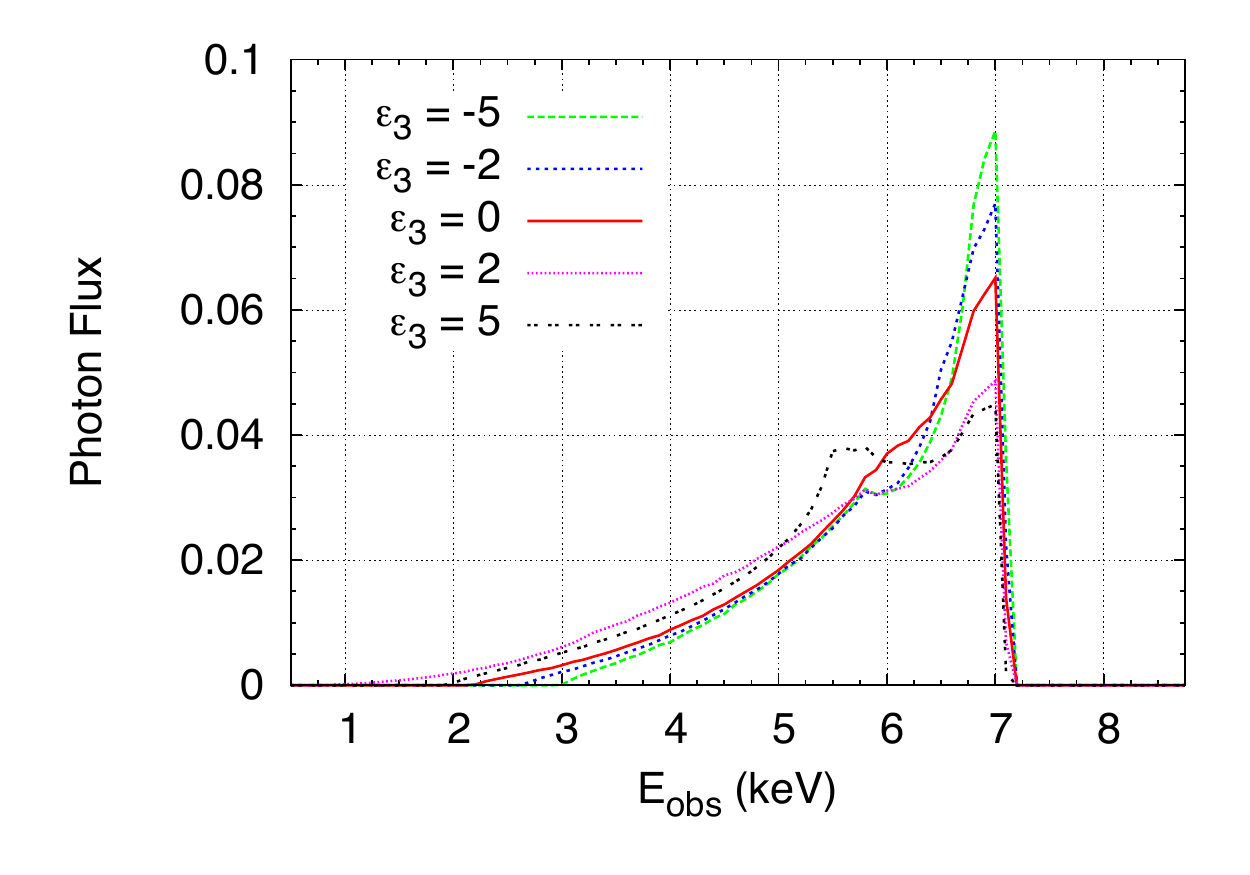}
\end{center}
\vspace{-0.5cm}
\caption{Impact of the model parameters on the iron line profile: viewing angle $i$ (top left panel), emissivity index $q$ (top right panel), spin parameter $a_*$ (bottom left panel), and JP deformation parameter $\epsilon_3$ (bottom right panel). When not shown, the values of the parameters are: $i = 45^\circ$, $q = 3$, $a_*=0.7$, and $\epsilon_3 = 0$. The outer radius is $r_{\rm out} = r_{\rm ISCO} + 100$~$M$. $E_{\rm obs}$ is the photon energy measured by the observer far from the BH. The vertical axis is the number flux of photons in arbitrary units. The iron line profiles are computed with the code described in~\citet{iron-c1}. \label{f-iron1}}
\end{figure*}

\begin{figure*}
\begin{center}
\includegraphics[type=pdf,ext=.pdf,read=.pdf,width=7.0cm]{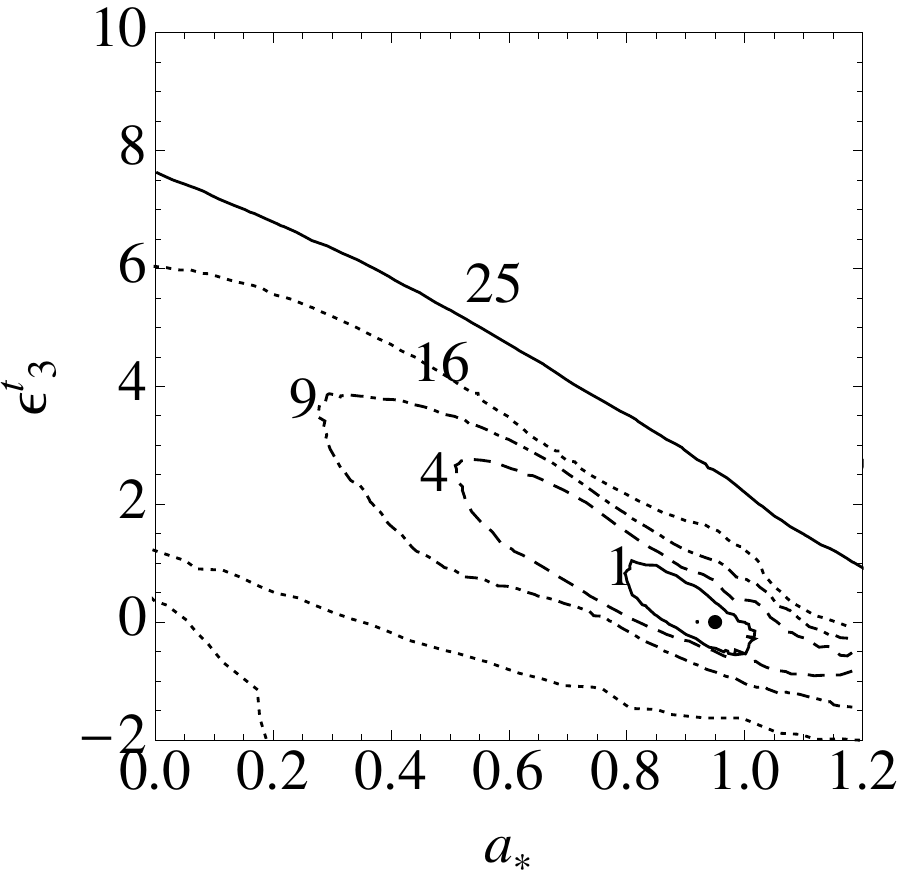}
\hspace{0.8cm}
\includegraphics[type=pdf,ext=.pdf,read=.pdf,width=7.0cm]{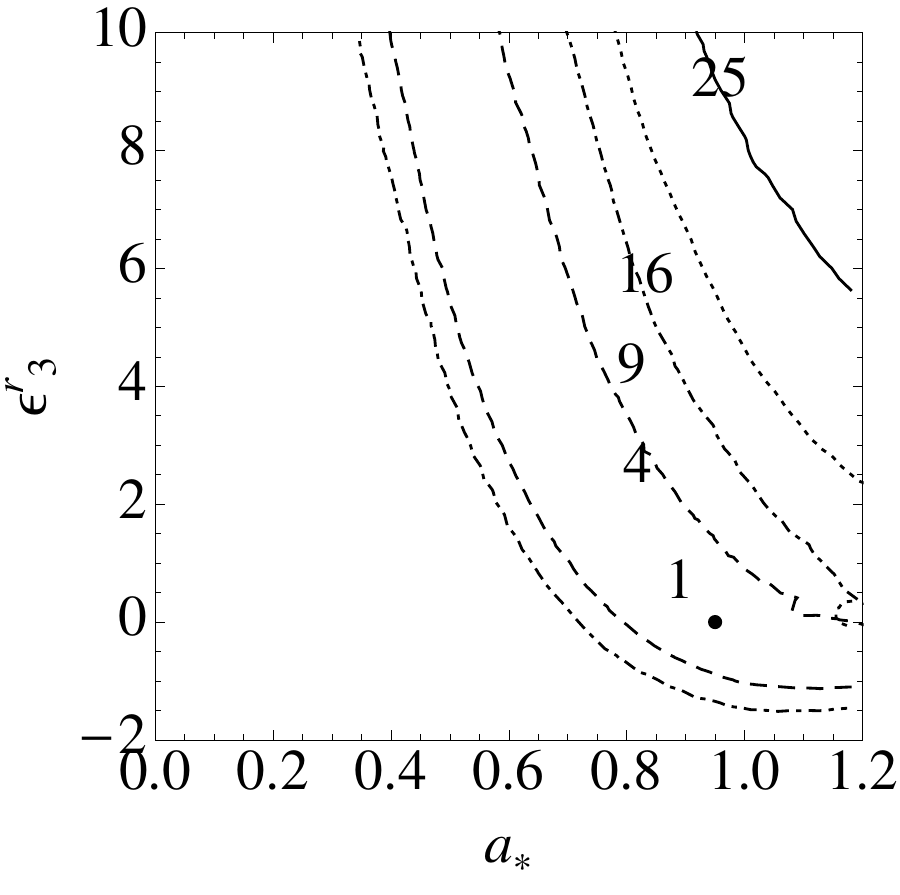}\\ 
\vspace{0.8cm}
\includegraphics[type=pdf,ext=.pdf,read=.pdf,width=7.0cm]{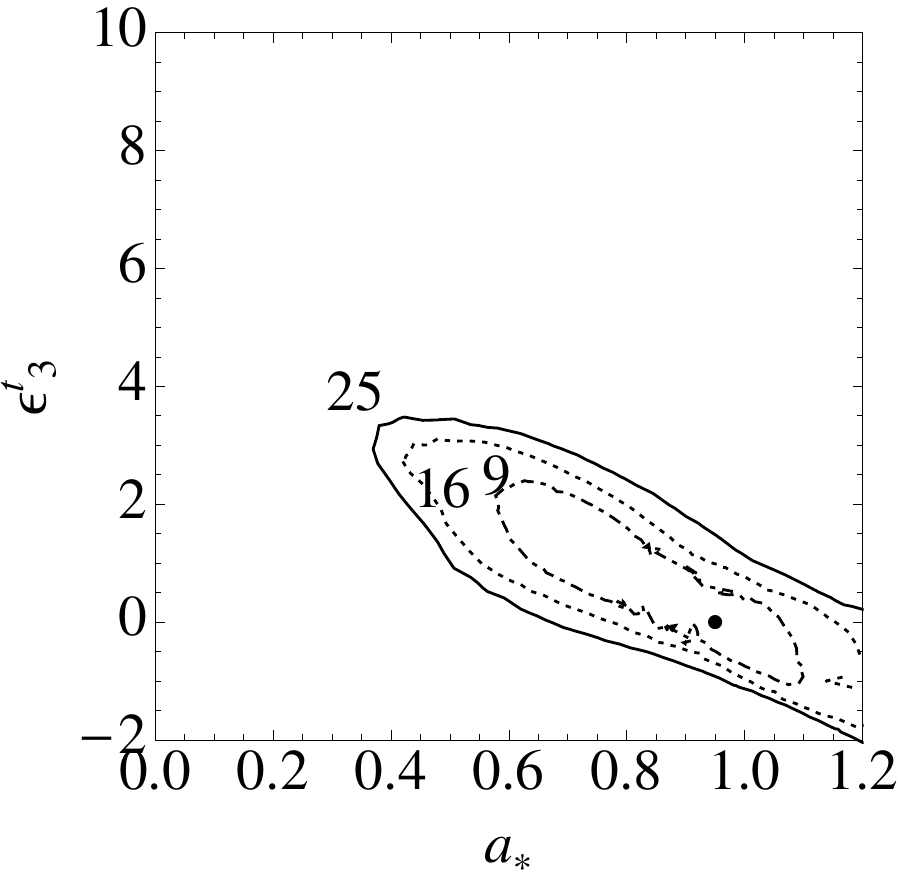}
\hspace{0.8cm}
\includegraphics[type=pdf,ext=.pdf,read=.pdf,width=7.0cm]{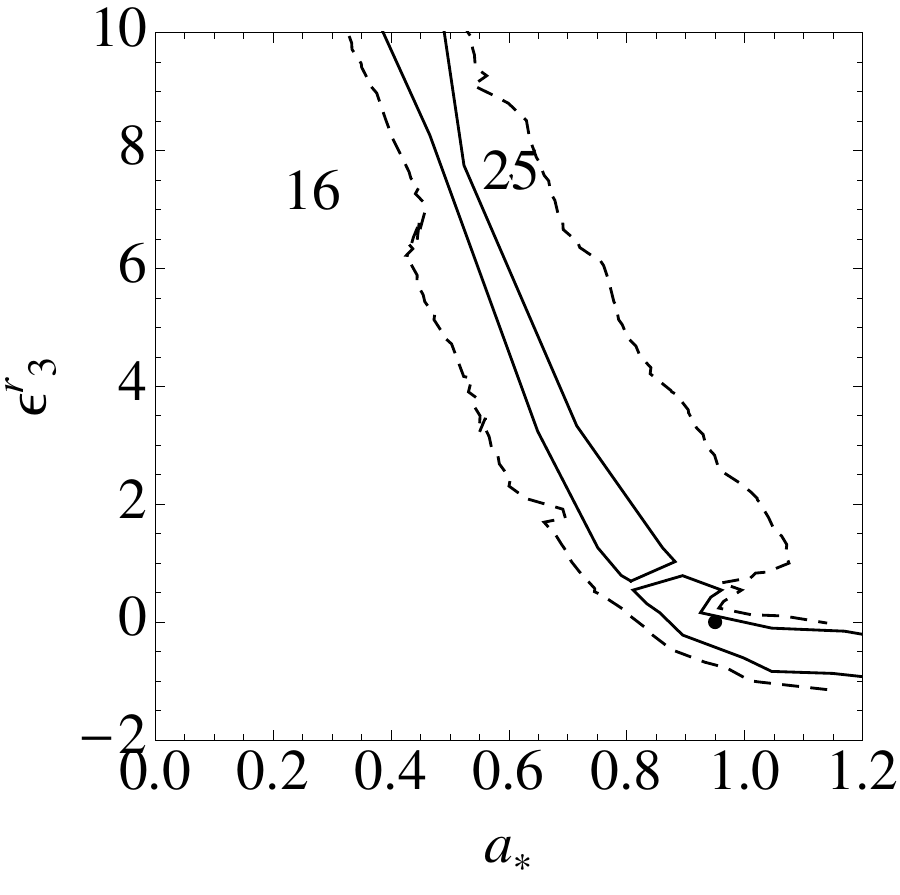}
\end{center}
\caption{$\Delta\chi^2$ contours with $N_{\rm line} = 10^3$ (top panels) and $N_{\rm line} = 10^4$ (bottom panels) from simulations of measurements of iron line profiles. The background geometry is the CPR metric with $\epsilon_3^t$ and $\epsilon^r_3$ as the only possible non-vanishing deformation parameters. The reference model is a Kerr BH with spin parameter $a_*' = 0.95$ and inclination angle $i' = 70^\circ$. In the left panels, we allow for a non-vanishing $\epsilon^t_3$ while we assume $\epsilon^r_3 = 0$. In the right panels we illustrate the converse measurement, namely $\epsilon^t_3 = 0$ while $\epsilon^r_3$ can vary. The ratio between the continuum and the iron line photon flux as well as the photon index of the continuum are also free parameters in the fit. The labels along the contour levels refer to the value of $\Delta\chi^2$. See the text for more details. From~\citet{jjc3}.}
\label{fig-iron-jjc}
\end{figure*}

\subsection{Iron line spectroscopy \label{ss-iii}}

The illumination of a cold disk by a hot corona produces a reflection component as well as some spectral lines by fluorescence in the X-ray spectrum of the source. The most prominent line is usually the iron K$\alpha$ line, which is at 6.4~keV in the case of neutral atoms and it shifts up to 6.97~keV for H-like iron ions. This line is intrinsically narrow in frequency, while the one observed in the X-ray spectrum of BH candidates appears broadened and skewed. The interpretation is that the line is strongly altered by relativistic effects, which produce a very characteristic profile first studied in~\citet{iron1}. This interpretation is currently well supported by reverberation measurements~\citep{utt-14}. The iron line is the strongest feature aside of the continuum, see Fig.~\ref{f-iron0}, and it can potentially provide quite detailed information on the spacetime geometry close to the BH candidate. This technique relies on fits of the whole reflected spectrum, but the spin measurement (or possible tests of the Kerr metric) is mainly determined by the iron line. For this reason the technique is often called the iron line method.

The shape of the line is primarily determined by the background metric, the geometry of the emitting region, the disk emissivity, and the disk's inclination angle with respect to the line of sight of the distant observer. In the Kerr background, the relativistic emission line profile emitted by an accretion disk illuminated by an X-ray corona with arbitrary geometry is typically parametrized by the BH spin $a_*$, the inner and the outer edges of the emission region $r_{\rm in}$ and $r_{\rm out}$, and the viewing angle $i$. The local spectrum $I_{\rm e}$ (the power radiated per unit area of emitting surface per unit solid angle per unit frequency) is determined by the reflection processes and the geometry of the system. Assuming axisymmetry, $I_{\rm e}$ depends on the photon energy, the emission radius, and possibly on the emission angle (the angle of propagation of the photon with respect to the normal to the disk) if the emission is not isotropic. The local spectrum may be modeled as a power-law $I_{\rm e} \propto r^{-q}$, where the emissivity index $q$ is a constant to be determined by the fit. A more sophisticated choice is to assume an intensity profile $I_{\rm e} \propto r^{-q}$ for $r < r_{\rm break}$ and $I_{\rm e} \propto r^{-3}$ for $r > r_{\rm break}$, where $q=3$ correspond to the Newtonian limit in the lamppost geometry at large radii far from the X-ray source. With this choice, we have two free parameters for the emissivity profile, $q$ and $r_{\rm break}$. If we assume a point-like hot corona located on the axis of the accretion disk and just above the BH, it is possible to compute the emissivity index $q = q(r)$ according to the height of the corona $h$~\citep{dauser}. Galactic BH binaries have line of sight velocities of just a few hundred km/s, which makes their relative motion negligible in the spectrum. In the case of AGN, the cosmological redshift of the source can instead be important for some objects and it can be an additional parameter of the model.

While the continuum-fitting method requires independent measurements of the mass, the distance, and the viewing angle of the source, the iron line analysis does not require them. Mass and distance have no impact on the line profile, as the physics is essentially independent of the size of the system, while the viewing angle can be inferred during the fitting procedure from the effect of the Doppler blueshift.

As in the continuum-fitting method, the iron line measurement assumes that the accretion disk can be described by the Novikov-Thorne model, and  crucially depends on the inner edge of the disk being located at the ISCO radius. The latter point can be tricky for the iron line analysis, because there is controversy over this assumption in hard states, where ``hard'' refers to the fact that the hard power-law component is the dominant feature in the spectrum. These states are those most frequently employed in reflection modeling.  Specifically, it is argued whether the disk is in fact truncated at some radius larger than the ISCO~\citep{v2-isco1,v2-isco2}. At present, the evidence indicates that if truncation is present, it is likely to be mild (a factor of a few).

In the case of stellar-mass BH candidates, one usually selects sources with high luminosity or sources in the hard-intermediate state, in which there are indications that the inner edge of the disk is closer to the compact object than in the hard states at low luminosities~\citep{v2-isco1}. In the case of supermassive BH candidates, this technique is more widely used, see the third column in Tab.~\ref{tab2}. In part, this is because we never have a precise estimate of the Eddington scaled accretion luminosity, due to the large uncertainties in the bolometric correction and mass estimates. This works both ways so that it is hard to say whether we are in the thin disk range or not. In part, it is because we cannot choose a different spectral state due to the much longer timescales of supermassive BHs than stellar-mass BHs.

If one were to fit data for a system in which the disk were truncated, and make the usual assumption that the inner-edge was at the ISCO, the fit would then incorrectly underestimate the value of spin (relative to the resulting Kerr prediction if it were truncated at the ISCO radius). The same question is at play for using reflection to test the Kerr metric with data from faint hard-states. In the case of too high accretion luminosities, the inner part of the disk may instead be geometrically thick, and this would lead to overestimate the BH spin (or, otherwise, get a wrong constraint on the deformation parameter). Some very high values of the BH spins reported in Tab.~\ref{tab2} are thus to be taken with caution~\citep{iron2}.

Moreover, this is not the end of the story. As clearly shown in~\citet{dauser} within the lamppost set-up, a spin measurement is only possible when the corona is close to the BH candidate. In particular, the characteristic low energy tail of the iron line used to measure the inner edge of the disk can be produced only in the case of a compact corona close to a fast-rotating BH candidate~\citep{dauser}. In the absence of the low energy tail, we may have either an extended corona far from the compact object or a slow-rotating BH candidate.

Bearing in mind these tricky points, the iron K$\alpha$ line is used to measure the BH spin under the assumption of the Kerr background and it can potentially be used to test the Kerr metric. For a review, see e.g.~\citet{iron2}. This technique can be used for both stellar-mass and supermassive BH candidates, because the iron line profile does not depend on the mass $M$. So it is currently the only available approach to probe the metric around supermassive BH candidates with the existing data. Current spin measurements of stellar-mass BH candidates with the iron line and under the assumption of the Kerr metric are reported in Tab.~\ref{tab1}. A summary of spin measurements of supermassive BH candidates is shown in Tab.~\ref{tab2}\footnote{In some cases, there exist a few independent measurements of the spin of the same source in the literature. Measurements from different groups are often consistent, but sometimes they are not; the discrepancy is more likely due to systematic effects not fully under control (see the references in the last column for the details).}.

The impact of the model parameters on the iron line profile is shown in Fig.~\ref{f-iron1}, where the emissivity profile has been modeled with a simple power-law $I_{\rm e} \propto r^{-q}$, and $q$ is constant. The iron line profile has a more complicated structure than the thermal spectrum of thin disks. However, as in the disk's thermal spectrum, the main effect of the spin parameter and of possible deviations from the Kerr geometry is often similar: both $a_*$ and the deformation parameter change the ISCO radius. This in turn affects the extension of the low energy tail of the line. There is thus a parameter degeneracy, as shown in~\citet{iron-jp}. As in the case of the continuum-fitting method, the technique can also be used to test the Kerr metric~\citep{iron-t,iron-jp,iron-c1}.

The iron line is potentially a more powerful tool than the continuum-fitting method to test the Kerr metric, in the sense that in the presence of high quality data it is typically possible to get independent estimates of the spin and the deformation parameter, while with the continuum-fitting method high quality data may not be enough to break the parameter degeneracy. \citet{iron-jp} estimated the required precision that observations with future X-ray missions have to achieve in order to measure potential deviations from the Kerr metric with the iron line. They found that a precision of about 5\% can constrain the absolute value of the JP deformation parameter $\epsilon_3$ to be smaller than 1 if the source is a Kerr BH with spin parameter $a_* > 0.5$ and the viewing angle is $30^\circ$ or $60^\circ$. The constraining power increases for fast-rotating Kerr BHs, because the inner edge of the disk is closer to the compact object.

\citet{jjc1,jjc2,jjc3} presented a study on the possibility of constraining the JP and CPR deformation parameters. In the case of a bright AGN, a good observation can have $N_{\rm line} \approx 10^3$ photons in the iron K$\alpha$ line. In the case of a bright stellar-mass BH candidate, $N_{\rm line}$ may be up to two orders of magnitude higher, say $N_{\rm line} \approx 10^5$. However, in stellar-mass BH candidates the low energy tail of the iron line overlaps with the thermal component of the accretion disk, the ionization is higher and less easily modeled, and Compton broadening plays a non-negligible role. All these effects make the brighter signal of stellar-mass BH lines messier to model. In \citet{jjc1,jjc2,jjc3}, we simulated iron lines in Kerr and non-Kerr backgrounds to estimate plausible constraints on the JP and CPR deformation parameters to be obtained from observations. We employed the standard analysis technique common throughout X-ray astronomy: the simulated spectra include Poisson noise and forward-folded fit after rebinning\footnote{Forward-folded fitting is the term commonly used in X-ray astronomy to generically indicate the fitting procedure. The actual spectrum measured by an instrument (in units of counts per spectral bin) can be written as
\be
C(h) = \tau \int R(h,E) \, A(E) \, s(E) \, dE \, ,
\ee
where $h$ is the spectral channel, $\tau$ is the exposure time, $R(h,E)$ is the redistribution matrix (essentially the response of the instrument), $A(E)$ is the effective area, and $s(E)$ is the intrinsic spectrum of the source. In general, the redistribution matrix cannot be inverted, and therefore it is not possible to obtain the intrinsic spectrum $s(E)$. Forward-folded fitting means that one assumes a theoretical model for $s(E)$ with a set of parameter values, convolves the model with the instrument response, and its result (called the folded spectrum) is compared with the observed spectrum through a goodness-of-fit statistical test. The process is repeated by changing the parameter values to find the best fit. For example, this approach is not necessary at optical wavelengths, where the redistribution matrix can be inverted, and therefore one can directly fit the theoretical model with the observed intrinsic spectrum of the source. Rebinning means that one groups data with a low count number to increase the number of counts per bin. For more details, see, e.g., \citet{v4-ass}.} to ensure that the count distribution in each spectral channel is well approximated by a Gaussian distribution, as required to use the $\chi^2$ as goodness-of-fit test.

Fig.~\ref{fig-iron-jjc} shows the $\Delta\chi^2$ contours with $N_{\rm line}=10^3$ (top panels) and $N_{\rm line} = 10^4$ (bottom panels) assuming as a reference model a Kerr BH with spin parameter $a_* = 0.95$ and observed with a viewing angle $i=70^\circ$. The spacetime geometry is described by the CPR metric and we have five free parameters, namely the spin $a_*$, the deformation parameter ($\epsilon^t_3$ or $\epsilon^r_3$), the viewing angle $i$, the ratio between the continuum and the photon iron line flux, say $K$, and the photon index of the continuum $\Gamma$. $\chi^2$ is thus minimized over $i$, $K$, and $\Gamma$. In the left panels, $\epsilon^t_3$ is a free parameter and $\epsilon^r_3 = 0$. In the right panels, $\epsilon^t_3=0$ and $\epsilon^r_3$ can vary. It is evident that $\epsilon^t_3$ is relatively easy to constrain, while $\epsilon^r_3$ is much more difficult. Even in the case of a fast-rotating Kerr BH observed with a large inclination angle, it is impossible to constrain $\epsilon^r_3$, in the sense that large positive values of $\epsilon^r_3$ cannot be ruled out. The difference in the constraints on $\epsilon^t_3$ and $\epsilon^r_3$ can be understood noting that $\epsilon^t_3$ strongly affects the position of the ISCO. The impact of $\epsilon^r_3$ on the ISCO is weaker (it enters only in $g_{t\phi}$). $\epsilon^r_3$ mainly affects the propagation of the photons in the spacetime, but the impact on the time-integrated iron line measurement is weak. More details on the constraints on the CPR deformation parameters from the iron line can be found in \citet{jjc2,jjc3}.

As the photon count number increases, the constraints become stronger. It is not clear what the best current observations can provide and work on real data is in progress. It is however important to note that current theoretical models are quite simple. The emissivity profile, for instance, is often modeled with two power-law indices and a breaking radius, which is clearly an approximation. Even if we presently had excellent data with a large number of photons in the iron line, it would be presumably impossible to get reliable constraints on the Kerr metric. More realistic modeling would be necessary to prevent systematic effects from becoming dominant. In the end, any measurement is as good as its theoretical model, and current iron line models are phenomenological and oversimplify the astrophysical picture.

\subsubsection{Comparison between continuum-fitting and iron line measurements}

In the case of stellar-mass BH candidates, there are some objects in which the spin has been measured by both the continuum-fitting and the iron line methods, and the estimate is usually consistent, see Tab.~\ref{tab1}. Let us note, however, that this is not enough to claim that these objects are Kerr BHs, because both techniques are mainly sensitive to the position of the inner edge of the disk and even in the presence of a non-Kerr background they should provide a very similar estimate of the spin parameter when it is assumed to be the Kerr metric~\citep{cfm-iron}. The agreement between the two techniques is however a good result to believe in the robustness of the two measurements. The disagreement in the case of some objects is surely due to systematic effects and cannot be solved by postulating a non-Kerr metric. Only in the case of very precise measurements, depending on the actual metric around the compact object, can the difference between the continuum-fitting and the iron line measurements be attributed to deviations from the Kerr solution~\citep{cfm-iron}.

\begin{figure*}
\begin{center}
\includegraphics[type=pdf,ext=.pdf,read=.pdf,width=8.5cm]{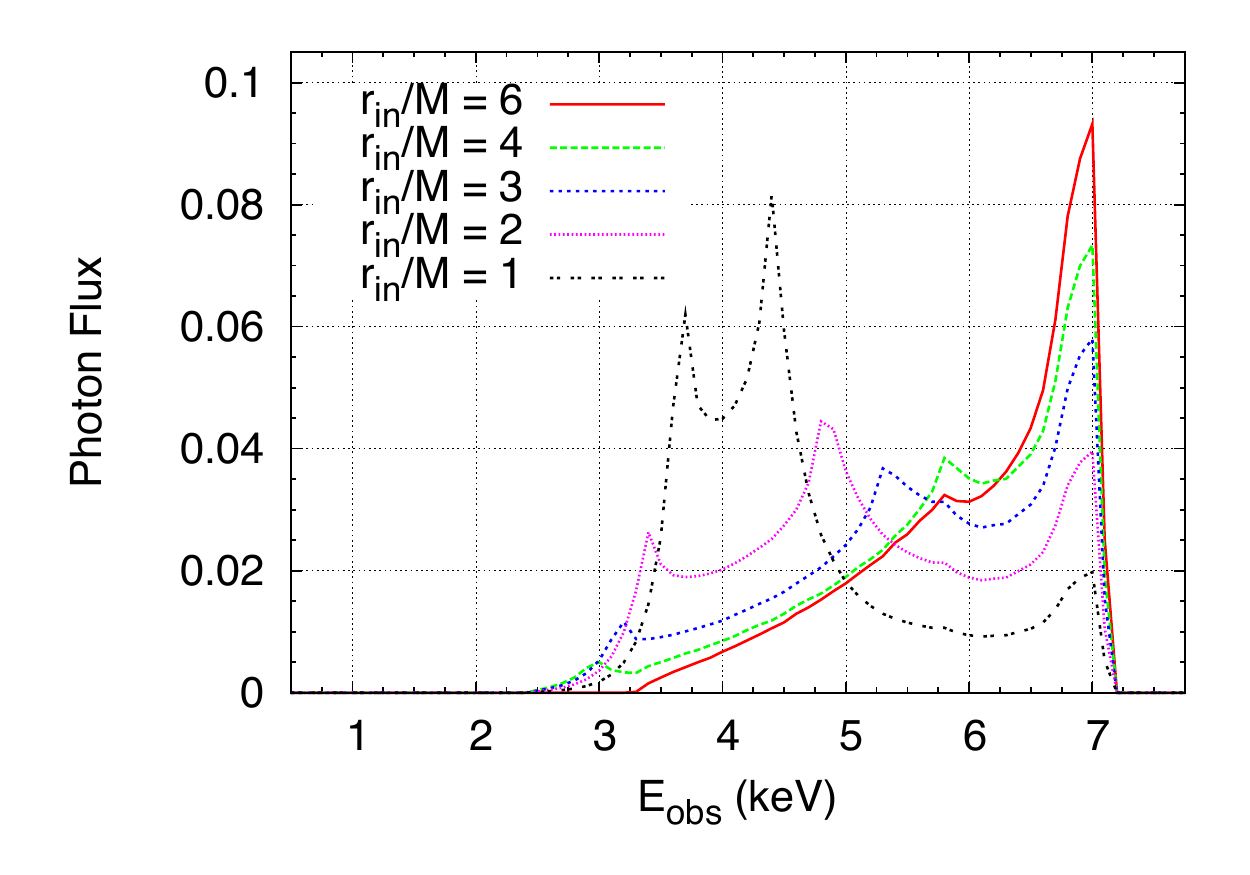}
\includegraphics[type=pdf,ext=.pdf,read=.pdf,width=8.5cm]{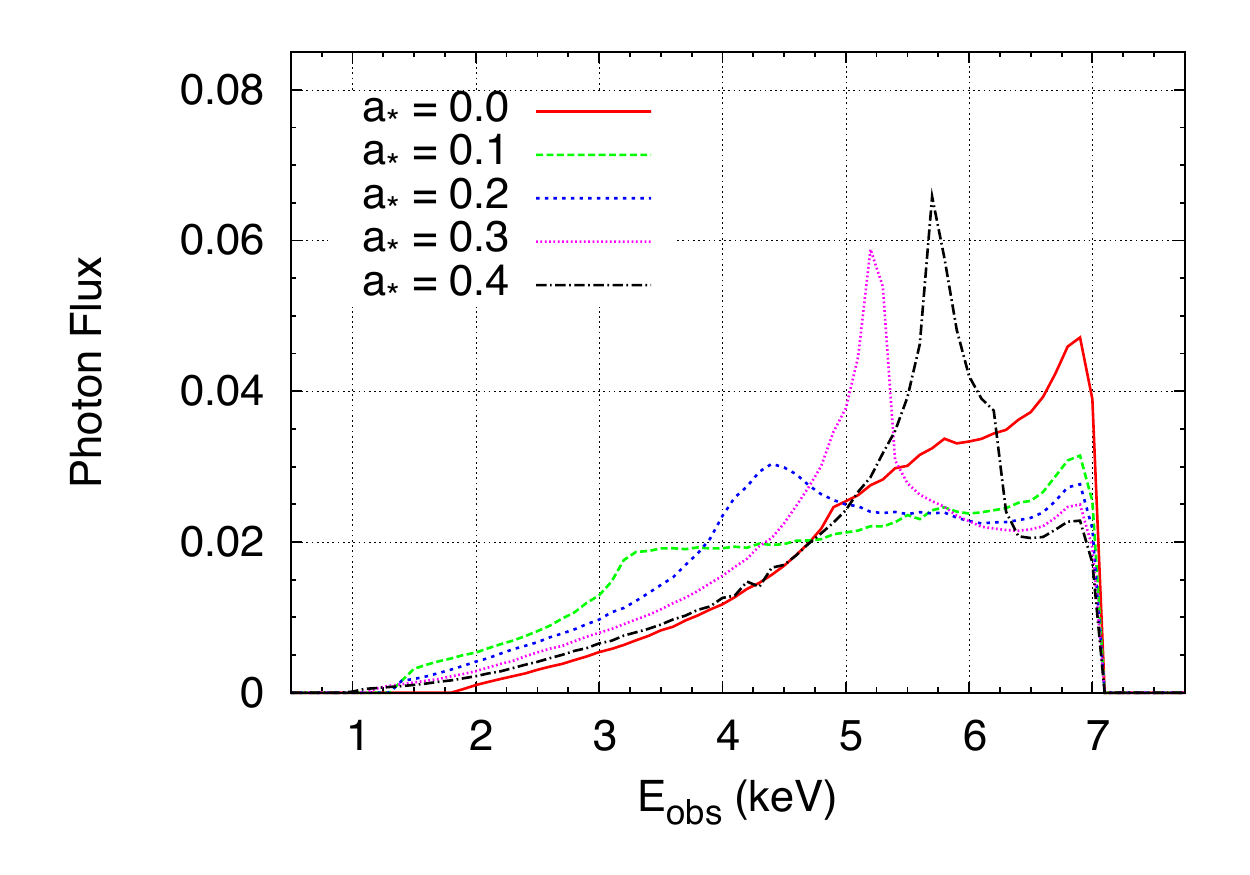}
\end{center}
\vspace{-0.5cm}
\caption{Iron line profiles produced from the accretion disks around a class of exotic compact stars (left panel) and a certain family of traversable wormholes (right panel). $E_{\rm obs}$ is the photon energy measured by the observer far from the BH. The vertical axis is the number flux of photons in arbitrary units. See the text for more details. The iron line profiles are computed with the code described in~\citet{iron-c1}, employing the metrics discussed in~\citet{exotic1} and \citet{exotic2}. \label{f-iron4a}}
\end{figure*}

\begin{figure}
\begin{center}
\includegraphics[type=pdf,ext=.pdf,read=.pdf,width=8.5cm]{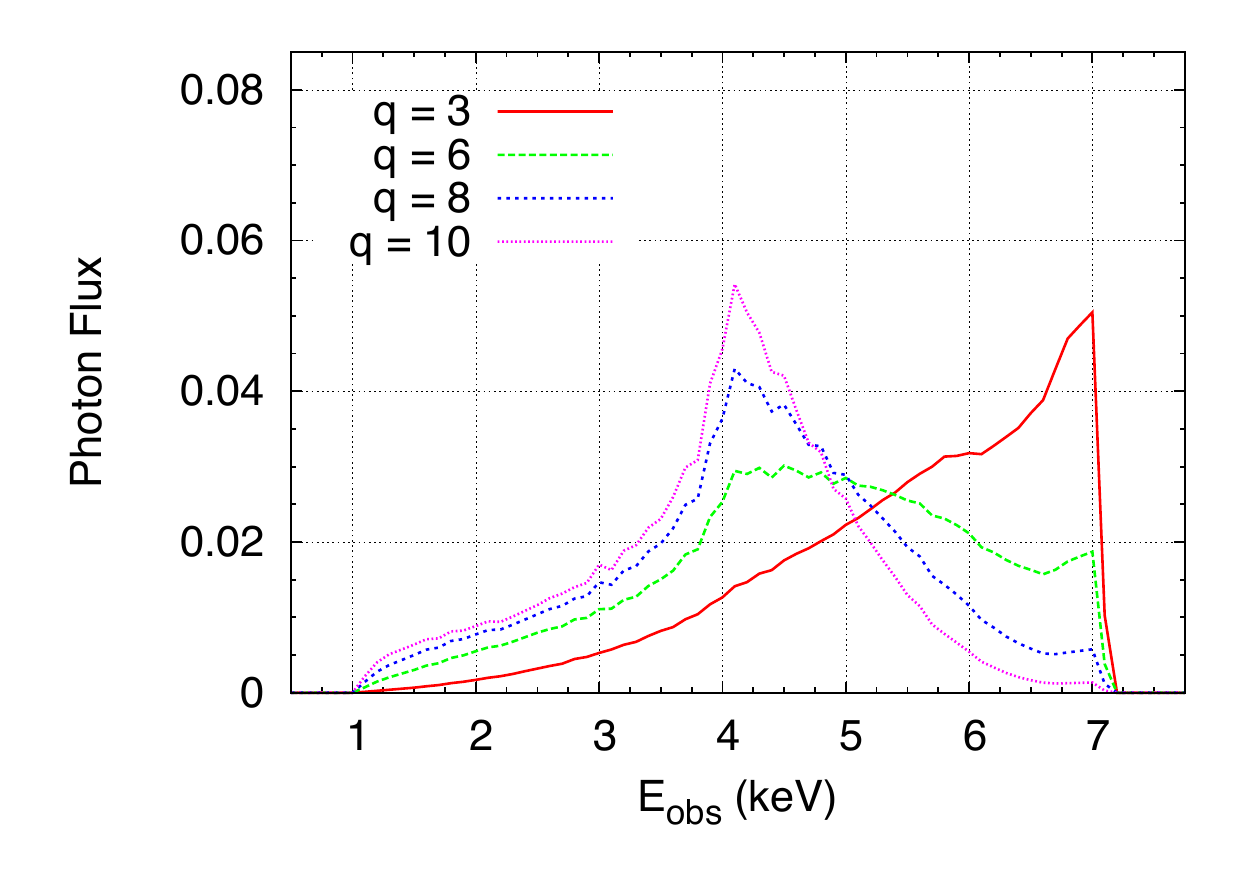}
\end{center}
\vspace{-0.5cm}
\caption{Examples of iron line profile from a Kerr BH with spin parameter $a_* = 0.95$ and observed from an inclination angle $i = 45^\circ$. The emissivity is $I_{\rm e} \propto (r_{\rm break}/r)^q$ for $r < r_{\rm break} = 5$~$M$ and $I_{\rm e} \propto (r_{\rm break}/r)^3$ for $r > r_{\rm break}$. See the text for more details. The iron line profiles are computed with the code described in~\citet{iron-c1}. \label{f-iron4b}}
\end{figure}

\subsubsection{Exotic candidates}

Note that the iron line can immediately rule out some exotic BH alternatives, or at least constrain their properties. This is possible because in some spacetimes the iron line profile should have qualitatively different features that are not observed in real data or, vice versa, their iron line cannot mimic that observed in real data. This permits one to conclude that these objects cannot be the BH candidates of the Universe.

Exotic compact stars represent a large class of BH alternatives. One usually assumes that general relativity holds, but the mass of the compact object can exceed the famous 3~$M_\odot$ limit of a neutron star thanks to special equations of state. Boson stars also belong to this group of objects~\citep{b-stars1,b-stars2} and they may also have a mass of millions or billions $M_\odot$, so that they may be the supermassive BH candidates at the center of galaxies. These spacetimes have typically no ISCO, namely equatorial circular orbits are always stable. One can set the inner edge of the disk at some radius $r_{\rm in}$. Since the gravitational redshift of the photons emitted in the inner part of the accretion disk is never very strong, the iron line profile cannot have the characteristic low energy tail of the iron line expected in the case of very fast-rotating BHs and observed in the spectra of several BH candidates~\citep{exotic1}. In the case of fast-rotating BHs, the radiation emitted from the inner part of the accretion disk is strongly redshifted. This produces an extended low energy tail in the iron line. In the case of compact stars, the gravitational redshift is usually not very strong even at very small radii. Examples of iron line profiles from the accretion disk of exotic compact stars are shown in the left panel in Fig.~\ref{f-iron4a}. See~\citet{exotic1} for more details. This argument does not imply the existence of a horizon, and consequently does not completely rule out scenarios with exotic compact stars, but at least it requires that the gravitational redshift close to BH candidates is extremely strong. It is not far from that expected in the vicinity of an apparent/event horizon, and this is not the case in typical exotic star models.

Another example of strange candidates is represented by the family of traversable wormholes discussed in~\citet{exotic2}. Their iron line profiles are shown in the right panel in Fig.~\ref{f-iron4a}. These objects have no event horizon, like the compact stars mentioned in the previous paragraph. Nevertheless, their iron line profile has a low energy tail. Here it is actually impossible not to have an extended low energy tail and the iron line of non-rotating wormholes looks like that from a fast-rotating Kerr BH~\citep{exotic2}. An interesting feature of these iron lines is the peak at low energies. Like in the exotic compact stars in~\citet{exotic1}, it is due to the photons emitted at very small radii, which can more easily escape to infinity than in the case of a BH spacetime. Can this feature be used as a smoking gun to rule out these spacetimes? It depends on the quality of the X-ray data and on the specific metric~\citep{v3-menglei}. In Fig.~\ref{f-iron4a}, the iron line profile has been computed assuming an emissivity $I_{\rm e} \sim r^{-3}$, corresponding to the Newtonian limit at large radii in the lamppost geometry. For such an emissivity profile, the iron lines in the Kerr background have no similar peaks, see Fig.~\ref{f-iron1}. However, in a correct relativistic lamppost geometry one should expect a much steeper emissivity function at smaller radii and for a source close to the compact object, as a result of strong light bending~\citep{dauser}. Observations seem to require a high, or even very high, value of $q$ at small radii~\citep{high-q-1,high-q-2}\footnote{However, this interpretation was criticized in \citet{high-q-3}.}. Fig.~\ref{f-iron4b} shows the iron line in a Kerr spacetime with $a_* = 0.95$ and $i = 45^\circ$, where the emissivity function is
\be
I_{\rm e} \propto \left\{
\begin{array}{rl}
\left(\frac{r_{\rm break}}{r}\right)^q & \text{if } r < r_{\rm break} \, , \\
\left(\frac{r_{\rm break}}{r}\right)^3 & \text{if } r > r_{\rm break} \, ,
\end{array} \right.
\ee
with $r_{\rm break} = 5$~$M$ and $q$ assumes different values. As shown in Fig.~\ref{f-iron4b}, such an emissivity profile could reproduce a peak at lower energies! The reason is that the emissivity is much higher at small radii than the simple case $r^{-3}$. Even if in the Kerr metric it is more difficult to escape from small radii, a much higher emissivity at small radii can balance the BH photon capture and have an iron line that looks like that of a traversable wormhole with lower emissivity at small radii. Eventually, only in the presence of high quality data is it possible to distinguish the difference between the astrophysical model and the metric and therefore between wormholes and BHs~\citep{v3-menglei}.

\begin{figure*}
\begin{center}
\includegraphics[type=pdf,ext=.pdf,read=.pdf,width=7.0cm]{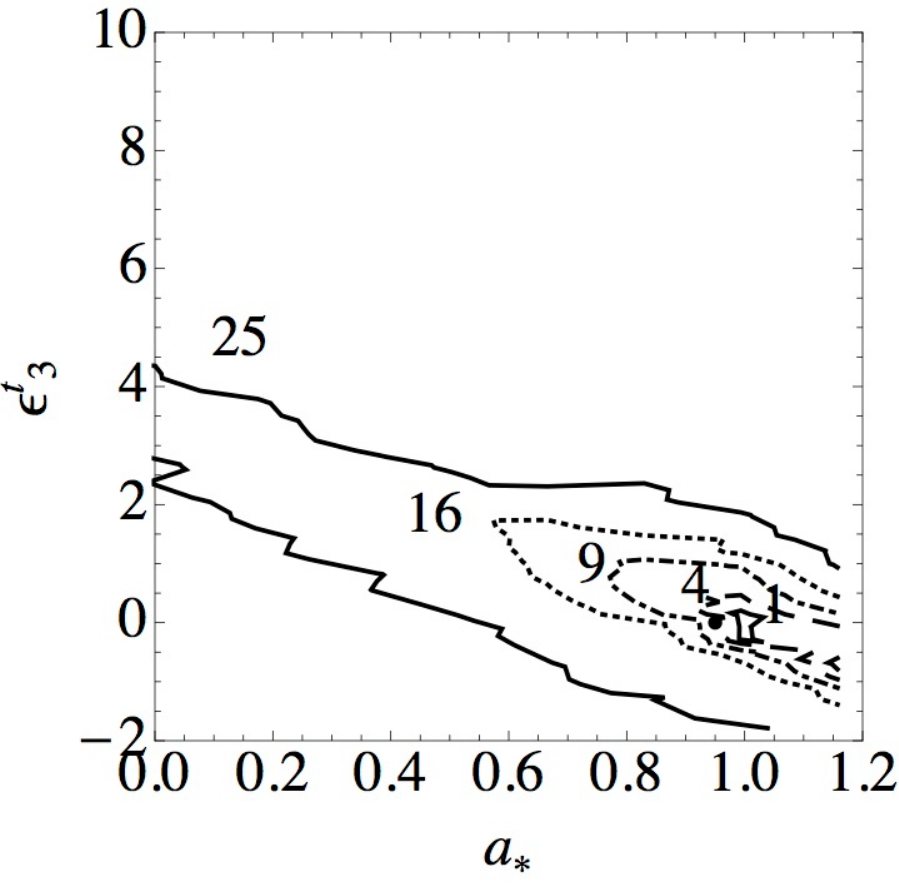}
\hspace{0.8cm}
\includegraphics[type=pdf,ext=.pdf,read=.pdf,width=7.0cm]{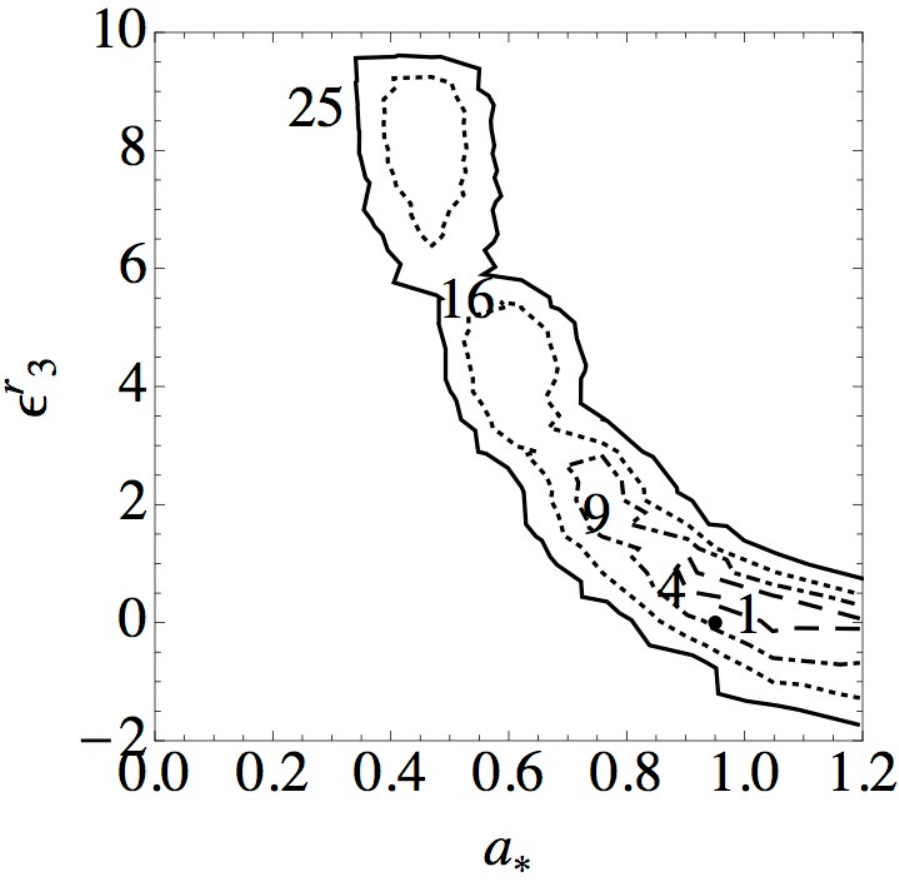}\\ 
\vspace{0.8cm}
\includegraphics[type=pdf,ext=.pdf,read=.pdf,width=7.0cm]{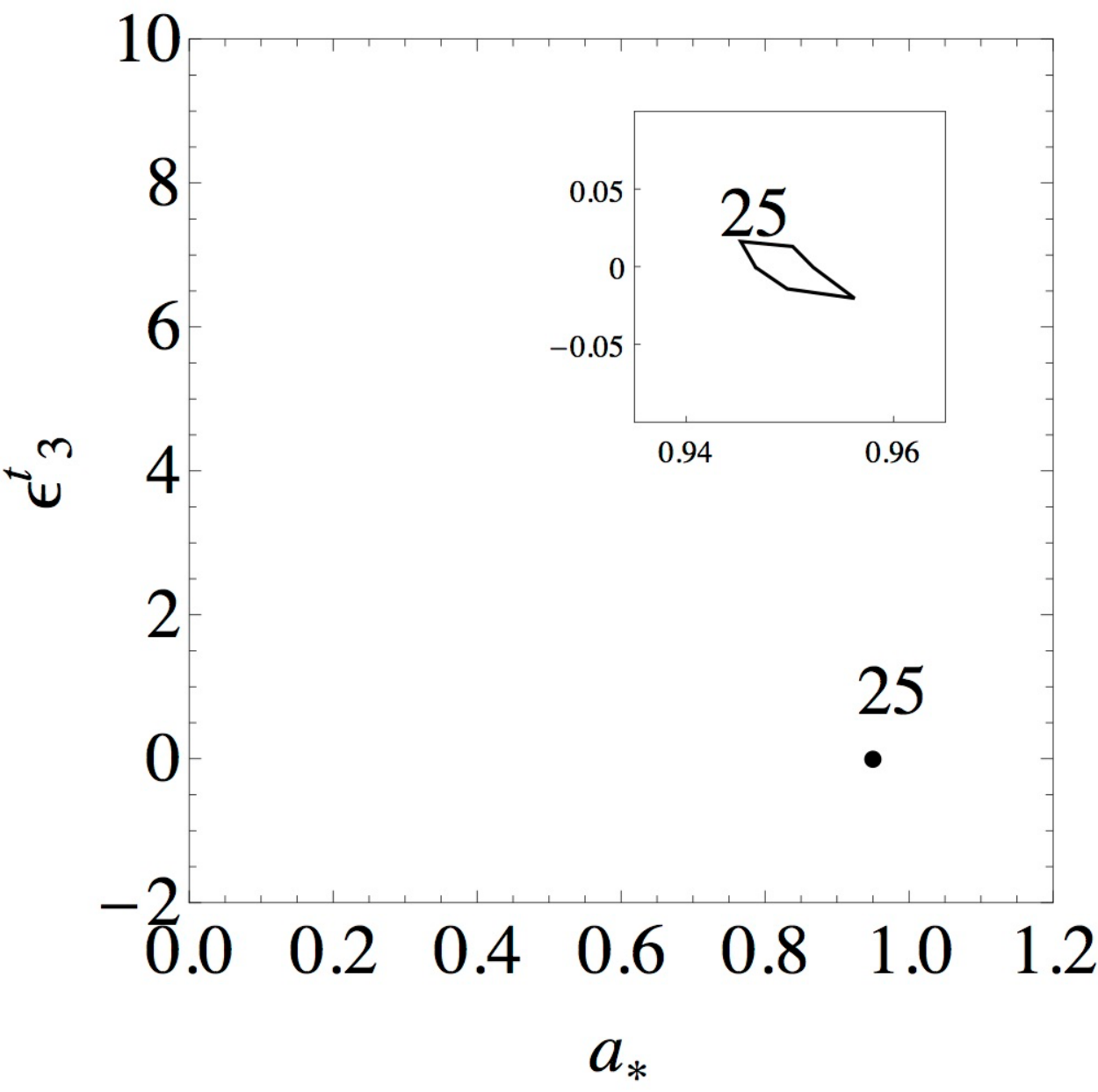}
\hspace{0.8cm}
\includegraphics[type=pdf,ext=.pdf,read=.pdf,width=7.0cm]{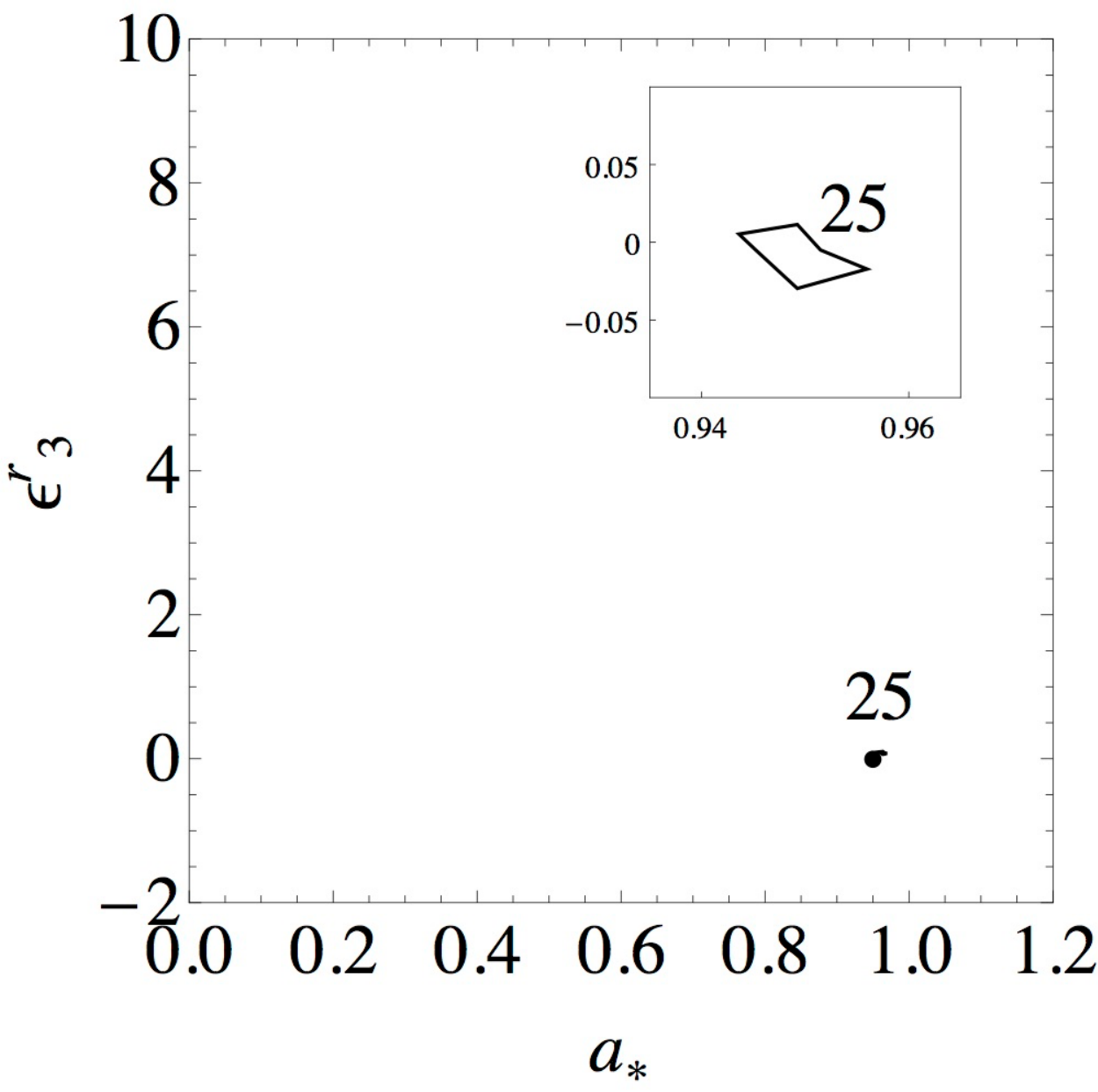}
\end{center}
\caption{Compare to Fig.~\ref{fig-iron-jjc}. $\Delta\chi^2$ contours with $N_{\rm line} = 10^3$ (top panels) and $N_{\rm line} = 10^4$ (bottom panels) from the analysis of the 2D transfer function. The reference model is a Kerr BH with spin parameter $a_*' = 0.95$ and inclination angle $i' = 70^\circ$. In the left panels, we allow for a non-vanishing $\epsilon^t_3$ and we assume $\epsilon^r_3 = 0$. In the right panels we consider the converse case, namely $\epsilon^t_3 = 0$ and $\epsilon^r_3$ can vary. The height of the source, the ratio between the continuum and the iron line photon flux, and the photon index of the continuum are also left as fit parameters. The labels along the contour levels refer to the value of $\Delta\chi^2$. See the text for more details. From~\citet{jjc3}.}
\label{fig-rev-jjc}
\end{figure*}

\subsection{Iron line reverberation \label{ss-iii2}}

Within the corona-disk model with lamppost geometry, the activation of a new flaring region in the corona illuminates the accretion disk and generates a time-dependent iron line profile due to the different propagation time for different photon paths~\citep{reverb1}. Reverberation refers to the iron line signal as a function of time in response to a flash of radiation from the corona. The resulting line spectrum as a function of both time and across photon energy is called the 2D transfer function, which is related to fundamental properties of the BH candidate and of the system geometry.

In the case of supermassive BH candidates, iron line reverberation measurements are currently possible and probably represent the most convincing argument in support of the fact the iron line originates from the inner part of the accretion disk, and therefore that its shape is determined by relativistic effects. For a review, see e.g.~\citet{utt-14}. However, with current X-ray facilities, because of the limited count rates in the iron line, reverberation measurements are not more powerful than the standard time-integrated ones in probing the spacetime around supermassive BH candidates. Future detectors with larger effective areas should be able to better study the temporal change in response to the activation of new flares and reverberation measurements can probably become a more powerful technique than time-integrated observations.

Iron line reverberation mapping in a non-Kerr background was investigated in~\citet{jjc1,jjc3}. Fig.~\ref{fig-rev-jjc} shows the constraints on the CPR deformation parameter $\epsilon^t_3$ and $\epsilon^r_3$ from simulations. These plots can be directly compared with those in Fig.~\ref{fig-iron-jjc} for a time-integrated iron line measurement. The reference model is the same, namely a Kerr BH with spin parameter $a_*=0.95$ and a viewing angle $i=70^\circ$. The top panels refer to observations with a photon number count in the iron line $N_{\rm line} =10^3$, and the bottom panels refer to the case $N_{\rm line} =10^4$. In the left panel, $\epsilon^t_3$ can vary and $\epsilon^r_3=0$ is frozen. In the right panels, we have the opposite cases, so $\epsilon^t_3=0$ and $\epsilon^r_3$ is free. The height of the source, the ratio between the continuum and the iron line photon flux, and the photon index of the continuum are also left as fit parameters.

The case $N_{\rm line} =10^3$ roughly corresponds to a good observation of a bright AGN with the current X-ray facilities. The reverberation measurement can constrain the background metric better than the time-integrated observation. However, it is still problematic to measure the CPR deformation parameter $\epsilon^r_3$. Reverberation mapping becomes much more powerful than the measurement of the time-integrated iron line when $N_{\rm line}$ increases. This is perfectly understandable. With a low photon number count, the large number of channels in the reverberation approach dilutes the photon count per channel and the intrinsic noise of the source frustrates the additional time information in the measurement. In the simulations with $N_{\rm line} =10^4$, we find that the reverberation measurement can constrain the CPR deformation parameter $\epsilon^r_3$. Interestingly, as discussed in~\citet{jjc3}, this is true even if the source is a slow-rotating Kerr BH observed from a low inclination angle. This is probably possible because $\epsilon^r_3$ affects the photon propagation in the background metric. Time-integrated observations are not very sensitive to it, while the time information of the photon propagation allows one to constrain $\epsilon^r_3$.

Time-resolved measurements may also be possible from the observation of an X-ray AGN eclipse~\citep{martin}. In principle, one could still exploit the variability of the source in order to probe different regions of the spacetime at different times, and better separate the relativistic effects occurring near the BH candidate. However, \citet{v3-cjb1} showed that this is not the case and an eclipse measurement does not have the advantage of a reverberation one. The difference between the two approaches is related to the capability of separating photons from different parts of the disk. In the reverberation approach, photons emitted from different regions are detected at different times. In the eclipse scenario we have the opposite case, namely one studies the properties of the radiation from every region of the accretion disk from the non-detection of the photons from that patch.

\subsection{Quasi-periodic oscillations \label{ss-qpos}}

Quasi-periodic oscillations (QPOs) are a common feature in the X-ray flux of stellar-mass BH candidates. They appear as peaks in the X-ray power density spectra of the source. They are thought to be a very promising tool for the future to get precise information on the spacetime geometry around BH candidates. However, there is currently no consensus on which mechanism is responsible for their production or even if it is a single mechanism or multiple ones. Many scenarios have been proposed, and in particular there are relativistic precession models~\citep{qpo1,qpo2}, diskoseismology models~\citep{qpo3}, resonance models~\citep{qpo4a,qpo4b}, p-mode oscillations of accretion tori~\citep{qpo5}. Other mechanisms may also be possible. Interestingly, in most scenarios the frequencies of the QPOs are directly related to the characteristic orbital frequencies of a test-particle (orbital or Keplerian frequency $\nu_\phi$, radial epicyclic frequency $\nu_r$, and vertical epicyclic frequency $\nu_\theta$), which are only determined by the background metric and are thus independent of the complicated astrophysical processes of the accretion. For this reason, QPOs are a promising technique to probe the metric around BH candidates. Moreover, the frequencies of QPOs can be measured with high accuracy, and therefore they can potentially provide more precise measurements than other techniques like the continuum-fitting and the iron line methods.

Different models provide a different measurement of the parameters of the background metric, which means that QPO data cannot be used to test fundamental physics at this time. However, there is already some work exploring the possibility of using QPOs to test the Kerr metric~\citep{qpos,qpojp,qpoc1,qpoa,qpoc2,qpo-rome}. Even if QPO data can potentially provide very accurate measurements, there is a fundamental degeneracy among the spin parameter and possible deviations from the Kerr solution, so one can typically only obtain a narrow allowed region on the spin parameter -- deformation parameter plane.

\citet{qpojp} discussed the constraining power of the diskoseismology model and of the 1:2 resonance model involving the Keplerian and the radial epicyclic frequencies in the framework of the non-Kerr metric of \citet{tests-gw1}. In the diskoseismology scenario, the pair of high-frequency QPOs observed in the X-ray flux of some BH binaries can be identified as the lowest order gravity (g-modes) and corrugation modes (c-modes). Within the Kerr metric, the measurements of the two frequencies would provide the values of the BH mass and spin because the lowest order modes would occur near the ISCO, so the radius is fixed. When we want to test the Kerr metric and we have a non-vanishing deformation parameter, we need an independent measurement of the mass, and we can get a value for the spin and the deformation parameter. In the case of the 1:2 resonance model, there is a degeneracy among the mass, the spin, and possible deviations from Kerr. A possible non-vanishing deformation parameter can only be constrained in the presence of independent measurements of the mass and the spin of the compact object.

In \citet{qpoc1}, I considered a number of different resonance models to constrain the JP deformation parameter $\epsilon_3$. Consistently with \citet{qpojp}, any resonance model can measure just one number of the spacetime geometry. In \citet{qpoc1}, the constraints of the resonance model for the BH candidates GRO~J1655-40, XTE~J1550-564, and GRS~1915+105 were combined with the dynamical measurement of their mass and with their constraints from the continuum-fitting method. Assuming the Kerr metric, no resonance model provides a spin measurement consistent with the continuum-fitting method for the three BH candidates at the same time. In the case of a non-vanishing $\epsilon_3$, the 3:1 resonance model involving the two epicyclic frequencies can be consistent with the continuum-fitting method estimates for the three objects.

In~\citet{qpoc2}, I studied the constraints on the JP deformation parameter $\epsilon_3$ for the BH candidate in GRO~J1655-40 assuming the relativistic precession model of \citet{qpo2}. GRO~J1655-40 is a special source because it is the only object for which we have a detection of three simultaneous QPOs. Assuming that the three QPOs are associated with oscillations of the fluid flow at the same radial coordinate, in the Kerr metric there are three unknown quantities (the BH mass, the BH spin, and the radial coordinate of the fluid oscillation) and it is possible to solve the system. In the case of a non-Kerr metric with only one non-vanishing deformation parameter, we need an independent measurement of the BH mass. In the case of GRO~J1655-40, there are two main measurements~\citep{gro-m1,sh06}, which are not consistent each other. If we adopt the mass estimate of~\citet{gro-m1}, the constraint on $\epsilon_3$ is consistent with the Kerr metric, but the estimate of the BH spin is not consistent with that from the continuum-fitting method. If we choose the mass estimate in~\citet{sh06}, the measurements of the continuum-fitting method and the QPOs may be consistent for a non-vanishing $\epsilon_3$.

\begin{figure*}
\begin{center}
\includegraphics[type=pdf,ext=.pdf,read=.pdf,width=8.9cm]{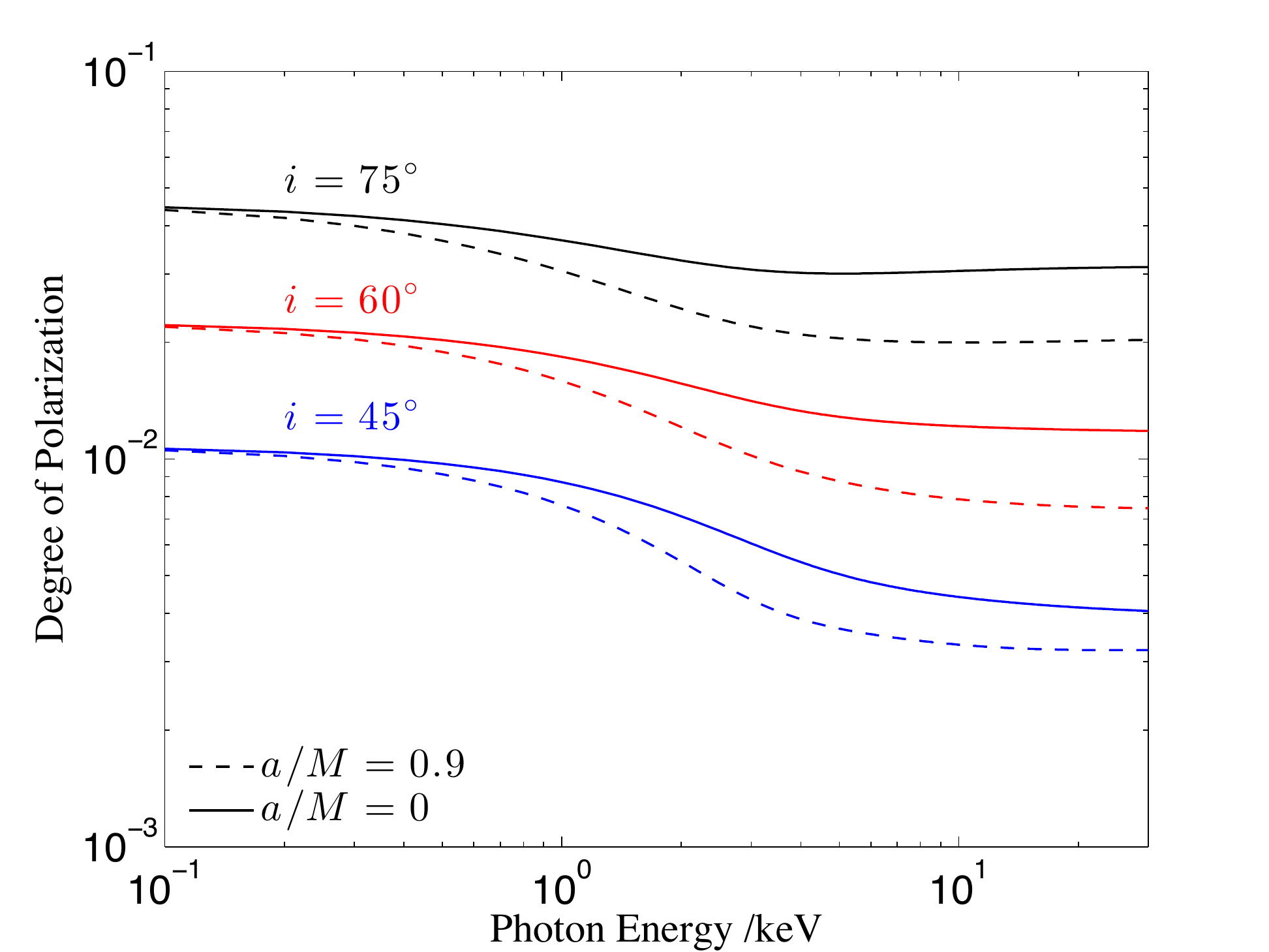}
\includegraphics[type=pdf,ext=.pdf,read=.pdf,width=8.9cm]{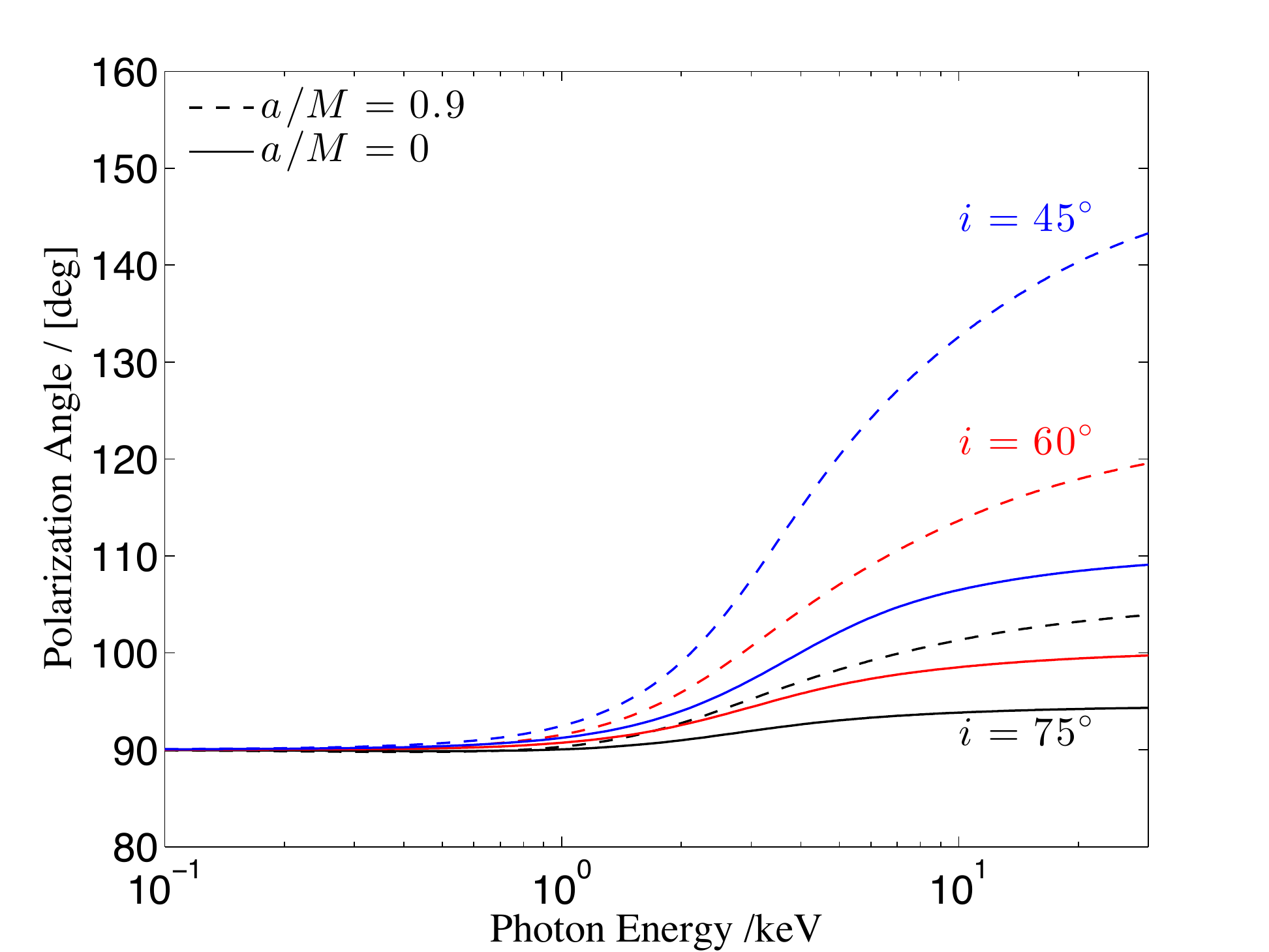}
\end{center}
\caption{Polarization degree (left panel) and polarization angle (right panel) as a function of the photon energy for a Schwarzschild BH (solid lines) and a Kerr BH with spin parameter $a/M = 0.9$ (dashed lines) and a viewing angle $i = 45^\circ$, $60^\circ$, and $75^\circ$. From~\citet{polar4}, under the terms of the Creative Commons Attribution License.}
\label{f-polar1}
\end{figure*}

\subsection{X-ray polarization \label{ss-polar}}

The measurement of the polarization of the thermal radiation of thin accretion disks may become a new technique to study stellar-mass BH candidates. This kind of measurement is not possible today, because there are no X-ray polarimetric missions. However, it will hopefully be possible in the near future, for instance with the missions eXTP, XIPE, IXPE, and PRAXYS.

The thermal radiation of a thin accretion disk is initially unpolarized, but it gets polarized at the level of a few percent due to Thomson scattering of X-ray radiation off free electrons in the disk's atmosphere. Because of relativistic effects (light bending and non-trivial parallel transport in curved spacetime) more pronounced in the vicinity of the compact object, the degree and the angle of polarization of photons generated in the inner part of the accretion disk deviate from the Newtonian predictions, see Fig.~\ref{f-polar1}. Assuming the Kerr background, X-ray spectropolarimetric observations of the thermal component could provide a measurement of the BH spin and of the inclination angle of the disk with respect to the line of sight of the observer~\citep{polar1,polar2}.

As in the previous techniques, even the polarization measurement may be used to test the Kerr metric~\citep{polar3,polar4}. At present, there are only some preliminary studies about its constraining capabilities. In \citet{polar4}, we found that a polarization measurement cannot test the Kerr metric better than the continuum-fitting method, and it is definitively worse than high quality data of the iron line. The problem is still the strong correlation between the estimate of the spin and possible deviations from the Kerr solution.


\section{The special case of S\lowercase{gr}A$^*$ \label{s4-sgra}}

As discussed in the previous section, the main problem to test the Kerr metric is the parameter degeneracy. The spectrum of a Kerr BH can be usually reproduced quite well by non-Kerr objects with different values of the model parameters. To break the parameter degeneracy, it is usually helpful to have different measurements of the same BH candidate. If these measurements are sensitive to different relativistic effects, we may combine the observations and constrain possible deviations from the Kerr geometry.

SgrA$^*$, the supermassive BH candidate at the center of the Galaxy, may soon become quite an ideal object to test the Kerr metric. While there are currently no observations suitable to test this BH candidate, we expect a number of unprecedented data with new facilities in the near future~\citep{falcke13}. The combination of these measurements is a very promising approach to test the nature of SgrA$^*$, see e.g.~\citet{sgra-c,sgra-tj}. A recent review on tests of the Kerr metric with SgrA$^*$ is~\citet{v2-rev-tim}. However, at present we do not know if SgrA$^*$ has all the features to be an optimal source for testing the Kerr metric (e.g. high spin parameter and large viewing angle).

\begin{figure}
\begin{center}
\includegraphics[type=pdf,ext=.pdf,read=.pdf,width=8.0cm]{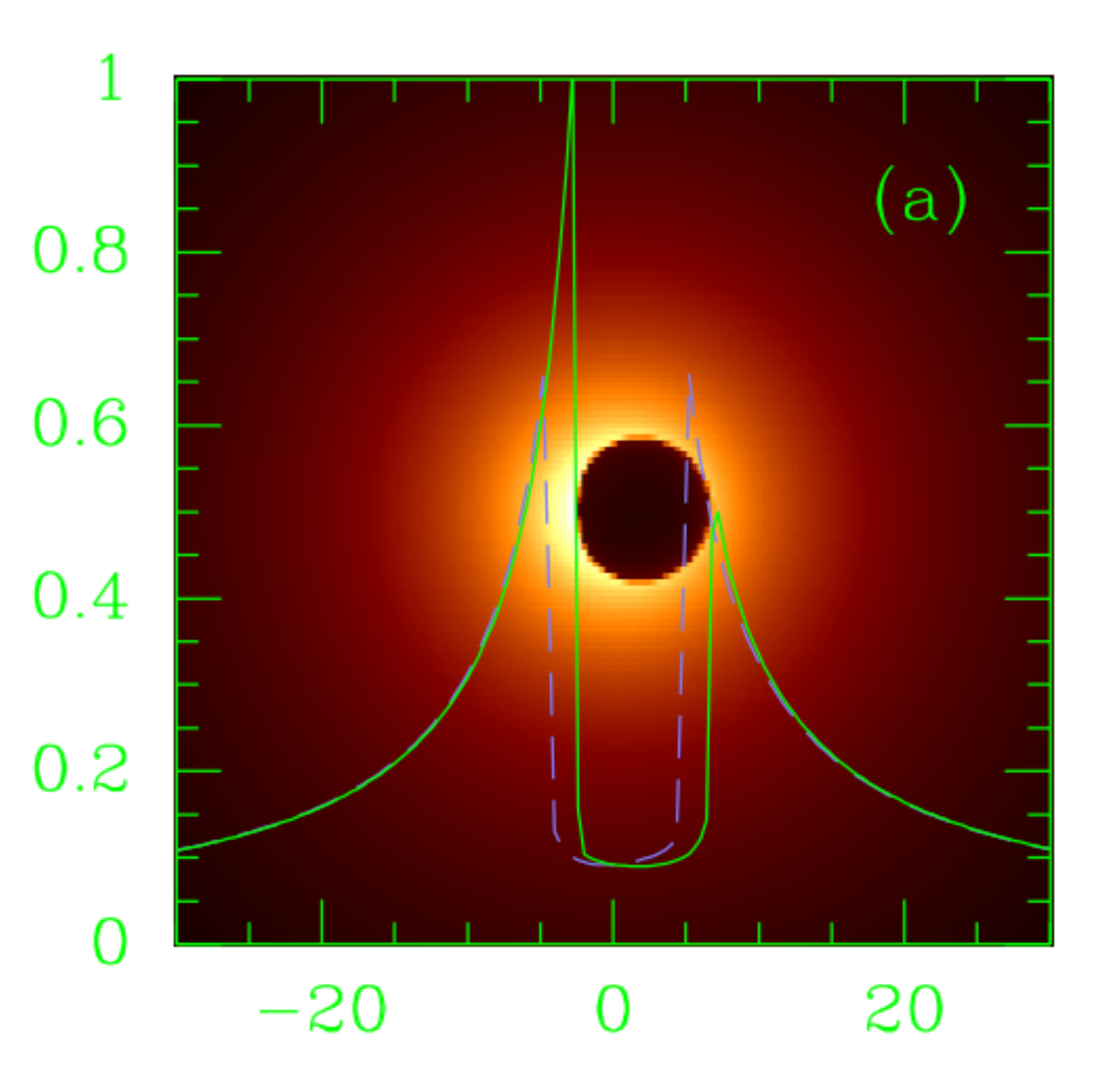}
\end{center}
\caption{Ray-tracing calculations of the direct image of a Kerr BH surrounded by an optically thin emitting medium. The dark area is the BH shadow and its boundary corresponds to the apparent photon capture sphere. From~\citet{shadow-0}. \copyright AAS. Reproduced with permission.}
\label{f-shadow1}
\end{figure}

\begin{figure}
\begin{center}
\includegraphics[type=pdf,ext=.pdf,read=.pdf,width=5cm]{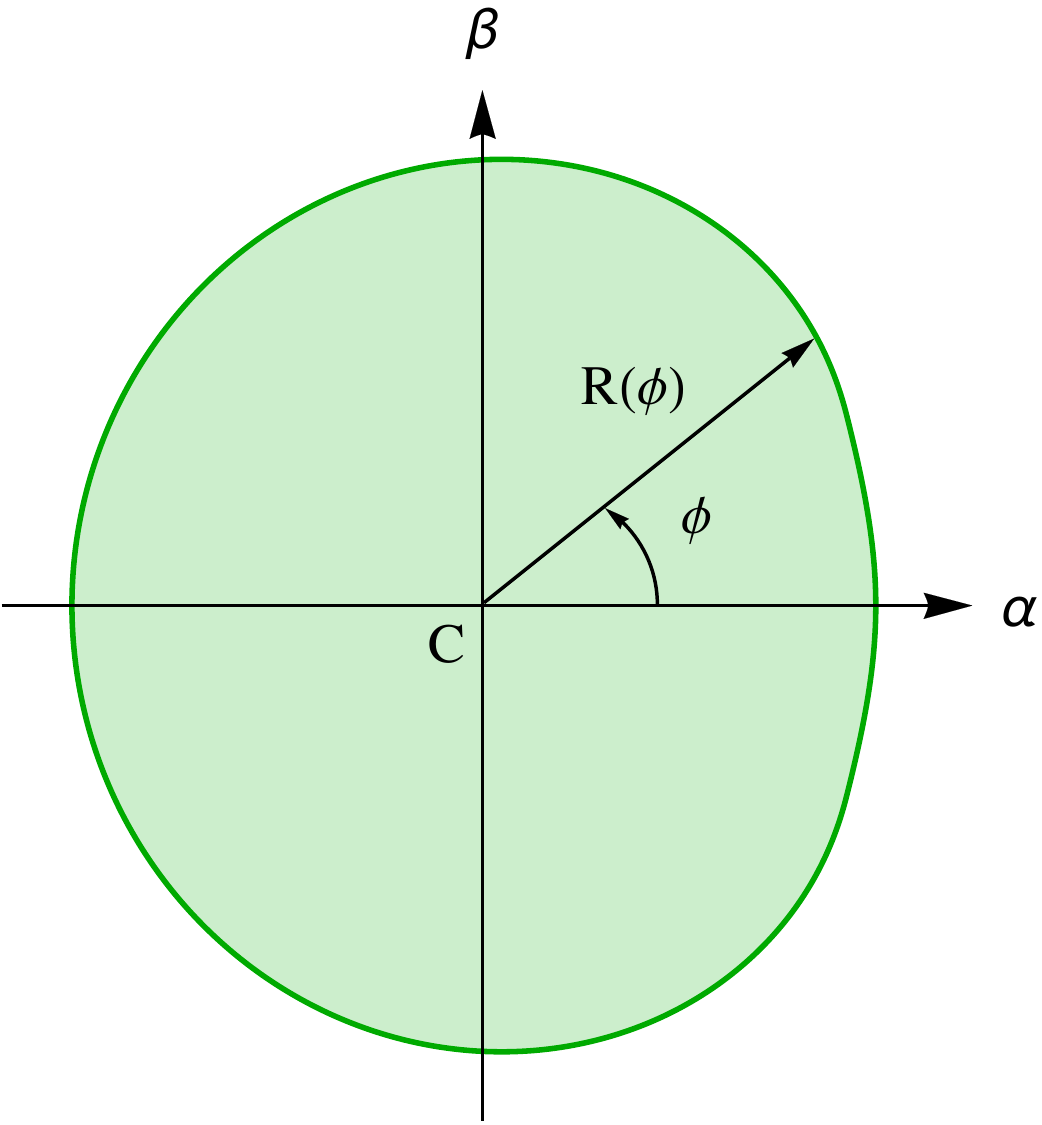}
\end{center}
\caption{The function $R(\phi)$ is defined as the distance between the center C and the boundary of the shadow at the angle $\phi$ as shown in this picture. See the text for more details. From~\citet{masoumeh}, under the terms of the Creative Commons Attribution License.}
\label{f-shadow2}
\end{figure}

\begin{figure*}
\begin{center}
\includegraphics[type=pdf,ext=.pdf,read=.pdf,width=8.5cm]{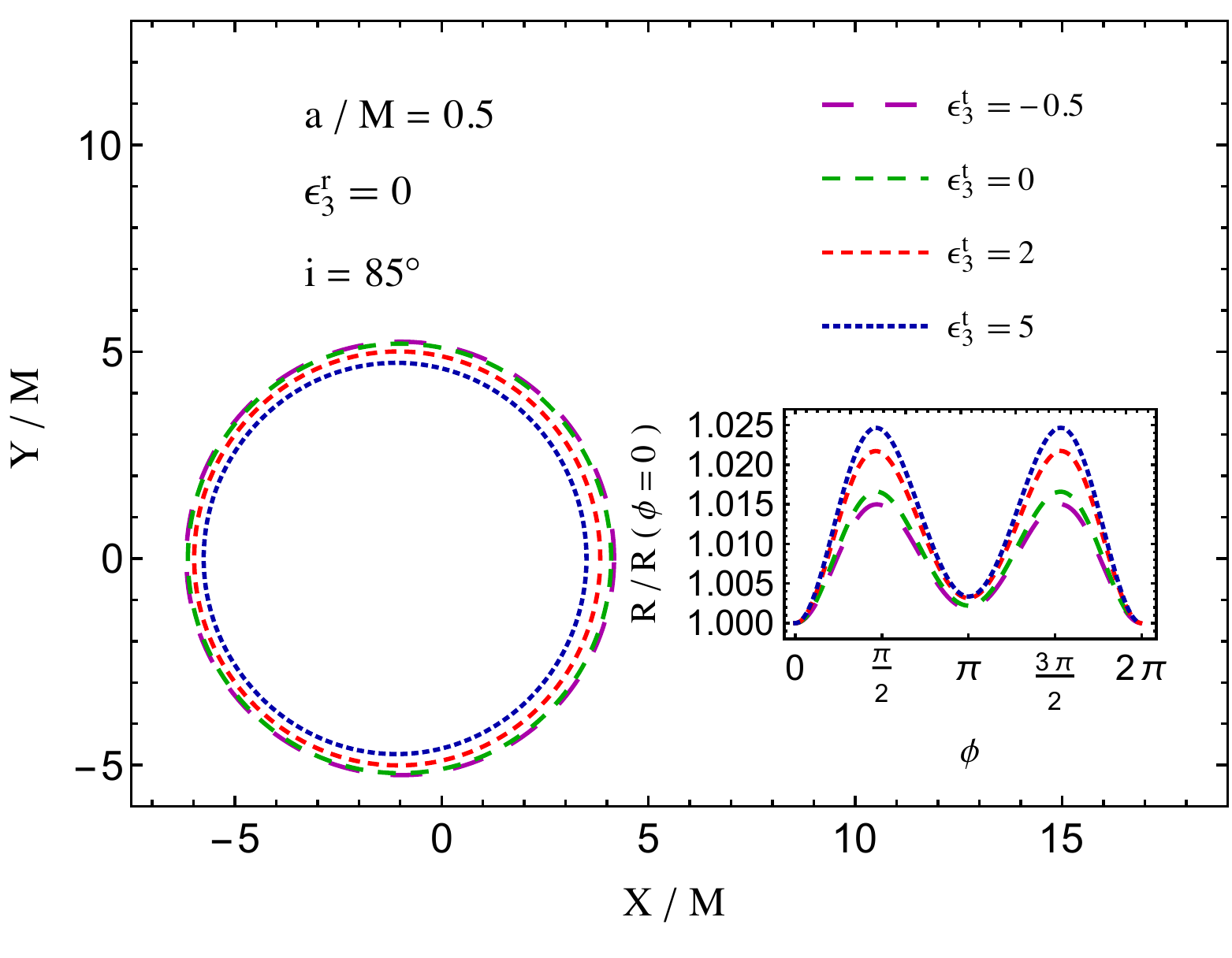}
\hspace{0.5cm}
\includegraphics[type=pdf,ext=.pdf,read=.pdf,width=8.5cm]{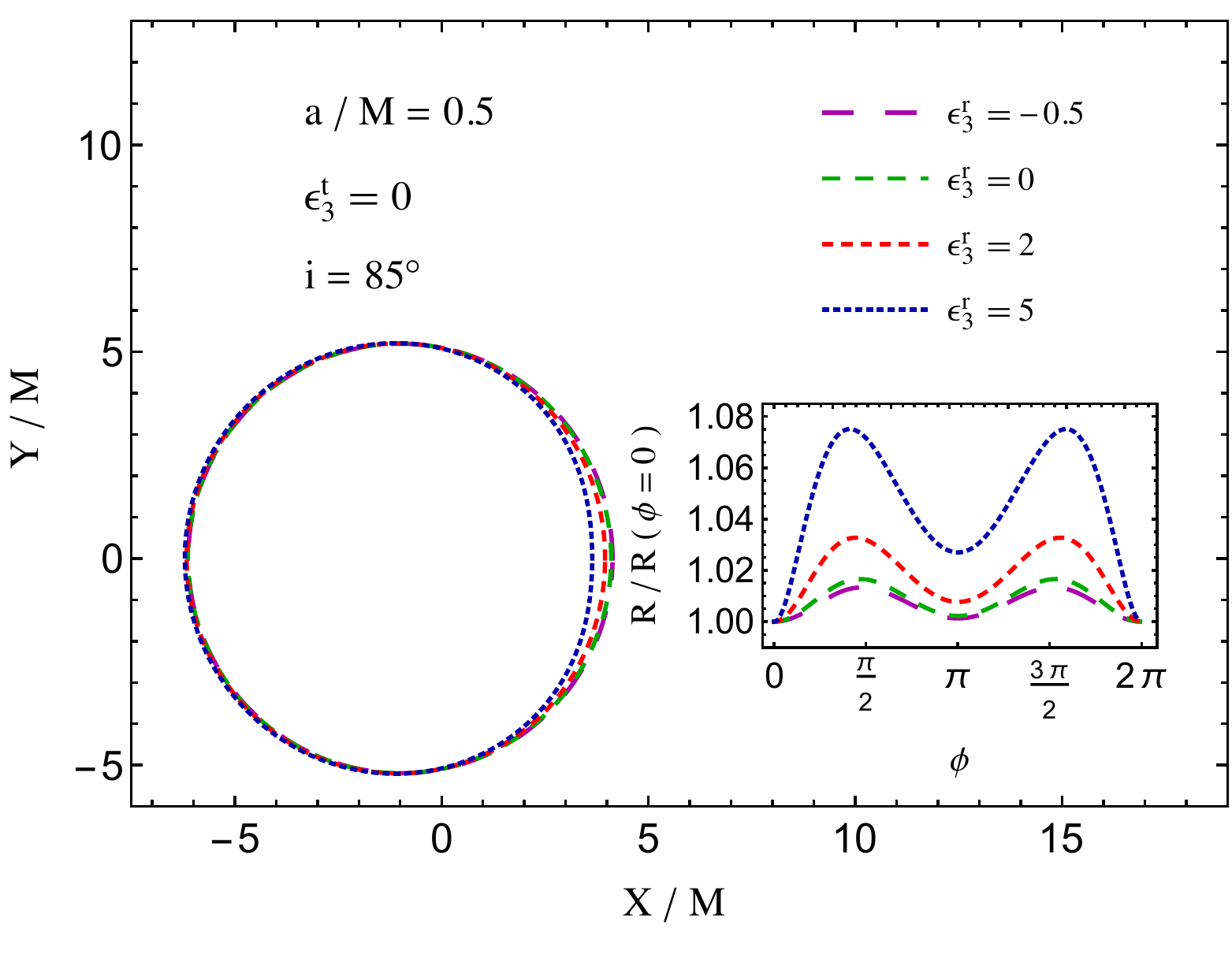} \\
\vspace{0.5cm}
\includegraphics[type=pdf,ext=.pdf,read=.pdf,width=8.5cm]{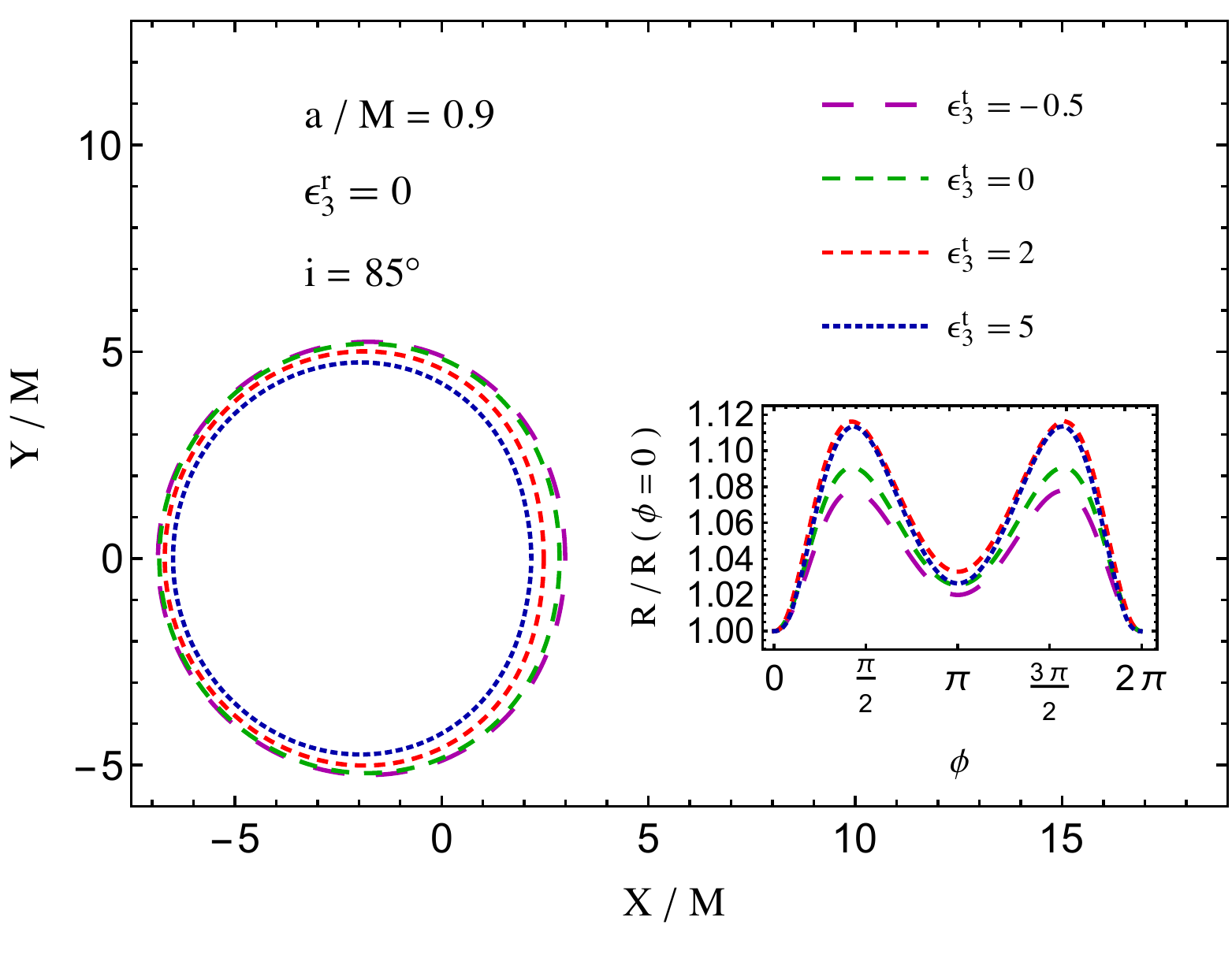}
\hspace{0.5cm}
\includegraphics[type=pdf,ext=.pdf,read=.pdf,width=8.5cm]{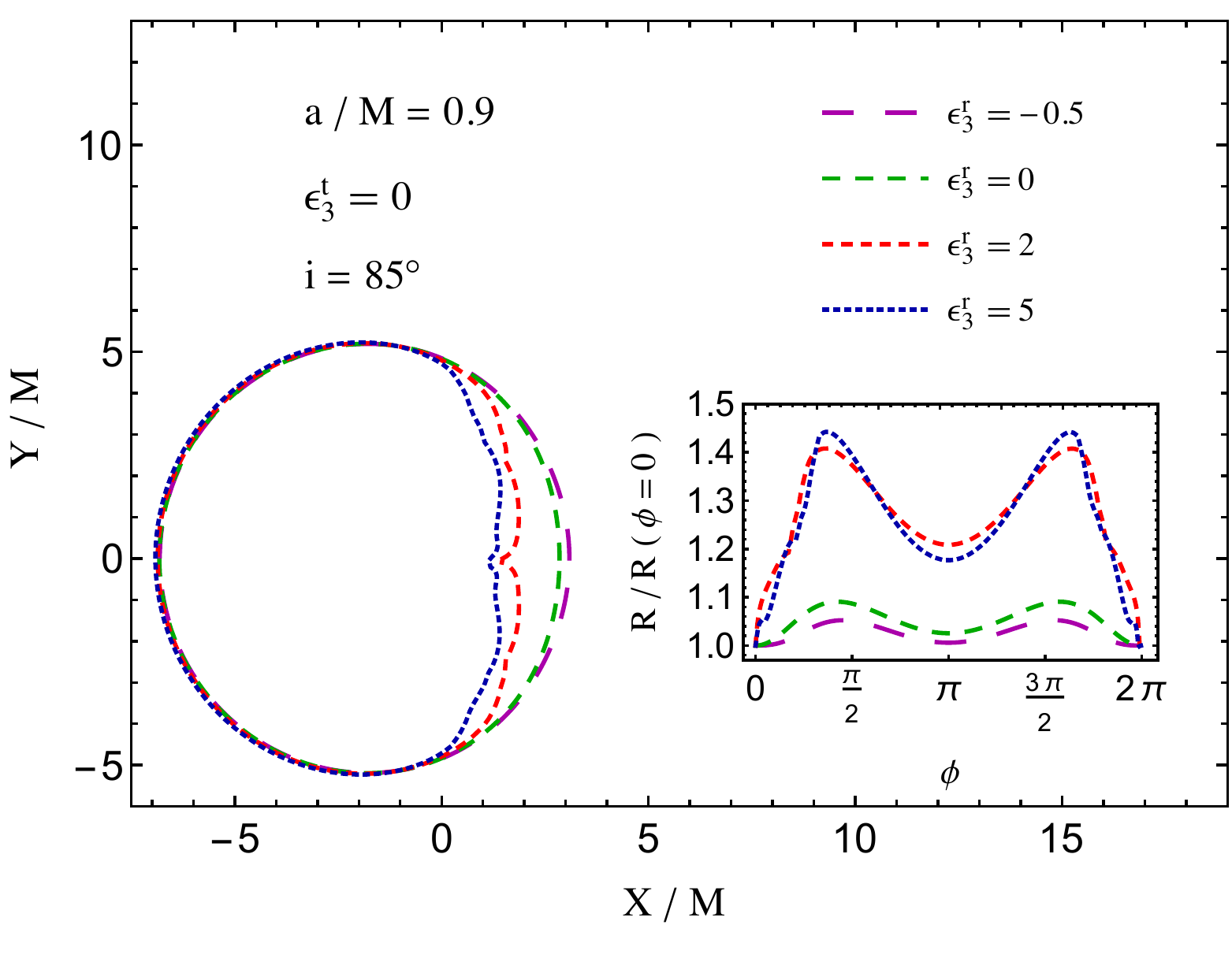} \\
\end{center}
\caption{BH shadows and $R$ functions for CPR BHs with different values of the spin parameter $a_*$, the deformations parameters $\epsilon^t_3$ and $\epsilon^r_3$, and the inclination angle $i$. From~\citet{masoumeh}, under the terms of the Creative Commons Attribution License.}
\label{f-shadow3}
\end{figure*}

\subsection{Black hole shadow}

The direct image of the accretion flow around a BH usually shows a dark area over a bright background. Such a dark area is commonly called the BH shadow~\citep{shadow-0}, even if the name may be a little bit misleading. If the BH is surrounded by an optically thin emitting medium, the boundary of the shadow corresponds to the photon capture sphere as seen by a distant observer. The ray-tracing calculations of the direct image of a Kerr BH surrounded by an optically thin emitting medium are shown in Fig~\ref{f-shadow1}; here the spin parameter is $a_*=0.998$ and the viewing angle is $i=45^\circ$. While the intensity map of the image depends on the properties of the accretion structure and on the emission mechanisms, the boundary of the shadow is only determined by the spacetime metric and the viewing angle of the observer. An accurate measurement of the direct image of the accretion flow around a BH candidate can thus test the spacetime geometry around the compact object.

Sub-millimeter very long baseline interferometry (VLBI) facilities should be able to resolve the shadow of SgrA$^*$ in the next few years~\citep{doeleman}. While it is currently not clear the effects of inevitable astrophysical complications and the level of accuracy that can be reached in the measurement of the boundary of the image of the photon capture sphere in the observer's sky, there has been much work to explore the possibility of testing the Kerr metric with the detection of a shadow and to calculate the shadows of non-Kerr BHs~\citep{shadow-ss,shadow-jp,shadow-a1,shadow-a2,shadow-a3,shadow-aaa1,shadow-aaa2,sh-c1,sh-c2,sh-c3,naoki,masoumeh,sh-wh-br,zilong041,wei13,kaz111}.

In order to infer the spin and the deformation parameters from a possible precise detection of the shadow of a BH, it is necessary to have a formalism to describe the boundary of the shadow\footnote{In what follows, strictly speaking we consider the boundary of the image of the photon capture sphere in the observer's sky, which has a well defined border. If one fires a photon inside such a boundary, the photon is captured by the BH; if the photon is fired outside, it can then escape to infinity. In order to use such a description to analyze real data, it would be necessary to define a precise link between the boundary of the dark area in the direct image (real or simulated) of the accretion flow and the boundary of the image of the photon capture sphere in the observer's sky. The intensity map of the image depends on the properties of the accretion flow and of the observational facilities (in particular on the wavelength of the observation), and there are no such data available today.}. A simple approach was proposed in~\citet{masoumeh} and illustrated in Fig.~\ref{f-shadow2} [a more sophisticated method was presented in~\citet{a-r-a}]. First, we find the ``center'' C of the shadow. Its Cartesian coordinates on the image plane of the observer are 
\be
X_{\rm C} &=& \frac{\int \rho(X,Y) X dX dY}{\int \rho(X,Y) dX dY} \, , \nonumber\\
Y_{\rm C} &=& \frac{\int \rho(X,Y) Y dX dY}{\int \rho(X,Y) dX dY} \, ,
\ee
where $\rho (X,Y) = 1$ inside the boundary of the shadow (which is a closed curve) and $\rho (X,Y) = 0$ outside. Assuming a reflection-symmetric spacetime, the shadow is symmetric with respect to the $X$-axis and we can define $R(0)$ as the shorter segment between C and the shadow boundary along the $X$-axis. Defining the angle $\phi$ as shown in Fig.~\ref{f-shadow2}, $R(\phi)$ is the distance between the point C and the boundary at the angle $\phi$. The function $R(\phi)/R(0)$ completely characterizes the shape of the BH shadow. Here we consider $R(\phi)/R(0)$ instead of $R(\phi)$ because the latter cannot be measured with good precision, as it would require an accurate measurement of the distance and of the mass of the BH, which is not available at the moment. Even the exact positions of the shadow on the image plane of the observer cannot be used to test the Kerr metric, because it is difficult to precisely identify the center $X=Y=0$ of the source.

Fig.~\ref{f-shadow3} shows some examples of shadows of CPR BHs and the associated $R(\phi)/R(0)$ function. In the top panels, the spin parameter is $a_* = 0.5$, while in the bottom panels it is $a_* = 0.9$. The inclination angle is always $i = 85^\circ$, which is high and can thus maximize the relativistic effects. In the left panels, $\epsilon^t_3$ changes and $\epsilon^r_3 = 0$ is frozen. In the right panels, we have the opposite case and $\epsilon^t_3 = 0$ and $\epsilon^r_3$ can vary. It is evident that $\epsilon^t_3$ mainly affects the size of the shadow, which increases (decreases) if $\epsilon^t_3$ decreases (increases). $\epsilon^r_3$ alters the shape of the shadow on the side of corotating orbits, while there are no appreciable effects in the other parts of the boundary of the shadow. The peculiar boundary appearing for $\epsilon^r_3 = 2$ and 5 in the bottom right panel is due to the non-trivial horizons of these BHs~\citep{pat1}.

Note that VLBI observations do not directly image the accretion flow, and therefore they cannot directly measure the shape of the boundary of the shadow discussed in this section. They instead sample the Fourier space conjugate to the sky image at a finite number of points. The boundary of the shadow can be obtained after image reconstruction. However, as mentioned, it is not clear whether a precise determination of the boundary of the shadow at the level necessary to test the Kerr metric is eventually possible, because systematic effects may prevent it.

\begin{figure}
\vspace{0.8cm}
\begin{center}
\includegraphics[type=pdf,ext=.pdf,read=.pdf,width=8.0cm]{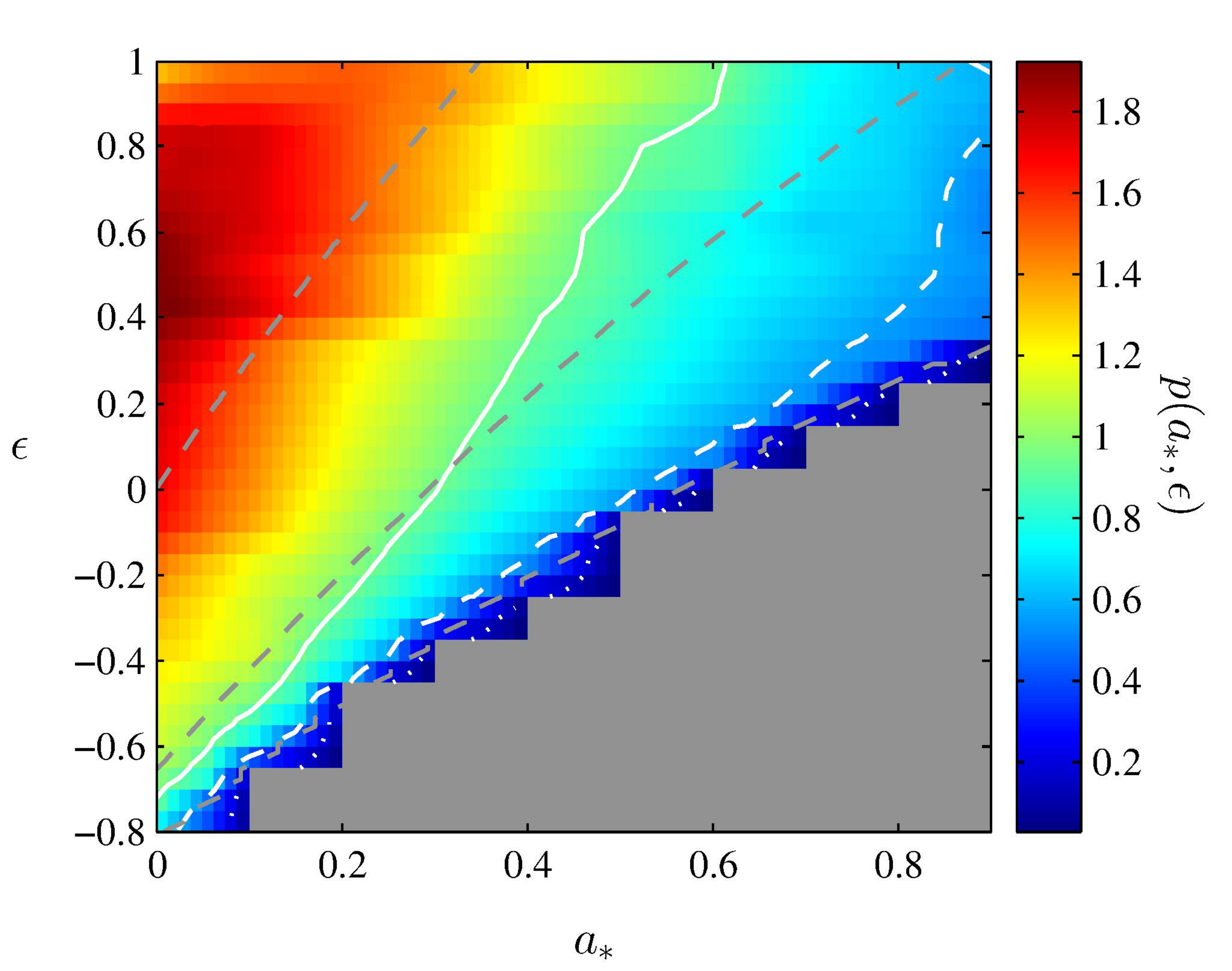}
\end{center}
\caption{Posterior probability density of spin and deformation parameters, marginalized over all other quantities, from the existing EHT data. The solid, dashed, and dotted white lines correspond, respectively, to the 1-, 2- , and 3-$\sigma$ boundaries. The dashed gray lines correspond (from top to bottom) to spacetimes with $r_{\rm ISCO}/M = 6$, 5, and 4. The grayed region in the lower right corresponds to metric with $r_{\rm ISCO}/M < 4$ and it is neglected because the calculations may be affected by some pathological properties of this non-Kerr metric. Here $\epsilon$ is the deformation parameter of the non-Kerr metric of \citet{tests-gw1}. From \citet{brod14}. \copyright AAS. Reproduced with permission.}
\label{f-shadow6}
\end{figure}

The Event Horizon Telescope (EHT)\footnote{http://www.eventhorizontelescope.org/} is a project involving mm and sub-mm observatories equipped with VLBI instrumentation to get high resolution images of the accretion flow around supermassive BH candidates at 230 and 345~GHz~\citep{fish13}. One of the main goals of this experiment is the observation of the shadow of SgrA$^*$. The existing mm-VLBI observations have been done with only three stations (respectively in Hawaii, California, and Arizona). Employing a radiatively inefficient accretion flow model, \citet{brod14} have explored the capability of present observations to constrain possible deviations from the Kerr geometry. They used the quasi-Kerr metric of \citet{tests-gw1}. In their simulations, we cannot see the exact shape of the apparent photon capture sphere because at the wavelengths accessible to mm-VLBI it is  partially obscured by the optically thick structure on the approaching side of the accretion flow. The result of this analysis is shown in Fig.~\ref{f-shadow6}.

\begin{figure}
\begin{center}
\vspace{-2.8cm}
\includegraphics[type=pdf,ext=.pdf,read=.pdf,width=7.0cm]{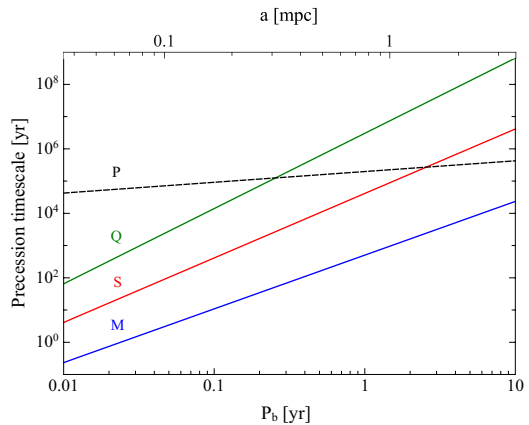}
\end{center}
\vspace{-0.5cm}
\caption{Precession timescale from the mass ($M$), the spin ($S$), the quadrupole moment ($Q$), and stellar perturbation ($P$) for a pulsar orbiting SgrA$^*$ as a function of orbital period $P_b$, assuming an orbital eccentricity of 0.5 and $10^3$ objects, all with mass $M = M_\odot$, within 1~mpc around SgrA$^*$. From~\citet{pulsar2}. \copyright AAS. Reproduced with permission. }
\label{f-pulsar}
\end{figure}

\subsection{Accurate measurements in the weak field \label{v2-weak}}

The nature of SgrA$^*$ can be potentially investigated even with accurate measurements of the spacetime metric at relatively large radii. In this case, the gravitational field is weak, $M/r \ll 1$, and we can adopt an approach similar to the PPN formalism of Solar System experiments. If a spacetime is stationary, axisymmetric, asymptotically flat, Ricci-flat outside the source, and analytic about the point at infinity, its metric in the region outside the source can be expanded in terms of mass moments $M_\ell$ and current moments $S_\ell$~\citep{mmm1,mmm2}\footnote{The expansion in multipole moments is also possible when the spacetime is not axisymmetric, but in this case the mass and the current moments of order $\ell$ are tensors. If the spacetime is axisymmetric, there are some simplifications, and the mass and the current moments of any order $\ell$ are completely determined by two scalars, namely $M_\ell$ and $S_\ell$. In the case of tests of the Kerr metric in the weak field, it is common to assume that the spacetime is axisymmetric, because it sounds a physically plausible hypothesis and simplifies the problem. Let us also note that some assumptions may not hold in some relevant cases. For instance, BH solutions in alternative theories of gravity may not be Ricci-flat, and an example is the case of BHs in Einstein-dilaton-Gauss-Bonnet gravity. The analyticity assumption may also not hold, as in the case of the presence of a massive scalar field with a Yukawa type solution.}. In the case of reflection symmetry, the odd $M$-moments and the even $S$-moments are identically zero, so that the leading order terms are the mass $M_0 = M$, the spin angular momentum $S_1 = J$, and the mass quadrupole moment $M_2 = Q$. In the case of a Kerr BH, the metric is completely determined by $M$ and $J$, and all the moments $M_\ell$ and $S_\ell$ are locked to the mass and the spin by
\be
M_\ell + i S_\ell = M \left( i \frac{J}{M} \right)^\ell \, ,
\ee
where $i$ is the unit imaginary number. In particular, the mass quadrupole term is $Q = - J^2/M$.

This approach has the advantage that it is quite general and relies on the small number of assumptions. Even the requirement of the Ricci-flat spacetime may approximatively hold in many cases. Here the spin measurement is really a spin measurement, and it should not be correlated to possible deviations from the Kerr solution in the near horizon region because the latter corresponds to higher order corrections. Such a measurement could be combined with a measurement in the strong gravity regime, which is usually a constraint on the spin and possible deviations from the Kerr geometry. In some favorable cases, it could be possible to determine also the mass quadrupole moment $Q$. This could permit us to test the Kerr metric at the quadrupole term, because in the case of a Kerr BH one must recover $Q = - J^2/M$. Higher order terms can unlikely be tested, but clean measurements of $J$ and $Q$ would instantly be very helpful if combined with observations in the strong gravity field.

\subsubsection{Radio pulsars}

It is thought that a population of radio pulsars is orbiting SgrA$^*$ with a short orbital period and there is already an intense work to detect these objects~\citep{pulsar-r}. For instance, \citet{v3-pulsar-gc} argue that there may be $\sim 200$~pulsars in the inner parsec region (orbital period $\lesssim 10^4$~yrs). Accurate timing observations of a radio pulsar orbiting SgrA$^*$ in a very close orbit ($\lesssim 1$~yr) would allow a precise measurement of the mass, the spin, and -- in exceptional cases -- even of the mass quadrupole moment of the supermassive BH candidate at the center of our Galaxy if the system is sufficiently free of external perturbations~\citep{pulsar2}.

Because of the high electron density in the ionized gas at the center of the Galaxy, this kind of observation must be made at much higher frequencies than those normally used for pulsar timing, which further challenges these measurements. A more serious problem is the possible presence of a population of stars or BHs orbiting very close to SgrA$^*$. The presence of these bodies may strongly affect, or even prevent, a clean measurement of the spin and the quadrupole moment of SgrA$^*$ with the radio pulsar. At the moment it is impossible to make predictions about the potentialities of future observations because we do not know the actual situation near SgrA$^*$.

Assuming a population of $10^3$ objects with a mass $M = M_\odot$ isotropically distributed within 1~mpc around SgrA$^*$, \citet{pulsar2} estimated the necessary orbital period of the pulsar to get a measurement of the mass, the spin, and the quadrupole moment of SgrA$^*$. Fig.~\ref{f-pulsar} shows the timescales of secular orbital precession for a pulsar orbiting SgrA$^*$ as a function of its orbital period: the contributions are from the mass monopole $M$ (pericenter advance), the spin $S$ (frame dragging), the quadrupole moment $Q$, and stellar perturbation $P$. The orbital eccentricity is assumed to be 0.5. The precession timescale of the pericenter advance is already lower than that of stellar perturbation for a 10~year orbital period, which means that the observation of a radio pulsar with an orbital period less than 10~years can already be used to estimate the BH mass $M$. The measurement of the spin would require an orbital period less than 0.5~years to have the contribution from frame dragging significantly above that from stellar perturbation. The measurement of the quadrupole moment $Q$ requires an orbital period less than 0.1~years. These estimates have to be taken as a general guide and the actual situation may be different. For instance, a population of 10~$M_\odot$ BHs in this region may completely spoil the measurement of the parameters of the metric around SgrA$^*$, as well as a significant anisotropy in the distribution of these bodies may challenge it.

Assuming a not too optimistic situation, in which we can observe a radio pulsar in an orbit of several months, timing observations could measure the mass and the spin of SgrA$^*$. Since the pulsar is in the weak field of SgrA$^*$, this would be a clean measurement of the spin parameter $a_*$, namely independent of the higher order multipole moments of the spacetime. First, such a measurement should satisfy the Kerr bound $|a_*| \le 1$, because otherwise SgrA$^*$ could not be a Kerr BH. Second, the spin measurement could be combined with other observations of the strong field (shadow, hot spot, etc.) in which there is typically a strong correlation between the estimate of the spin and possible deviations from the Kerr solution to break such a degeneracy. In this manner we could test the Kerr metric.

\subsubsection{Normal stars}

Even normal stars in compact orbits can be used to probe the weak gravitational field of SgrA$^*$ and measure its spin parameter and, possibly, its mass quadrupole moment. The idea was proposed in~\citet{will2} and further explored in~\citet{will3,will4}.

If SgrA$^*$ is rotating fast, the observation of at least two stars with an orbital period of 0.1~years or less and in orbits with a high eccentricity, say $\sim 0.9$, may provide a measurement of the mass, the spin, and the mass quadrupole moment of SgrA$^*$ and thus test the Kerr nature of this object at the level of the quadrupole term. Today we know stars with an orbital period as short as 10~years. These objects are still too far from SgrA$^*$ and currently no relativistic effects in their orbit are observed (but they should be observed in the near future). However, observational facilities like GRAVITY~\citep{gravity}, with the capability of accurate astrometric measurements at the level of 10~$\mu$as close to SgrA$^*$, may observe stars with a sufficiently short orbital period to test the Kerr metric.

\citet{will2} proposed to test the Kerr metric by measuring the precession of the orbital plane of these stars. The advance of the pericenter of these stars is dominated by the mass term of the supermassive object, while the contribution of the spin and the quadrupole moment would be subdominant and difficult to estimate. On the contrary, in the weak field limit, the precession of the orbital plane is only determined by the spin (through the frame-dragging) and the mass quadrupole moment; see \citet{will2} for the details. Here the dominant contribution comes from the spin, but in the case of sufficiently compact orbits it is also possible to infer the mass quadrupole moment $Q$. As in the pulsar case, a measurement of the spin could already be useful in combination with other measurements probing the strong gravity field. In the case a quadrupole moment measurement is also available, one can check {\it a posteriori} whether it satisfies the Kerr relation $Q = - J^2/M$.

Recently, \citet{z-l-y-15} showed that measurements of the spin of SgrA$^*$ are possible even with stars with orbital configurations similar to those already known, as in the case of long-term high precision observations.

\subsection{Hot spots}

\begin{figure}
\vspace{0.8cm}
\begin{center}
\includegraphics[type=pdf,ext=.pdf,read=.pdf,width=7.0cm]{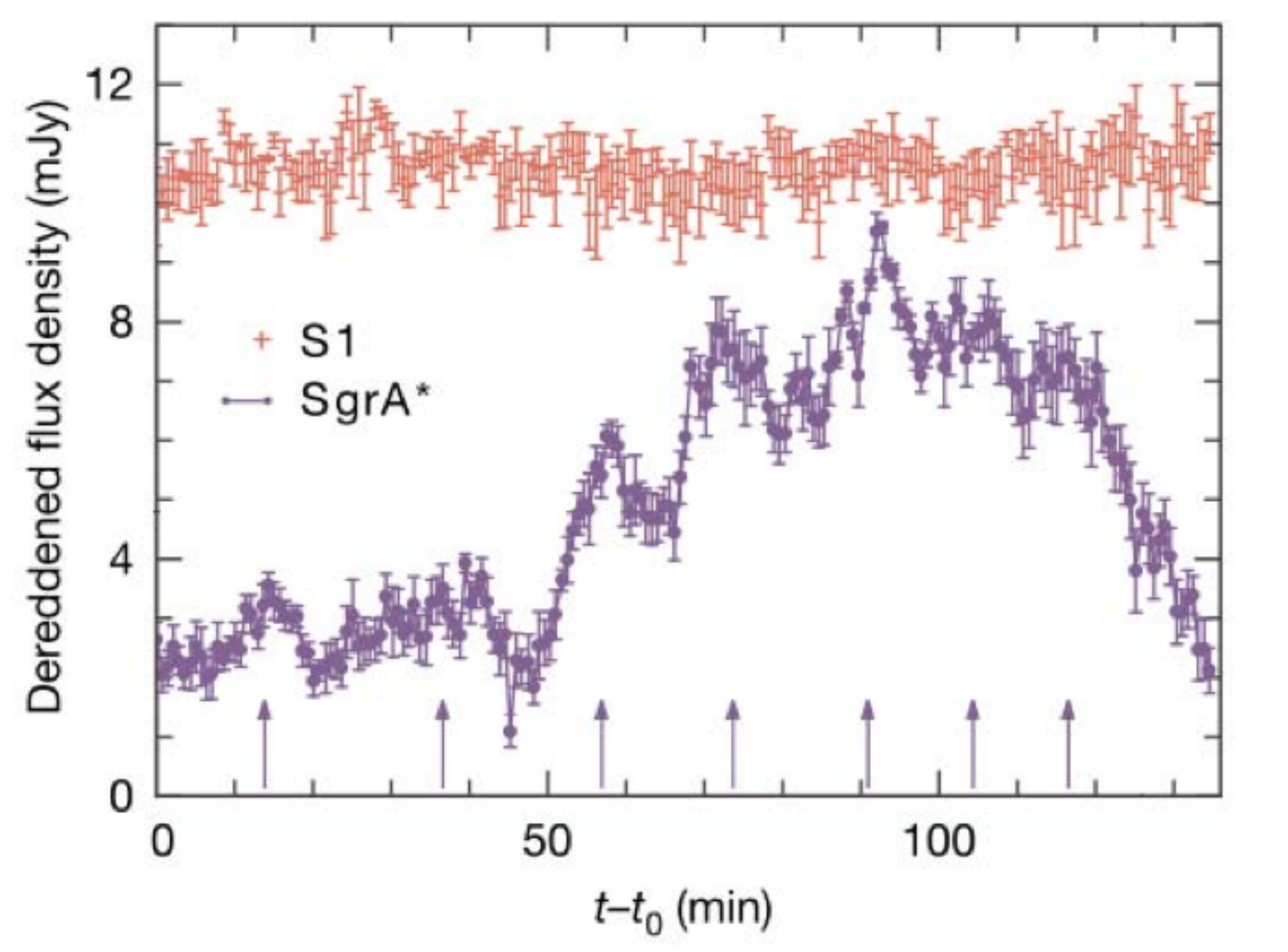}
\end{center}
\caption{Light curve of a NIR flare of SgrA$^*$ with the characteristic quasi-periodic substructure with a timescale of about 20~minutes. The arrows at the bottom indicate the peaks of the substructure. Reprint with permission from \citet{flares1} and \citet{flares}. }
\label{f-flares}
\end{figure}

SgrA$^*$ exhibits powerful flares in the X-ray, near-infrared, and sub-millimeter bands~\citep{flares1,flares2,flares3}. During a flare, the flux increases up to a factor of 10. There are a few flares per day. Every flare lasts 1-3~hours and has a quasi-periodic substructure with a timescale of about 20~minutes, see Fig.~\ref{f-flares}. These flares may be associated with blobs of plasma orbiting near the ISCO of the supermassive BH candidate~\citep{hot-spot-m}, even if current observations cannot exclude other explanations~\citep{nhs1,nhs2,nhs3}. Temporary clumps of matter should indeed be common in the region near the ISCO radius~\citep{hot-spot-s} and, if so, they may be studied by the GRAVITY instrument for the ESO Very Large Telescope Interferometer (VLTI)~\citep{gravity}.

\begin{figure}
\vspace{0.8cm}
\begin{center}
\includegraphics[type=pdf,ext=.pdf,read=.pdf,width=8.0cm]{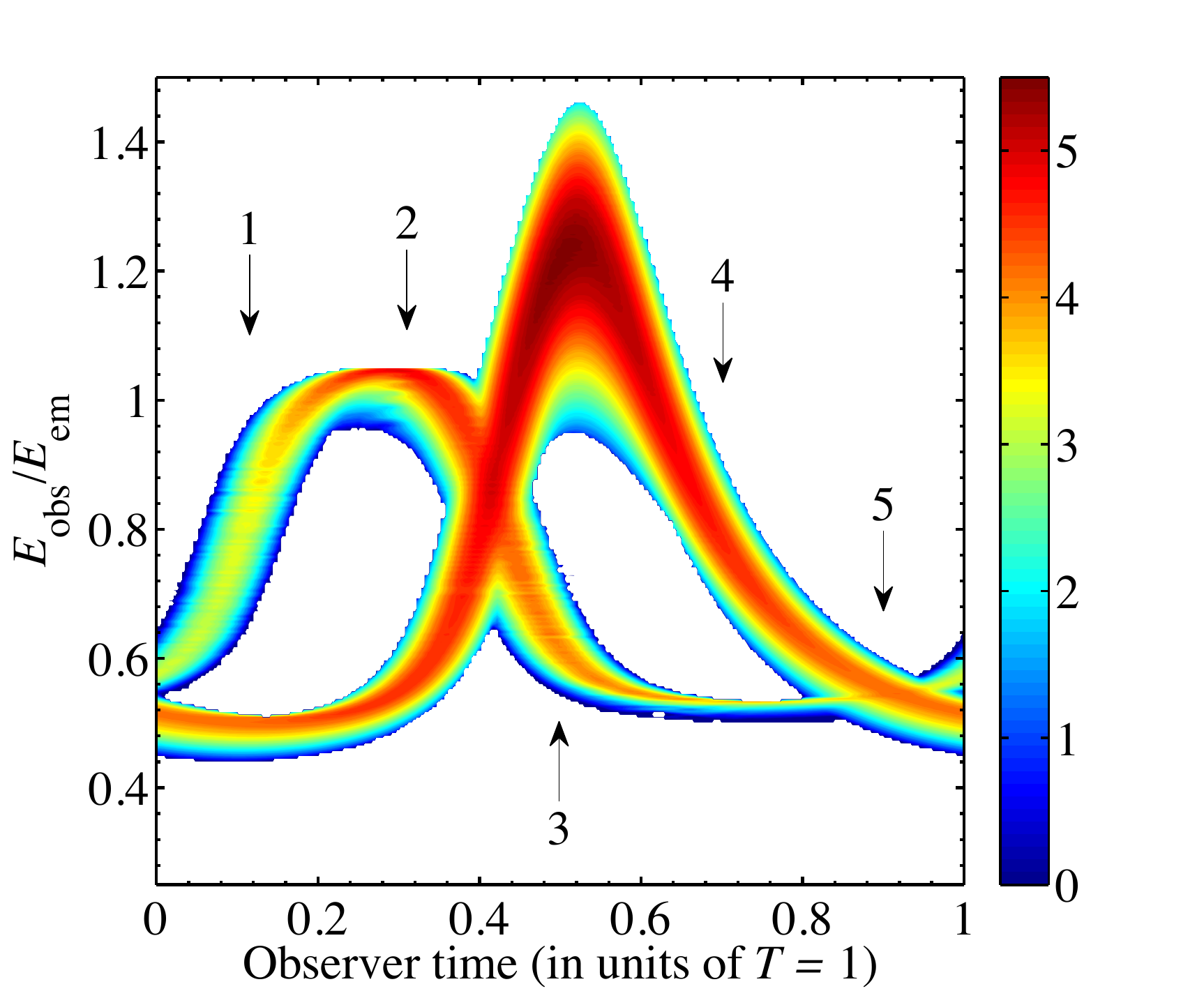}
\end{center}
\caption{Spectrogram of a hot spot orbiting a Schwarzschild BH at the ISCO radius from an inclination angle $i = 60^\circ$. The vertical axis is the ratio between the photon energy measured by a distant observer and the photon energy at the emission point. The color scale indicates the energy flux (in arbitrary units). We see two tracks because one is the spectrogram of the primary image, the other one is for the secondary image. The points labeled 1-5 refer to another figure in the original paper and can be ignored here. From \citet{hotspot-f1}. }
\label{f-hotspot1}
\end{figure}

The radiation emitted by a blob of plasma orbiting the strong gravity region of a BH candidate is significantly affected by the metric of the spacetime and can potentially be used to test the Kerr metric~\citep{hotspot-f1}. Fig.~\ref{f-hotspot1} shows the spectrogram, namely the spectrum as a function of time, of a monochromatic blob of plasma orbiting the ISCO of a Schwarzschild BH and observed at a viewing angle of $i = 60^\circ$. An accurate measurement of the spectrogram of a similar blob of plasma would be surely an ideal tool to test the metric around SgrA$^*$. However, in the reality the situation is much more complicated. The astrophysical model (shape and size of the blob of plasma, spectrum of the blob of plasma in its rest-frame, etc.) is usually much more important than the small features associated with the relativistic effects characterizing the background metric~\citep{hotspot-f1}.

At the moment, it is not clear if tests of the Kerr metric are possible with this approach. In some spacetimes, the photon capture radius can be significantly different from that of Kerr BHs, and in this case it is possible to identify specific signatures of these metrics~\citep{hotspot-f3,hotspot-f2}. In general, it seems more likely that relativistic effects cannot really be identified and eventually the radiation from a blob of plasma can only provide an estimate of the orbital frequency. Since the observed period of the quasi-periodic substructure of the flares of SgrA$^*$ ranges from 13 to about 30~minutes, the orbital radius of these hot spots should change and be at a radius larger than the ISCO. For a 4~million~$M_\odot$ Kerr BH, the ISCO period ranges from about 30~minutes ($a_* = 0$) to 4~minutes ($a_* = 1$ and corotating orbit). The shortest period ever measured is $13\pm2$~minutes, and it may be an upper bound for the ISCO period. In the Kerr metric, such a measurement translates into the spin estimate $a_* \ge 0.70 \pm 0.11$~\citep{trippe07}.

\begin{figure}
\vspace{0.8cm}
\begin{center}
\includegraphics[type=pdf,ext=.pdf,read=.pdf,width=8.0cm]{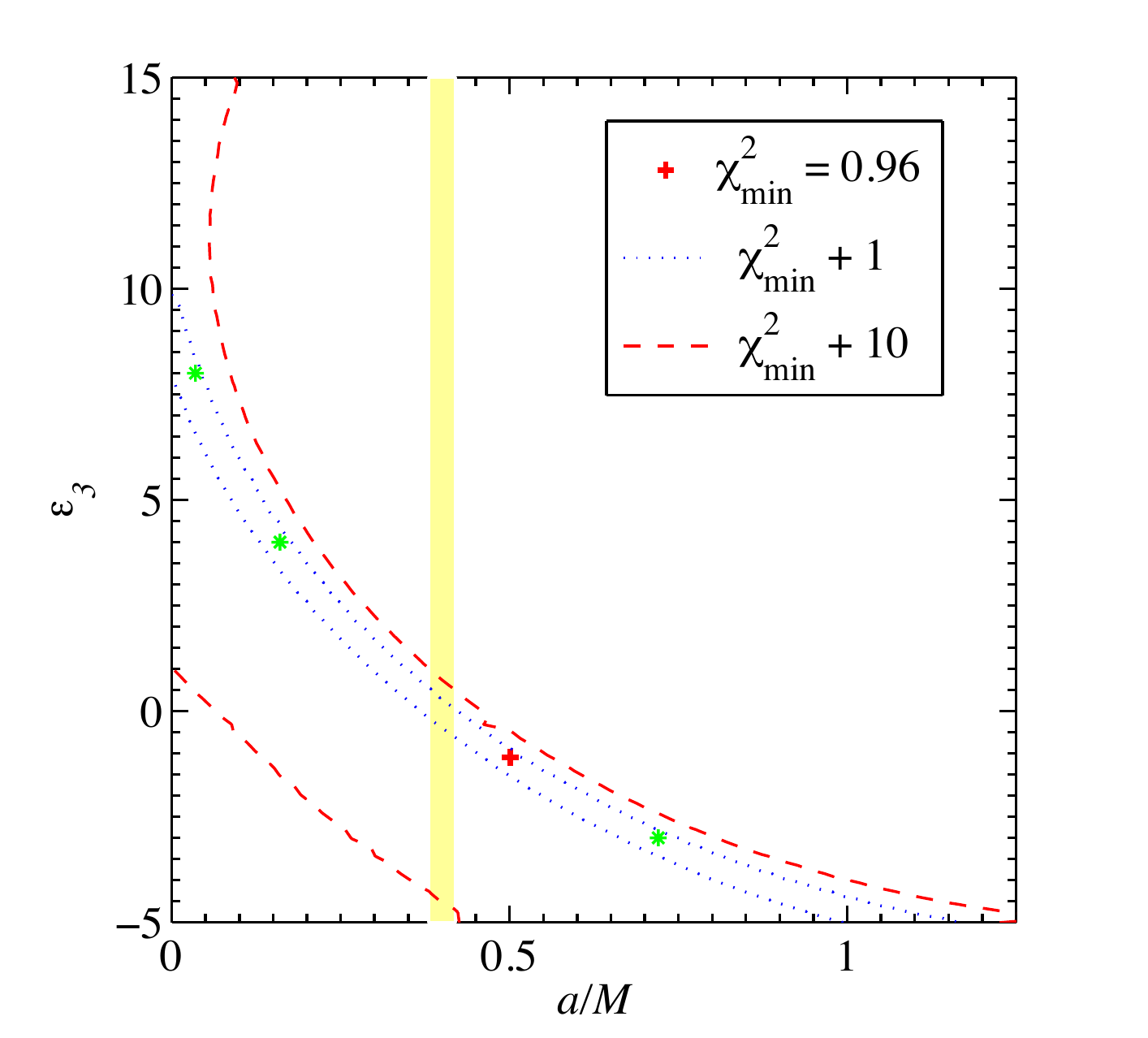}
\end{center}
\caption{Assuming that the hot spot is at the ISCO radius, the measurement of its light curve can only select the spacetimes with the same orbital frequency. In this plot, the reference model employs a Kerr BH with spin parameter $a_* = 0.4$. The allowed regions are those between the two blue dotted lines ($\Delta\chi^2 < 1$) and the two red dashed lines ($\Delta\chi^2 < 10$). The possible measurement of the spin through a binary pulsar would provide the constraint given by the yellow area, and the combination of the two observations could potentially break the degeneracy between $a_*$ and $\epsilon_3$. See the text for more details. From \citet{hotspot-f1}.}
\label{f-hotspot2}
\end{figure}

In the case of a metric with a non-vanishing deformation parameter, there is a degeneracy between the estimate of the spin and possible deviations from the Kerr solution. An example of the possible constraints is shown in Fig.~\ref{f-hotspot2}. Such a degeneracy may be broken with another measurement, for instance a precise estimate of the spin parameter by the observation of a radio pulsar. The latter would be independent of the deformation parameter $\epsilon_3$ because the pulsar would probe the weak field limit, where the metric can be expanded in $M/r$. The leading order term in $\epsilon_3$ is subdominant with respect to the leading order term in $a_*=J/M^2$ and the pulsar data may not be able to measure $\epsilon_3$.

\subsection{Spectrum of the accretion structure}

Today, the two most popular and widely used techniques to probe the metric around BH candidates are the continuum-fitting and the iron line methods. Both approaches require geometrically thin and optically thick accretion disks and employ the Novikov-Thorne model. However, any accretion structure determined by the metric of the spacetime can potentially be used to test BH candidates.

The accretion structure around SgrA$^*$ seems to be a radiatively inefficient advection dominated accretion flow~\citep{adaf95}. There are a few different analytic models, with several variants, in the literature~\citep{x-rev}. In principle, one could use a model to fit the data and infer the free parameters of the model, including those related to the geometry of the spacetime. An explorative work in this direction was presented in~\citet{x-nan}, which employs the ion torus model of~\citet{x-1a,x-1b}.

This approach currently faces a few problems. First, we do not know the correct model, and different models presumably provide different results. Second, even simple analytical models have several free parameters that should be inferred by fitting the spectrum. The Novikov-Thorne model for thin disks follows from basic rules like the conservation of mass, energy, and angular momentum; the model is already quite constrained with a reasonable number of free parameters. The accretion structures necessary to describe the flow around SgrA$^*$ are more complicated and unconstrained. The advantage is that the spectrum of the accretion structure around SgrA$^*$ seems to have many features and this may break the degeneracy among the parameters of the model. For the moment, there is no accurate measurement of the full spectrum, and therefore it is impossible to constrain the model. However, this is presumably possible in the future.


\section{Other approaches \label{s5-other}}

\subsection{Tests in weak fields \label{s5-weak}}

SgrA$^*$ is not the only BH candidate for which we may get accurate measurements at relatively large radii. However, in the case of other BH candidates we may not be so fortunate as to have independent measurements in the strong gravity field. In order to test the Kerr metric with these objects, we need to have measurements good enough to determine $M$, $J$, and $Q$. One can then check {\it a posteriori} whether $Q = - J^2/M$, as expected in the case of a Kerr BH.

The ideal candidate for this kind of test is a pulsar binary in which the companion is a stellar-mass BH candidate~\citep{pulsar1}. A similar system is not known at the moment, but it should not be too rare and there is no reason to believe that it cannot be found in the future. It is also possible that the signal of a binary pulsar with a BH companion is already in the available radio data, but the data have not yet been analyzed. The identification of a new pulsar is a very time-consuming process. After the measurement of the period of the pulsar, it is just an issue of time and accurate measurements of the system can be obtained thanks to the fact that a pulsar is like a precise clock.

Even if the uncertainty is large in comparison with what could be possible with a pulsar binary, the measurement of the mass quadrupole moment of a BH candidate has been done in~\citet{valtonen1} and it could be relatively improved in the future~\citep{valtonen2}. The object is the supermassive BH candidate in the quasar OJ287. Optical observations show a quasi-periodic light curve characterized by two timescales, one of $\sim$12~years and another of $\sim$60~years. The interpretation is that the system is a binary BH, with the secondary BH orbiting the more massive primary one with an orbital period of $\sim$12~years and a periastron precession of $\sim$60~years~\citep{valtonen3}. The observed major outbursts occurring every $\sim$12~years are thought to be due to the secondary BH that crosses the accretion disk of the primary. Within this interpretation, \citet{valtonen1} employed a 2.5~PN (Post-Newtonian) accurate orbital dynamics to fit current observational data. Since the mass quadrupole moment interaction term enters at the 2~PN order, it is possible to constrain the mass quadrupole moment of the primary BH. Writing the quadrupole moment as $Q = - q (J^2/M)$, where $q = 1$ for a Kerr BH, current observational data provide the measurement~\citep{valtonen1} 
\be
q = 1 \pm 0.3 \, ,
\ee
namely the mass quadrupole moment is tested at the level of 30\%. In the next few years, this test could be improved at the level of 10\%~\citep{valtonen2}.

\subsection{Black hole jets}

Jets and outflows are quite a common feature of accreting compact objects. In the case of stellar-mass BH candidates, we observe two kinds of jets~\citep{rev-jets}. Steady jets are observed when the source is in the hard state. Transient jets typically appear when the source switches from the hard to the soft state.

The actual mechanism responsible for the formation of these jets is currently unknown. One of the most popular scenarios is the Blandford-Znajek mechanism~\citep{bz-jets}, in which magnetic fields threading the BH event horizon are twisted and can extract the rotational energy of spinning BHs producing an electromagnetic jet. Numerical simulations show that the process can be very efficient and strongly depends on the BH spin~\citep{mckinney,tchek1,tchek2}. The Blandford-Znajek mechanism is usually considered for the formation of steady jets. Other mechanisms do not involve the BH spin but still require magnetic fields to collimate the jets.

\begin{figure*}
\begin{center}
\hspace{-2cm}
\includegraphics[type=pdf,ext=.pdf,read=.pdf,width=9.5cm]{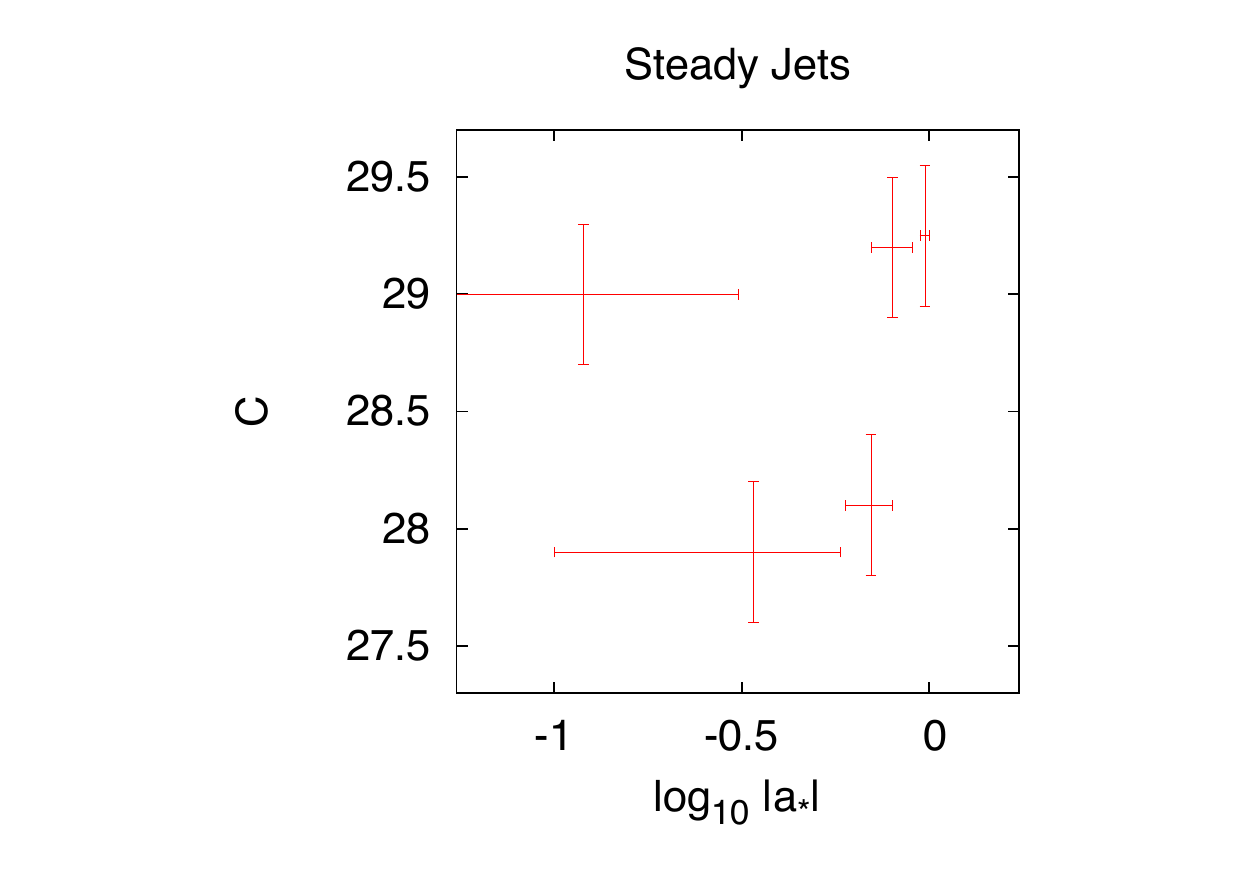}
\hspace{-2cm}
\includegraphics[type=pdf,ext=.pdf,read=.pdf,width=9.5cm]{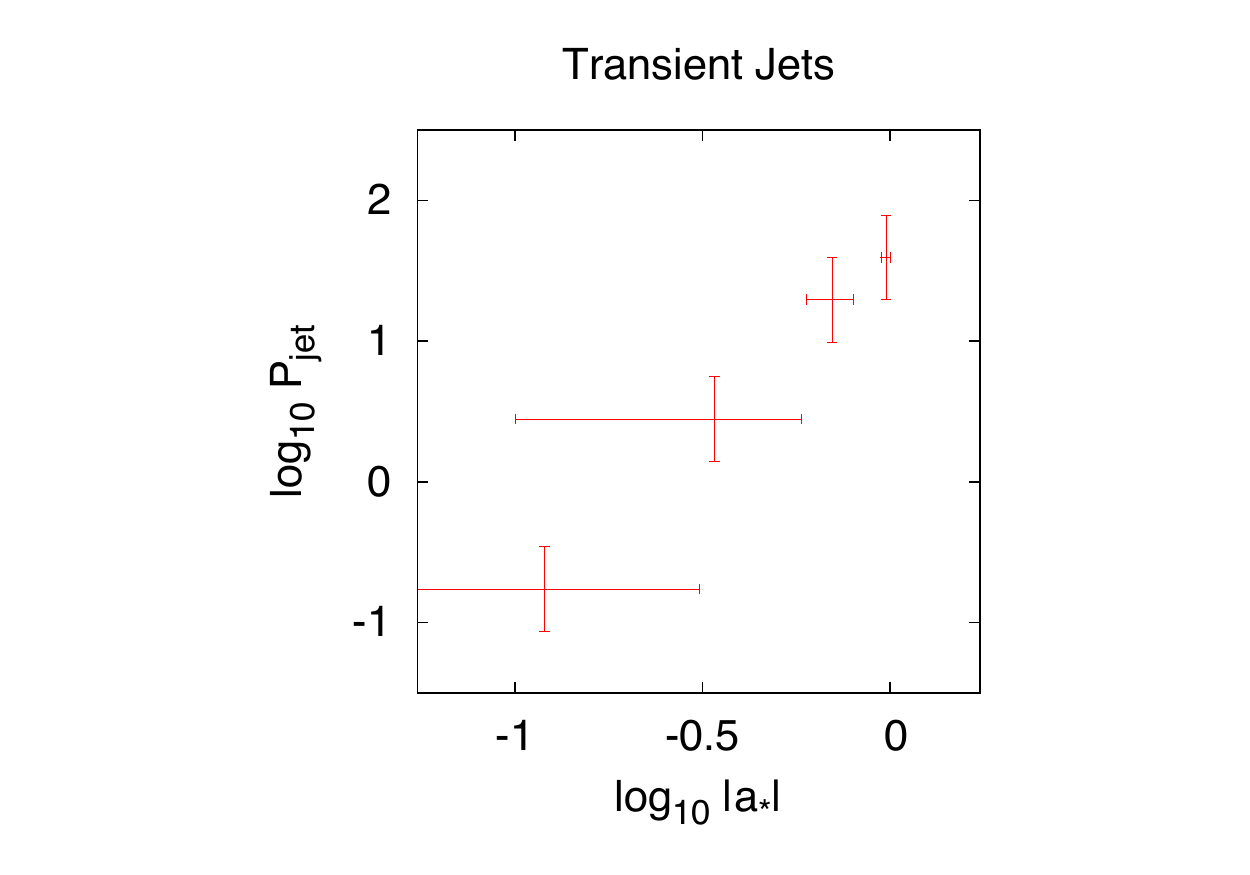}
\hspace{-2cm}
\end{center}
\vspace{-0.5cm}
\caption{Left panel: data reported in \citet{no-jets} to claim the absence of evidence for a correlation between the power of steady jets and BH spin measurements of the continuum-fitting method. Right panel: data reported in \citet{si-jets} to claim the evidence for a correlation between the power of transient jets and BH spin measurements of the continuum-fitting method.}
\label{f-jets}
\end{figure*}

Observations of a possible correlation between estimates of jet power and BH spin measurements are controversial. In \citet{no-jets}, the authors claimed that there is no evidence of a correlation between the jet power and the spin measurements of BH binaries reported in the literature with the continuum-fitting and the iron line methods. See also \citet{russell+} for more details. \citet{si-jets} proposed that the Blandford-Znajek mechanism may be responsible for the formation of transient jets and showed that there is a correlation between jet power and the most recent spin measurements with the continuum-fitting method. The left panel in Fig.~\ref{f-jets} shows the data used in \citet{no-jets} for steady jets and continuum-fitting spin measurements. The right panel shows the data reported in \citet{si-jets}.

The discrepancy between \citet{no-jets} and \citet{si-jets} can be easily understood. The two groups use different spin measurements and a different method to estimate the jet power. Moreover, as shown in Fig.~\ref{f-jets}, we only have a few measurements with large error bars. In the future, with a larger number of measurements of the jet power and more precise spin estimates, it will be possible to test the existence of a correlation between jet power and BH spin, see \citet{jack-jets}.

If the actual mechanism responsible for the formation of steady or transient jet is the Blandford-Znajek one, the measurement of the jet power could be used to estimate the BH spin if we assume the Kerr metric~\citep{jack-jets} or to test the Kerr metric otherwise~\citep{jet-c1,jet-c2,v4-jet}. The key-point is that the estimate of the jet power is (typically) quite independent of the nature of BH candidates, while the measurement of the spin does depend on the background metric. If the Blandford-Znajek mechanism is responsible for the formation of transient jets, as suggested in \citet{si-jets}, we can constrain possible deviations from the Kerr solution. If BH candidates were not Kerr BHs, the spin measurements would be different and we would lose the correlation between BH spin and jet power, see~\citet{jet-c1}. On the contrary, if the Blandford-Znajek mechanism is responsible for the formation of steady jets, the absence of a correlation between BH spin and jet power found in~\citet{no-jets} can be explained with the fact that the estimate of the BH spin is wrong, because it was obtained assuming the Kerr metric. As shown in~\citet{jet-c2}, if the continuum-fitting spin measurements are reanalyzed in a non-Kerr background it is possible to find a correlation between BH spin and jet power. At present these are just speculations based on a small number of data with large uncertainty, but they may provide some interesting results in the future.

\begin{figure*}
\begin{center}
\includegraphics[type=pdf,ext=.pdf,read=.pdf,width=8.0cm]{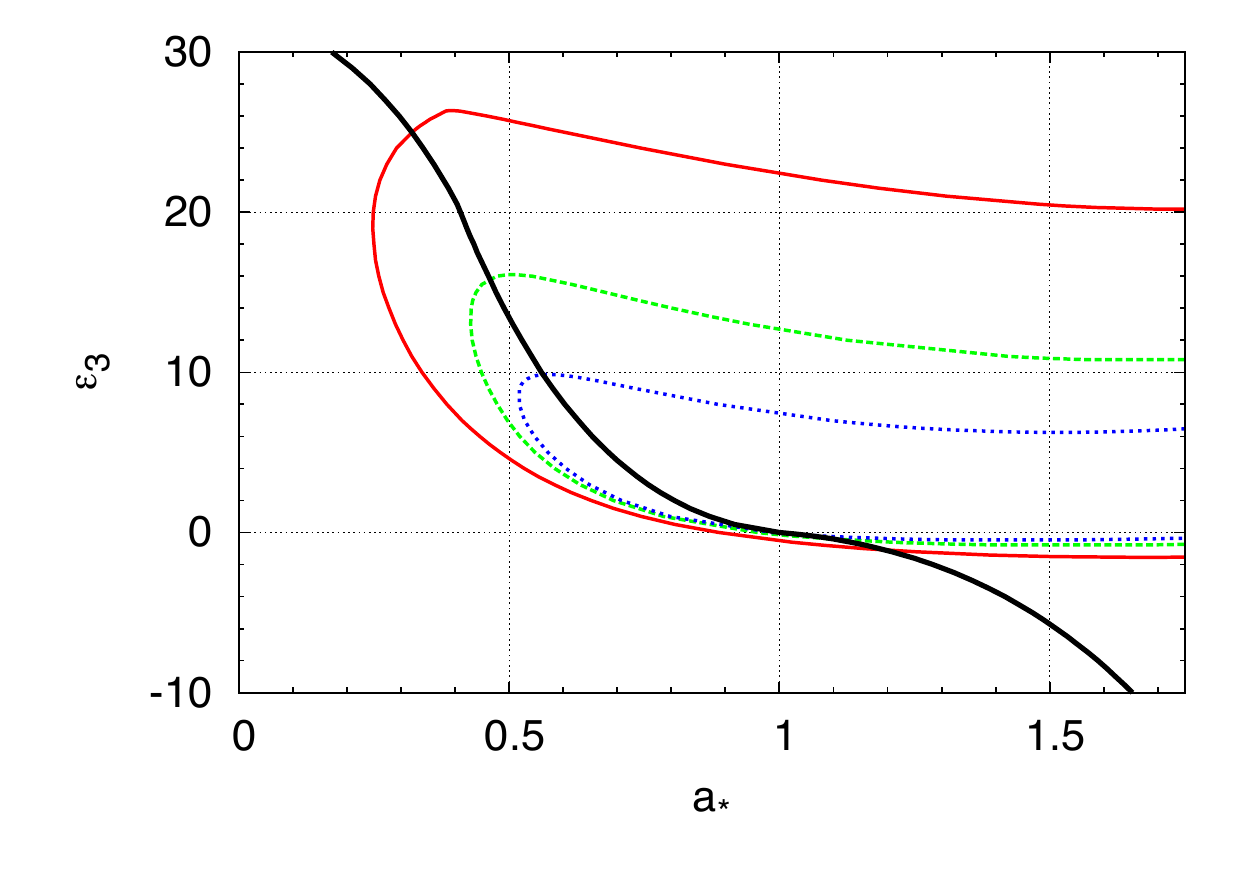}
\hspace{0.5cm}
\includegraphics[type=pdf,ext=.pdf,read=.pdf,width=8.0cm]{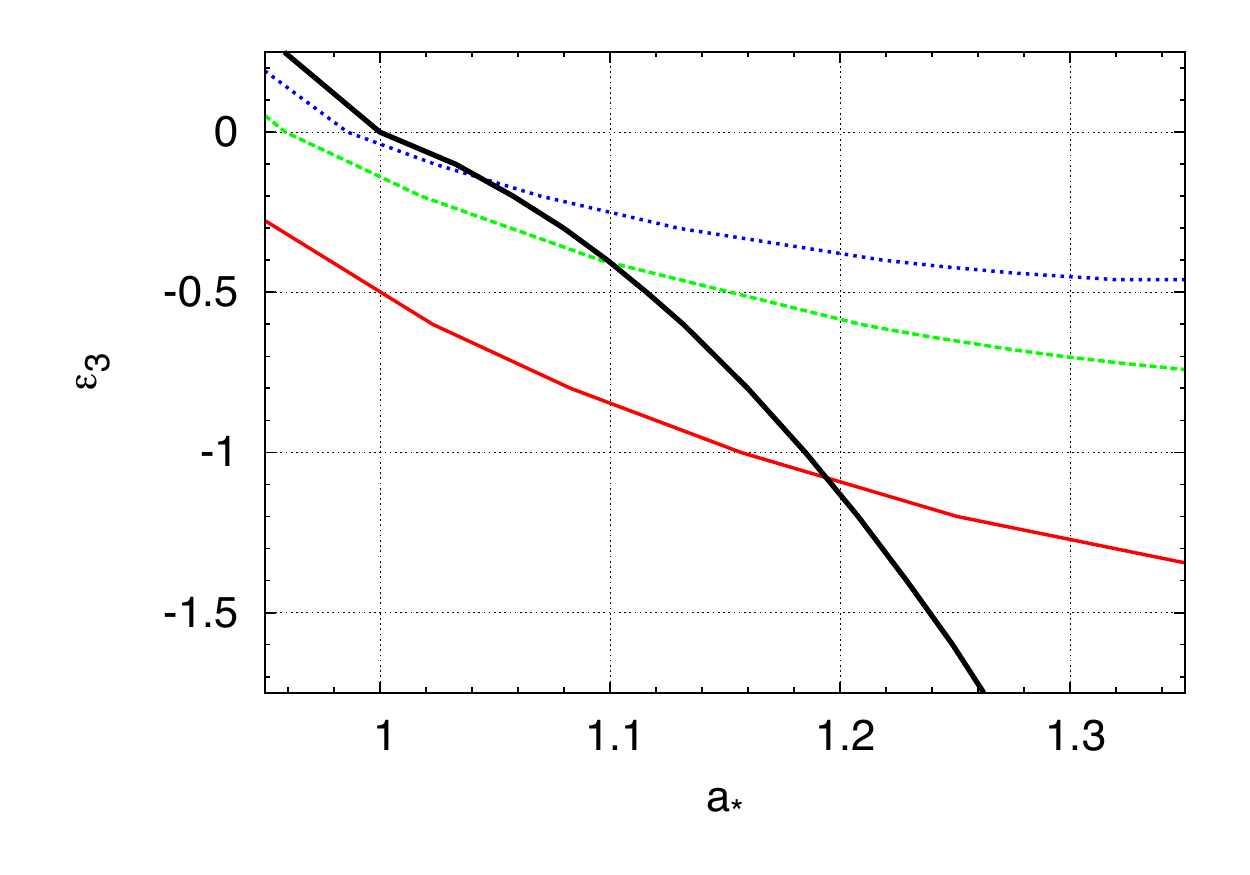}
\end{center}
\vspace{-0.5cm}
\caption{JP BHs with deformation parameter $\epsilon_3$. The black solid line corresponds to the equilibrium spin parameter $a_*^{\rm eq}$ for a thin disk, as it is inferred from Eq.~(\ref{eq-a}). The red solid, green dashed, and blue dotted curves are, respectively, the contour levels of the Novikov-Thorne radiative efficiency $\eta_{\rm NT} = 0.15$, 0.20, and 0.25. Supermassive BH candidates in galactic nuclei must be on the left of the black solid line and observations show that their radiative efficiency can be high or even very high. The allowed region is thus on the left of the black solid line and on the right of the $\eta_{\rm NT}$ contour levels. The right panel is the enlargement of the region of intersection of the curves with higher spin. See the text and the original paper \citet{spin-c4} for more details.}
\label{f-spin1}
\end{figure*}

\begin{figure*}
\begin{center}
\includegraphics[type=pdf,ext=.pdf,read=.pdf,width=8.0cm]{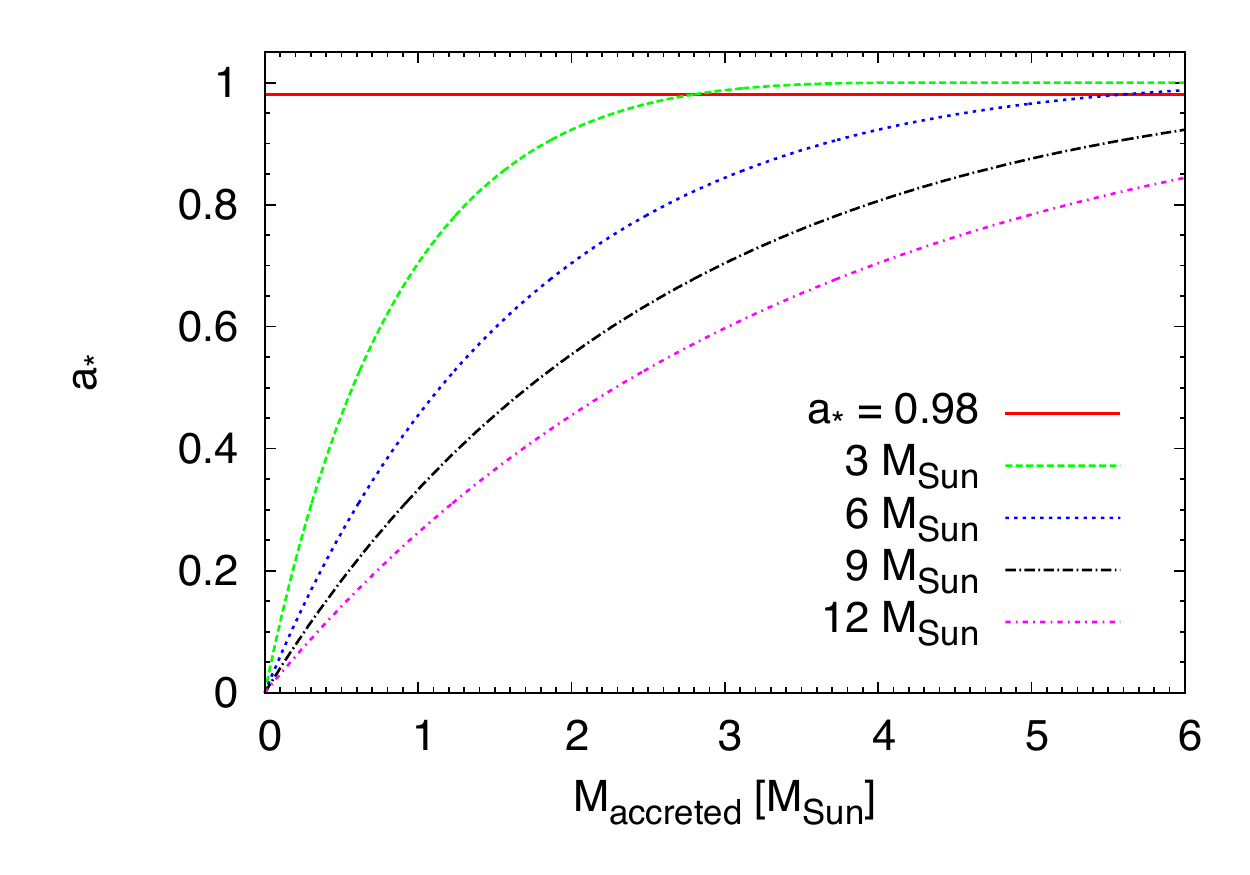}
\hspace{0.5cm}
\includegraphics[type=pdf,ext=.pdf,read=.pdf,width=8.0cm]{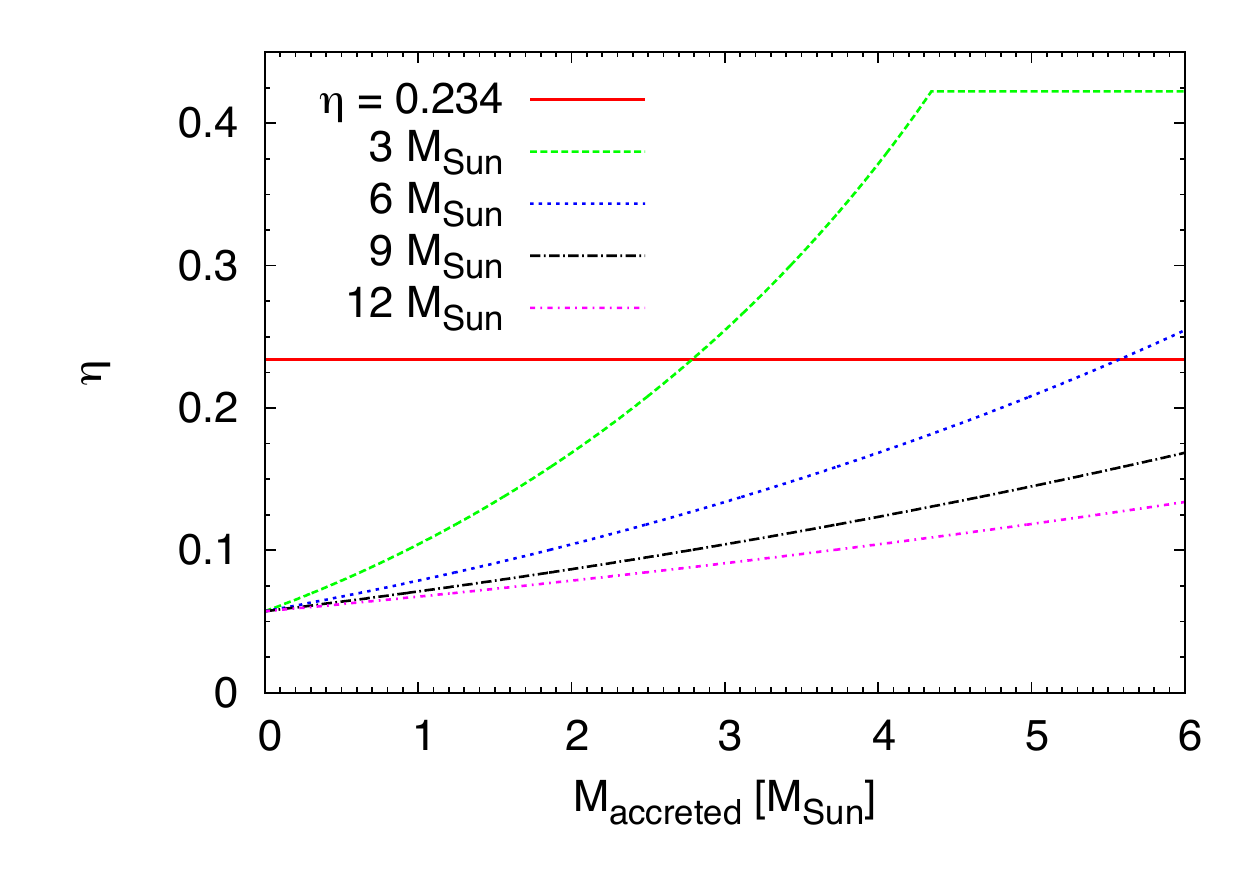} \\
\vspace{0.5cm}
\includegraphics[type=pdf,ext=.pdf,read=.pdf,width=8.0cm]{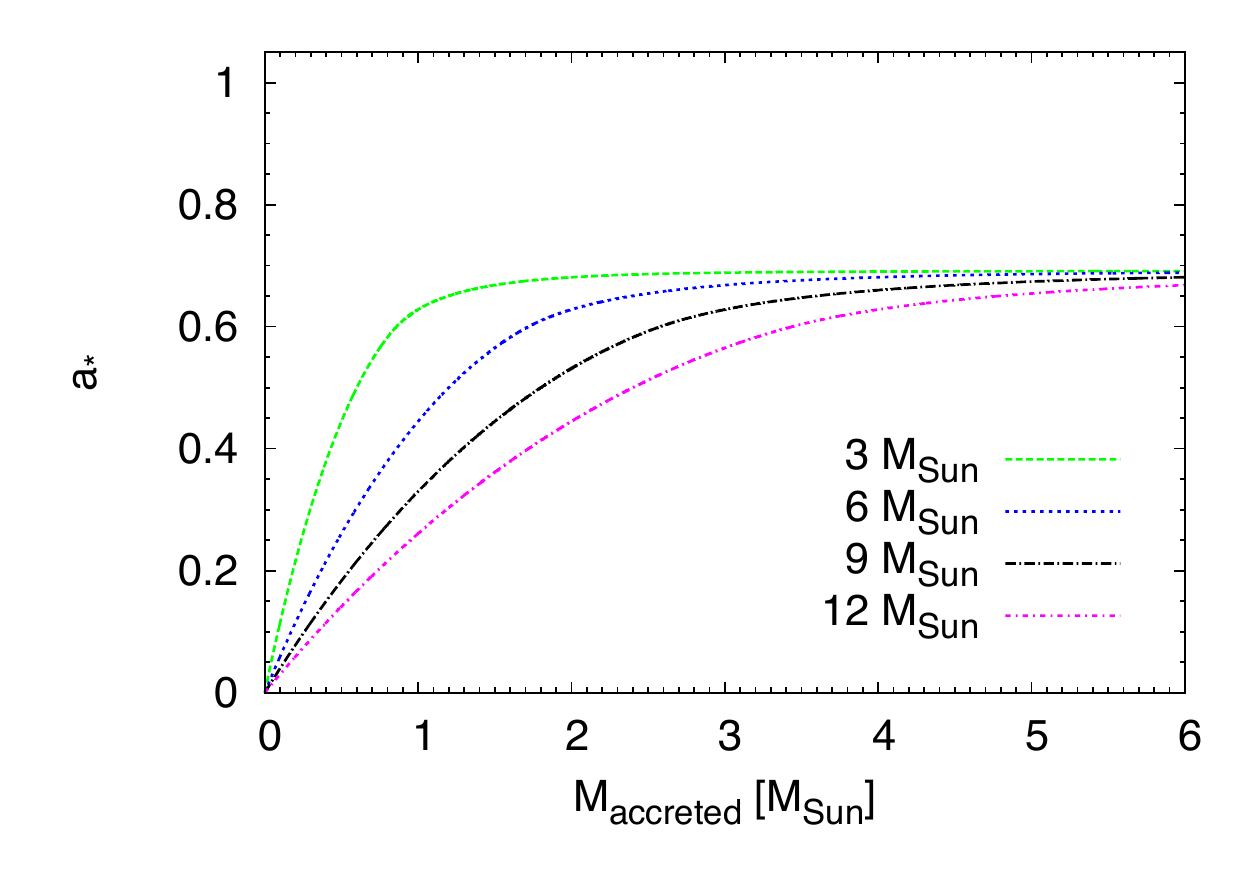}
\hspace{0.5cm}
\includegraphics[type=pdf,ext=.pdf,read=.pdf,width=8.0cm]{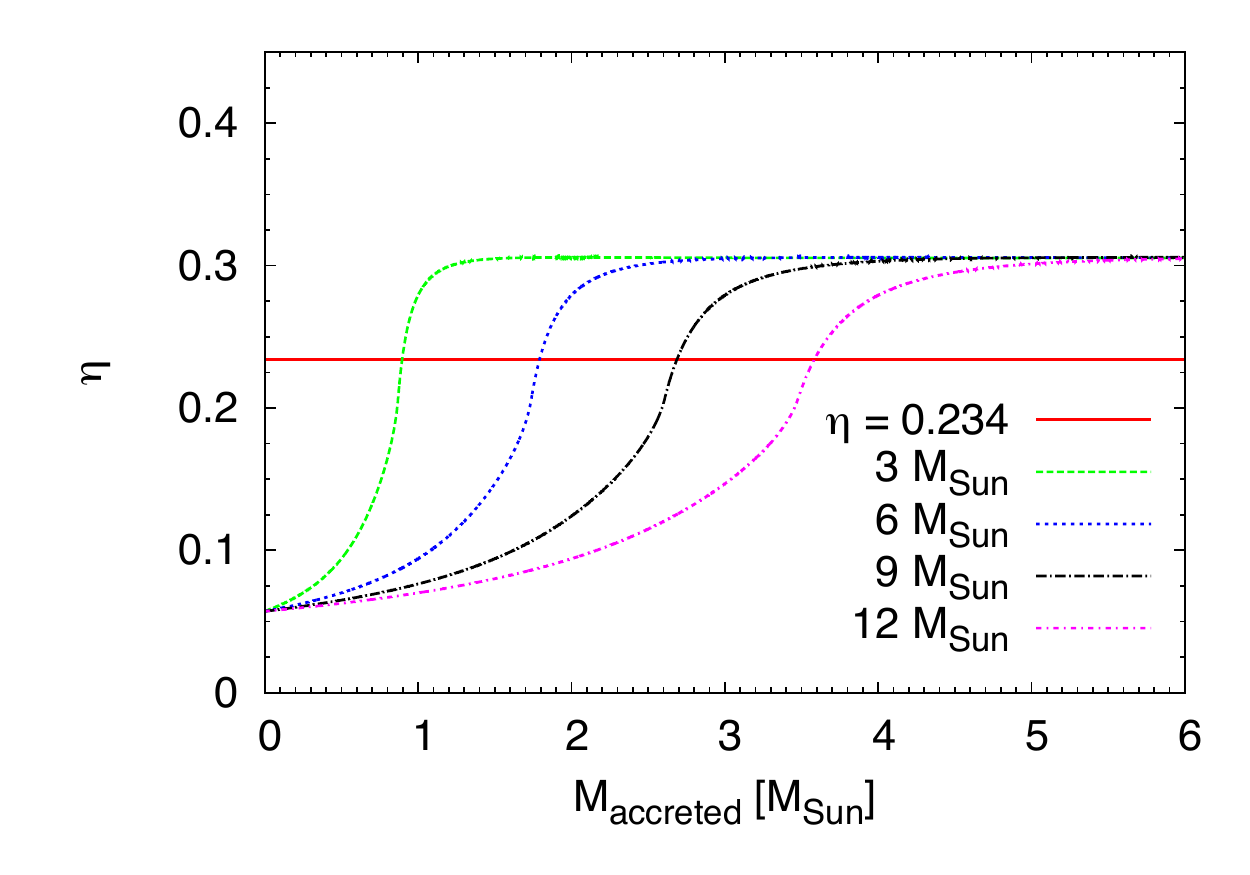}
\end{center}
\vspace{-0.5cm}
\caption{Evolution of the spin parameter $a_*$ (left panels) and of the Novikov-Thorne radiative efficiency $\eta_{\rm NT} = 1 - E_{\rm ISCO}$ (right panels) as a function of the amount of matter accreted onto an initially non-rotating BH for four different initial BH masses. The top panels are for Kerr BHs, the bottom panels for JP BHs with $\epsilon_3 = 10 \; a_*^2$. The horizontal red solid line indicates the spin parameter $a_* = 0.98$ in the top left panel, and the Novikov-Thorne radiative efficiency $\eta_{\rm NT} = 0.234$ (which requires $a_* = 0.98$ in the Kerr metric) in the right panels. From \citet{spin-c5}, under the terms of the Creative Commons Attribution License.}
\label{f-spin2}
\end{figure*}

\subsection{Evolution of the spin parameter}

As seen, there is a strong connection between the spin and possible deviations from the Kerr solution. In particular, most observations are sensitive to both parameters and their measurement is usually correlated. In actuality, the connection between the spin and the deformation parameters is even stronger. For instance, the Kerr metric with $|a_*| > 1$ can unlikely have astrophysical implications: as discussed in Subsection~\ref{naked}, it is not clear if a similar object can be created by a physical mechanism and, even if it could be created, the spacetime would be very unstable and should quickly evolve to something else. The same considerations may hold for non-Kerr BHs and other exotic compact objects. Depending on the specific spacetime metric and gravity theory, there may be a critical value of the spin parameter above which the scenario cannot be relevant in astrophysics, even in the case its gravity theory is right. Such a critical value may be either higher or lower than 1, because $|a_*| = 1$ does not correspond to any special property outside of the Kerr metric.

Neglecting possible instabilities, which would depend on the unknown gravity theory, one may get a rough estimate of the maximum BH spin as follows. Considering the astrophysical processes that can spin up and down a BH, the most efficient mechanism to create very fast-rotating bodies seems to be a prolonged accretion from a thin disk~\citep{berti08}. In the Novikov-Thorne model, the inner edge of the disk is at the ISCO radius. The gas's particles reach the ISCO and then immediately plunge onto the compact object with specific energy $E_{\rm ISCO}$ and specific angular momentum $L_{\rm ISCO}$, namely with their value at the ISCO radius. The mass $M$ and the spin angular momentum $J$ of the compact object change, respectively, by 
\be
\delta M &=& E_{\rm ISCO} \delta m \, , \\
\delta J &=& L_{\rm ISCO} \delta m \, ,
\ee
where $\delta m$ is the gas rest-mass. The evolution of the spin parameter turns out to be governed by the following equation~\citep{bardeen70}
\be\label{eq-a}
\frac{da_*}{d\ln M} = \frac{1}{M} 
\frac{L_{\rm ISCO}}{E_{\rm ISCO}} - 2 a_* \, ,
\ee
where here the small effect of the radiation emitted by the disk and captured by the object is neglected. Prolonged disk accretion is a very efficient mechanism to spin the compact object up till an equilibrium spin parameter $a_*^{eq}$, which is reached when the right-hand side of Eq.~(\ref{eq-a}) becomes zero. If, for some reason, the spin is $a_* > a_*^{eq}$, the accretion process spins the compact object down.

In the case of the Kerr metric, it is possible to integrate Eq.~(\ref{eq-a}) and find an analytic expression for the spin parameter $a_*$ as a function of the BH mass $M$~\citep{bardeen70}
\be\label{eq-a-v3}
\hspace{-0.6cm}
a_* =  
\begin{cases}
\sqrt{\frac{2}{3}}
\frac{M_0}{M} \left[4 - \sqrt{18\frac{M_0^2}{M^2} - 2}\right] 
 & \text{if } M \le \sqrt{6} M_0 \, , \\
1 & \text{if } M > \sqrt{6} M_0 \, ,
\end{cases}
\ee
assuming an initially non-rotating BH with mass $M_0$. In Eq.~(\ref{eq-a-v3}), the equilibrium spin parameter is $a_*^{eq} = 1$, which is reached after the BH has increased its mass by a factor of $\sqrt{6} \approx 2.4$~\citep{bardeen70}. If we include the effect of the radiation emitted by the disk and captured by the BH, we obtain the Thorne bound $a_*^{eq} \approx 0.998$~\citep{th74}, because radiation with angular momentum opposite to the BH spin has a larger capture cross-section.

One can repeat the calculations in a non-Kerr metric and find the corresponding equilibrium spin parameter $a_*^{eq}$, which may be either larger or smaller than 1~\citep{spin-c1,spin-c2}. In the case of the JP metric with non-vanishing $\epsilon_3$, the curve of $a_*^{eq}$ is the black solid line shown in Fig.~\ref{f-spin1}. If the object is on the left of the curve, an accretion disk spins it up to reach the equilibrium spin parameter. If the object is on the right of the curve, the accretion process spins the body down. As we can see from Fig.~\ref{f-spin1}, $a_*^{eq} > 1$ for $\epsilon_3 < 0$ and $a_*^{eq} < 1$ when $\epsilon_3 > 0$. Accretion from thick disks might be a little bit more efficient, but the difference would be small~\citep{z1}.

In the case of the supermassive BH candidates in galactic nuclei, the initial value of their spin parameter is irrelevant, as these objects have increased their mass by several orders of magnitude and their spin parameter has evolved accordingly. Prolonged disk accretion is the most efficient mechanism to get a high spin and therefore we can conclude that these objects cannot have today a spin parameter larger than $a_*^{eq}$~\citep{spin-c4}. At the same time, it is possible to provide a lower bound on their radiative efficiency, either as a mean radiative efficiency from the Soltan argument~\citep{soltan1,soltan2} or for specific sources~\citep{davis}. The radiative efficiency of these objects seems to be high, or even very high, but the exact estimate is more controversial. Fig.~\ref{f-spin1} shows the contour levels of the Novikov-Thorne radiative efficiency $\eta_{\rm NT} = 0.15$ (red solid curve), 0.20 (green dashed curve), and 0.25 (blue dotted curve). The observed high radiative efficiency constrains the spacetime geometry around supermassive BH candidates~\citep{cb2011e,cb2012e}. Assuming the conservative bound $\eta_{\rm NT} > 0.15$ and imposing that $a_* < a_*^{eq}$, from Fig.~\ref{f-spin1} we see that supermassive BH candidates are constrained to be in the region bounded on the left by the red solid curve of $\eta_{\rm NT} = 0.15$, and on the right by the black solid line of the equilibrium spin parameter. Moreover, it is possible to estimate an upper bound for the spin parameter of supermassive BH candidates, at the level of $|a_*| \lesssim 1.2$ (1.1, 1.05) if one assumes the constraint $\eta_{\rm NT} \gtrsim 0.15$ (respectively 0.20, 0.25), and this seems to be only very weakly dependent on the exact non-Kerr metric adopted for the calculations~\citep{spin-c4}.

In the case of stellar-mass BH candidates, it is commonly thought that the spin of the compact object is natal~\citep{kk99}; however, also see~\citet{fragos15}. The point is that stellar-mass BH candidates have a mass around 10~$M_\odot$. If the companion is a few $M_\odot$, the BH candidate cannot significantly change its mass and spin angular momentum even if it swallows the whole star. If the companion is heavy, its lifetime is very short, and it is impossible to transfer the necessary amount of material to spin the BH candidate up even in the case of accretion at the Eddington rate. In the end, a BH candidate cannot get more than a few $M_\odot$ from the companion star, and for a 10~$M_\odot$ object this is arguably not enough to appreciably change its spin parameter for sufficiently large values of spin~\citep{kk99}.

Moreover, while there are still uncertainties in the angular momentum transport mechanisms of the progenitors of stellar-mass BHs, it is widely accepted that the gravitational collapse of a massive star with solar metallicity cannot create fast-rotating remnants~\citep{wb06,yoon06}. However, this is not what we observe. Assuming the Kerr metric, we see BHs with spin close to 1, see Tab.~\ref{tab1}. For instance, the BH candidate in GRS~1915+105 has $a_* > 0.98$ and $M = 12.4 \pm 2.0$~$M_\odot$, while the stellar companion's mass is $M = 0.52 \pm 0.41$~$M_\odot$. In the case of high-mass X-ray binaries, the BH candidate in Cygnus~X-1 has $a_* > 0.98$ and $M = 14.8 \pm 1.0$~$M_\odot$, while the stellar wind from the companion is not an efficient mechanism to transfer mass.

A speculative possibility to explain this puzzle is to admit that BH candidates are not the Kerr BHs of general relativity~\citep{spin-c5}. The top panels in Fig.~\ref{f-spin2} show the evolution of the spin parameter (left panel) and of the Novikov-Thorne radiative efficiency (right panel) of a Kerr BH as a function of the accreted mass for a few different BH initial masses. At the beginning, the BH is assumed to be non-rotating. We can see that a Kerr BH with $M \sim 12$~$M_\odot$ and $a_* > 0.98$ (corresponding to a radiative efficiency $\eta_{\rm NT} > 0.234$, which is the actual quantity measured with the continuum-fitting method) had to swallow about 6~$M_\odot$. The bottom panels are as the top ones for a JP metric with $\epsilon_3 = k a_*^2$ and $k=10$. In this case, the object needs to get a smaller mass from the companion, about 2~$M_\odot$, to acquire a radiative efficiency to explain the Kerr measurement of its spin. More detailed calculations confirm that such an exotic scenario could explain this puzzle~\citep{spin-c5}. If it were possible to estimate the mass transfer to pass from a BH with low radiative efficiency to a BH with high radiative efficiency, it could be possible to test the Kerr metric.


\section{Comparison between tests with electromagnetic and gravitational wave observations \label{s-new}}

The announcement by the LIGO/Virgo Collaboration of the first direct detection of gravitational waves has opened a new window for testing gravity in the strong field regime~\citep{v3-ligo}. Gravitational wave experiments now promise to be able to perform precise tests within 5-10~years. It is natural to wonder how such a breakthrough can affect the attempts to test BH candidates with electromagnetic radiation discussed in this review paper.

First, it is worth noting that in general it is not straightforward to compare the capabilities of the gravitational wave and the electromagnetic approaches to test BH candidates. As already pointed out in Subsection~\ref{ss-v3-ir}, they measure different things. Electromagnetic tests are sensitive to the motion of particles (massive particles in the accretion disk and photons propagating from the emission point in the disk to the detection point in the flat faraway region). Gravitational wave tests are sensitive to the field equations of the gravity theory, and eventually one can study the evolution of perturbations on the background metric.

In some alternative theories of gravity, BHs are still described by the Kerr metric, but the emission of gravitational waves is typically different~\citep{kenrico}: in such a case, only the gravitational wave approach can test the model. The contrary is also possible. The existence of non-minimal interaction terms between the electromagnetic and the gravitational fields or, more in general, the presence of new fields leading to a violation of the weak equivalence principle may affect the motion of photons/particles without altering the emission of gravitational waves. In similar frameworks, only the electromagnetic approach can detect deviations from standard predictions.

Bearing in mind these fundamental differences between the gravitational wave and the electromagnetic approaches, some preliminary studies have already investigated the constraints that can be inferred by the LIGO data of GW150914 and compared with the constraining power of some electromagnetic techniques.

\citet{v3-yyp} considered a number of alternative theories of gravity and showed that the data of GW150914 can already strongly constraint some models. This depends on the specific gravity theory under consideration, because in some models the gravitational wave signal is very different from that expected in general relativity, while in other frameworks it is not.

\citet{v3-ring} pointed out that the ringdown signal from a binary coalescence -- like in the case of GW150914 -- cannot be seen as conclusive proof for the formation of an event horizon after the merger. They showed that universal ringdown waveforms indicate only the presence of light rings, which can be possessed even by very compact objects without horizon.

\citet{v3-kz} tried to test the Kerr metric, independently of the gravity theory, from the signal of GW150914. In the coalescence of a binary, we can distinguish three stages: the post-Newtonian inspiral, the merger, and the ringdown. The post-Newtonian inspiral may be the same in many alternative theories of gravity if the metric deviates from the Kerr solution only in the near horizon region (this is not universally true and depends on the specific theory of gravity). The merger is a very short and complicated stage. The ringdown phase may be the most suitable to test strong gravity. As a further simplification, \citet{v3-kz} studied the quasi-normal frequencies of a scalar field in the deformed background. In general relativity and in other theories of gravity, these frequencies are not much different from those of the gravitational waves derived from the field equations. However, this is not always true and there are also examples of gravity theories in which the scalar field quasi-normal frequencies can be quite different from those of the gravitational waves. The analysis in \citet{v3-kz} showed that it is difficult to constrain deviations from the Kerr solution from GW150914 because the measurement of the spin and the deformation parameter are correlated. This would remain true even in the presence of the detection of additional modes\footnote{The data of GW150914 are consistent with a single damped sinusoid, which can be naturally interpreted as either the $l=m=2$ mode or the $l=2$, $m=-2$ mode since the event was a binary BH merger.}.

\citet{v3-cjb2} and \citet{v3-sourabh} studied the constraining power of, respectively, iron line and QPOs for the metric discussed in \citet{v3-kz} in order to compare the electromagnetic and the gravitational wave approaches. The conclusions of these preliminary studies can be summarized as follows. In the presence of high-quality data and the correct astrophysical model, the iron line method can provide strong constraints on the spin and the deformation parameter. The technique may thus be able to compete, or to be complementary to, the gravitational wave approach. The necessary high-quality data may be already available in the case of BH binaries. The systematics is currently the main concern about this approach and there is not a common consensus if eventually one can really get reliable and accurate measurements of the background metric. QPOs can provide very precise measurements, but they are not able to break the degeneracy between the spin and possible deviations from the Kerr metric. The parameter degeneracy cannot be broken even imagining some very accurate frequency measurements in the future.


\section{Summary and conclusions \label{s6-sc}}

Astrophysical BH candidates can be naturally interpreted as the Kerr BHs of general relativity. However, a direct observational evidence that the spacetime geometry around these objects is really described by the Kerr solution is still lacking and, at the same time, a number of theoretical models suggest the possibility of new physics and macroscopic deviations from the Kerr background. In this paper, I have reviewed current attempts to test the Kerr metric with electromagnetic radiation. The spectrum of the accretion disk, but even of stars orbiting close to a supermassive BH candidate, has features that can be used to study the metric around these compact objects and thus test the Kerr BH hypothesis.

\subsection{Current constraints}

Today, we do not have strong constraints on the actual nature of BH candidates and on the spacetime geometry around them. However, current observations are consistent with the Kerr BH hypothesis and can rule out some alternative scenarios. In particular:
\begin{enumerate}
\item A number of observations are consistent with the fact that BH candidates have an event horizon~\citep{eea1,eeb1,eeb2,eec1,tournear}. There is no direct proof, because such a proof is by the definition of event horizon impossible, but many (not all) alternative scenarios in which these objects would not be BHs can be ruled out. 
\item Many compact objects made of exotic weakly-interacting matter can be ruled out because their spacetime cannot reproduce the characteristic low energy tail of the iron line profile expected in the X-ray spectrum of very fast-rotating BHs~\citep{exotic1}. The gravitational field around these exotic objects is never very strong (and indeed they have no horizon), so photons cannot be strongly redshifted as is instead expected when near BHs. Current data show that sources with iron line profiles with a long low energy tail are common and therefore it is possible to rule out a number of exotic compact objects.
\item The mass-quadrupole moment of the supermassive BH candidate in the quasar OJ287 has been measured and it is consistent with the Kerr prediction at the level of 30\%~\citep{valtonen1}. This bound is weak, because BHs of different types and from different gravity theories are usually quite similar. However, if we consider a compact object like a neutron star, the value of its mass-quadrupole moment is a few times that of a Kerr BH with the same mass and spin.
\item The data of the thermal spectrum of the accretion disks of the stellar-mass BH candidates in GRS~1915+105 and in Cygnus~X-1 exclude the possibility of large deviations from the Kerr solution~\citep{lingyao,cfm-cc}. These observations cannot constrain any kind of deformation, but they can safely rule out some deformations with a strong impact on the position of the ISCO radius.
\item Even in the case of supermassive BH candidates we can exclude some large deviations with a strong impact on the ISCO radius. Very deformed objects cannot have a high radiative efficiency. At the same time, one can constrain the maximum value of their spin parameter from considerations on the spin evolution. The combination of these two arguments is a limited allowed region on the spin parameter -- deformation parameter plane: eventually, very large deviations from the Kerr solution are not permitted~\citep{spin-c4}.
\end{enumerate}

\subsection{Prospectives for the future}

Current constraints mainly rule out the possibility that BH candidates are certain compact objects made of exotic matter and far from forming a horizon. Indeed, the gravitational redshift experienced by photons emitted close to the surface of these objects is surely very strong (points 1-2 in the previous subsection). Rotation does not make these objects as oblate as one should expect, for instance, in the case of a neutron star, and this is consistent with the fact that these objects are BHs (points 3-5). However, these constraints are typically unable to distinguish BHs from different theories of gravity. It is definitively challenging to reach accurate and stringent constraints on the metric around these objects, because it is necessary to have a very good astrophysical model. The current situation of present techniques and future prospectives can be summarized as follows:
\begin{enumerate}
\item The continuum-fitting method is probably the best technique available today to test BH candidates, but it can only be used for the stellar-mass ones. Its weak points are that mass, distance, and inclination angle of the source must be obtained from independent measurements (with current methods, the uncertainties are usually large and systematic effects are possible), and that corrections to the blackbody spectrum depend on certain atmosphere models. With more precise measurements of BH masses, distances, and viewing angle in the future, it is possible to obtain stronger constraints on the spin (if we assume the Kerr metric) or on the spin and possible deviations from the Kerr solution. However, the spectrum is typically degenerate with respect to the spin and the deformation parameters, so it is impossible to test the Kerr spacetime without independent measurements of the metric. This seems to be an intrinsic limitation of this method.
\item The analysis of the iron K$\alpha$ line is potentially quite a powerful tool to test the Kerr metric. It can be used for both stellar-mass and supermassive objects. The main problem is the astrophysical model and the systematic effects. Currently there is no unanimous consensus on the possibility of using this technique to get accurate and reliable measurements of the metric around BH candidates. Some assumptions are still to be proven, such as the fact that the inner edge of the disk is at the ISCO radius in the hard state. The intensity profiles currently used are based on phenomenological models. Even the lamppost geometry is one among other configurations. In the case of supermassive BH candidates and with current X-ray facilities, there is an additional limitation due to the low photon number count in the iron line. In the presence of the correct theoretical model and high quality data, the iron K$\alpha$ line may become a leading technique to test general relativity.
\item Since tests of the Kerr metric have to face the problem of a strong degeneracy among the spin and possible deviations from the Kerr background, the combination of measurements sensitive to different relativistic effects seems to be the key point to break the parameter degeneracy and test the Kerr BH hypothesis. From this point of view, SgrA$^*$ may be one of the best sources to test the Kerr nature of BH candidates. A number of new observations may be available in the near future and the combination of these measurements could test the nature of SgrA$^*$. 
\item QPOs, pulsars, jet power, etc. are potential tools for the future. Today we cannot use these techniques, either because the physical mechanism is not well understood, or because there are no current measurements, or the measurements are not yet good enough. Some of these techniques may eventually work and be useful to test the Kerr metric. Other approaches may remain just as an idea.
\end{enumerate}


\section*{Acknowledgments}

I thank Alejandro C\'ardenas-Avenda\~no, Matteo Guainazzi, Jiachen Jiang, Gleniese McKenzie, and James Steiner for useful discussions and suggestions. This work was supported by the NSFC grants No.~11305038 and No.~U1531117, the Shanghai Municipal Education Commission grant No.~14ZZ001, the Thousand Young Talents Program, Fudan University, and the Alexander von Humboldt Foundation.



\begin{thebibliography}{99}

\bibitem[Abbott et al.(2016a)]{v3-ligo} 
Abbott, B.~P., Abbott, R., Abbott, T.~D., et al.\ 2016a, Physical Review Letters, 116, 061102

\bibitem[Abbott et al.(2016b)]{v4-gw} 
Abbott, B.~P., Abbott, R., Abbott, T.~D., et al.\ 2016b, Physical Review Letters, 116, 241103

\bibitem[Abdujabbarov et al.(2013)]{kaz111} 
Abdujabbarov, A., Atamurotov, F., Kucukakca, Y., Ahmedov, B., \& Camci, U.\ 2013, \apss, 344, 429

\bibitem[Abdujabbarov et al.(2015)]{a-r-a} 
Abdujabbarov, A.~A., Rezzolla, L., \& Ahmedov, B.~J.\ 2015, \mnras, 454, 2423

\bibitem[Abramowicz \& Klu{\'z}niak(2001)]{qpo4a} 
Abramowicz, M.~A., \& Klu{\'z}niak, W.\ 2001, \aap, 374, L19

\bibitem[Abramowicz et al.(2002)]{hhh1} 
Abramowicz, M.~A., Klu{\'z}niak, W., \& Lasota, J.-P.\ 2002, \aap, 396, L31

\bibitem[Agol et al.(2002)]{v3-micro2} 
Agol, E., Kamionkowski, M., Koopmans, L.~V.~E., \& Blandford, R.~D.\ 2002, \apjl, 576, L131

\bibitem[Aliev et al.(2013)]{qpoa} 
Aliev, A.~N., Daylan Esmer, G., \& Talazan, P.\ 2013, Classical and Quantum Gravity, 30, 045010

\bibitem[Amarilla \& Eiroa(2012)]{shadow-a1} 
Amarilla, L., \& Eiroa, E.~F.\ 2012, \prd, 85, 064019

\bibitem[Amarilla \& Eiroa(2013)]{shadow-a2} 
Amarilla, L., \& Eiroa, E.~F.\ 2013, \prd, 87, 044057

\bibitem[Amarilla et al.(2010)]{shadow-a3} 
Amarilla, L., Eiroa, E.~F., \& Giribet, G.\ 2010, \prd, 81, 124045

\bibitem[Ang{\'e}lil et al.(2010)]{will4} 
Ang{\'e}lil, R., Saha, P., \& Merritt, D.\ 2010, \apj, 720, 1303

\bibitem[Arnaud et al.(2011)]{v4-ass} 
Arnaud, K., Smith, R., \& Siemiginowska, A.\ 2011, {\it Handbook of X-ray Astronomy} (Cambridge University Press, Cambridge, UK)

\bibitem[Astashenok et al.(2014)]{ns-atg} 
Astashenok, A.~V., Capozziello, S., \& Odintsov, S.~D.\ 2014, \prd, 89, 103509

\bibitem[Atamurotov et al.(2013a)]{shadow-aaa1} 
Atamurotov, F., Abdujabbarov, A., \& Ahmedov, B.\ 2013a, \apss, 348, 179

\bibitem[Atamurotov et al.(2013b)]{shadow-aaa2} 
Atamurotov, F., Abdujabbarov, A., \& Ahmedov, B.\ 2013b, \prd, 88, 064004

\bibitem[Azreg-A{\"i}nou(2011)]{azreg} 
Azreg-A{\"i}nou, M.\ 2011, Classical and Quantum Gravity, 28, 148001

\bibitem[Baiotti et al.(2005)]{luca1} 
Baiotti, L., Hawke, I., Montero, P.~J., et al.\ 2005, \prd, 71, 024035

\bibitem[Baiotti \& Rezzolla(2006)]{luca2} 
Baiotti, L., \& Rezzolla, L.\ 2006, Physical Review Letters, 97, 141101

\bibitem[Bambi(2011a)]{review1} 
Bambi, C.\ 2011a, Modern Physics Letters A, 26, 2453

\bibitem[Bambi(2011b)]{spin-c4} 
Bambi, C.\ 2011b, Physics Letters B, 705, 5

\bibitem[Bambi(2011c)]{spin-c2} 
Bambi, C.\ 2011c, \jcap, 5, 009

\bibitem[Bambi(2011d)]{spin-c1} 
Bambi, C.\ 2011d, EPL (Europhysics Letters), 94, 50002

\bibitem[Bambi(2011e)]{cb2011e} 
Bambi, C.\ 2011e, \prd, 83, 103003

\bibitem[Bambi(2012a)]{cfm-c2} 
Bambi, C.\ 2012a, \apj, 761, 174

\bibitem[Bambi(2012b)]{qpoc1} 
Bambi, C.\ 2012b, \jcap, 9, 014

\bibitem[Bambi(2012c)]{jet-c1} 
Bambi, C.\ 2012c, \prd, 85, 043002

\bibitem[Bambi(2012d)]{jet-c2} 
Bambi, C.\ 2012d, \prd, 86, 123013

\bibitem[Bambi(2012e)]{cb2012e} 
Bambi, C.\ 2012e, \prd, 85, 043001

\bibitem[Bambi(2013a)]{review2} 
Bambi, C.\ 2013a, The Astronomical Review, 8, 4

\bibitem[Bambi(2013b)]{hhh2} 
Bambi, C.\ 2013b, Scientific World Journal 2013, 204315 

\bibitem[Bambi(2013c)]{iron-c1} 
Bambi, C.\ 2013c, \prd, 87, 023007

\bibitem[Bambi(2013d)]{cfm-iron} 
Bambi, C.\ 2013d, \jcap, 8, 055 

\bibitem[Bambi(2013e)]{exotic2} 
Bambi, C.\ 2013e, \prd, 87, 084039

\bibitem[Bambi(2013f)]{sh-wh-br} 
Bambi, C.\ 2013f, \prd, 87, 107501

\bibitem[Bambi(2014a)]{cfm-cc} 
Bambi, C.\ 2014a, Physics Letters B, 730, 59

\bibitem[Bambi(2014b)]{cfm-cc2} 
Bambi, C.\ 2014b, \prd, 90, 047503

\bibitem[Bambi(2015a)]{qpoc2} 
Bambi, C.\ 2015a, European Physical Journal C, 75, 162

\bibitem[Bambi(2015b)]{sgra-c} 
Bambi, C.\ 2015b, Classical and Quantum Gravity, 32, 065005

\bibitem[Bambi(2015c)]{spin-c5} 
Bambi, C.\ 2015c, European Physical Journal C, 75, 22

\bibitem[Bambi \& Barausse(2011a)]{cfm-c1} 
Bambi, C., \& Barausse, E.\ 2011a, \apj, 731, 121

\bibitem[Bambi \& Barausse(2011b)]{bb2011} 
Bambi, C., \& Barausse, E.\ 2011b, \prd, 84, 084034

\bibitem[Bambi et al.(2012)]{sh-c3} 
Bambi, C., Caravelli, F., \& Modesto, L.\ 2012, Physics Letters B, 711, 10

\bibitem[Bambi et al.(2009)]{petrov} 
Bambi, C., Dolgov, A.~D., \& Petrov, A.~A.\ 2009, \jcap, 9, 013

\bibitem[Bambi \& Freese(2009)]{sh-c1} 
Bambi, C., \& Freese, K.\ 2009, \prd, 79, 043002

\bibitem[Bambi et al.(2016a)]{v3-cqg} 
Bambi, C., Jiang, J., \& Steiner, J.~F.\ 2016a, Classical and Quantum Gravity, 33, 064001

\bibitem[Bambi \& Lukes-Gerakopoulos(2013)]{glg} 
Bambi, C., \& Lukes-Gerakopoulos, G.\ 2013, \prd, 87, 083006

\bibitem[Bambi \& Malafarina(2013)]{exotic1} 
Bambi, C., \& Malafarina, D.\ 2013, \prd, 88, 064022

\bibitem[Bambi et al.(2014a)]{bmm} 
Bambi, C., Malafarina, D., \& Modesto, L.\ 2014a, European Physical Journal C, 74, 2767

\bibitem[Bambi et al.(2016b)]{v3-bmm2} 
Bambi, C., Malafarina, D., \& Modesto, L.\ 2016b, \jhep, 4, 147

\bibitem[Bambi et al.(2014b)]{disk} 
Bambi, C., Malafarina, D., \& Tsukamoto, N.\ 2014b, \prd, 89, 127302

\bibitem[Bambi \& Modesto(2011)]{pat1} 
Bambi, C., \& Modesto, L.\ 2011, Physics Letters B, 706, 13

\bibitem[Bambi \& Nampalliwar(2016)]{v3-sourabh} 
Bambi, C., \& Nampalliwar, S.\ 2016, arXiv:1604.02643

\bibitem[Bambi \& Yoshida(2010)]{sh-c2} 
Bambi, C., \& Yoshida, N.\ 2010, Classical and Quantum Gravity, 27, 205006

\bibitem[Barack \& Cutler(2007)]{tests-gw2} 
Barack, L., \& Cutler, C.\ 2007, \prd, 75, 042003

\bibitem[Barausse et al.(2010)]{super2} 
Barausse, E., Cardoso, V., \& Khanna, G.\ 2010, Physical Review Letters, 105, 261102

\bibitem[Barausse et al.(2014)]{v2-bcp} 
Barausse, E., Cardoso, V., \& Pani, P.\ 2014, \prd, 89, 104059

\bibitem[Barausse \& Sotiriou(2008)]{kenrico} 
Barausse, E., \& Sotiriou, T.~P.\ 2008, Physical Review Letters, 101, 099001

\bibitem[Bardeen(1970)]{bardeen70} 
Bardeen, J.~M.\ 1970, \nat, 226, 64

\bibitem[Bardeen \& Petterson(1975)]{alignment1} 
Bardeen, J.~M., \& Petterson, J.~A.\ 1975, \apjl, 195, L65

\bibitem[Bardeen et al.(1972)]{bpt72} 
Bardeen, J.~M., Press, W.~H., \& Teukolsky, S.~A.\ 1972, \apj, 178, 347

\bibitem[Beer \& Podsiadlowski(2002)]{gro-m1} 
Beer, M.~E., \& Podsiadlowski, P.\ 2002, \mnras, 331, 351

\bibitem[Belloni(2010)]{v2-tb10} 
Belloni, T.~M.\ 2010, Lecture Notes in Physics, Berlin Springer Verlag, 794, 53

\bibitem[Bergstr{\"o}m(2000)]{darkmatter} 
Bergstr{\"o}m, L.\ 2000, Reports on Progress in Physics, 63, 793

\bibitem[Berti et al.(2015)]{berti15} 
Berti, E., Barausse, E., Cardoso, V., et al.\ 2015, Classical and Quantum Gravity, 32, 243001

\bibitem[Berti et al.(2006)]{berti06} 
Berti, E., Cardoso, V., \& Will, C.~M.\ 2006, \prd, 73, 064030

\bibitem[Berti \& Volonteri(2008)]{berti08} 
Berti, E., \& Volonteri, M.\ 2008, \apj, 684, 822

\bibitem[Bertotti et al.(2003)]{v3-ppn1} 
Bertotti, B., Iess, L., \& Tortora, P.\ 2003, \nat, 425, 374

\bibitem[Blandford \& Znajek(1977)]{bz-jets} 
Blandford, R.~D., \& Znajek, R.~L.\ 1977, \mnras, 179, 433

\bibitem[Brenneman(2013)]{brenneman-review13} 
Brenneman, L.\ 2013, {\it Measuring the Angular Momentum of Supermassive Black Holes}, SpringerBriefs in Astronomy.~ISBN 978-1-4614-7770-9

\bibitem[Brenneman \& Reynolds(2006)]{BR06} 
Brenneman, L.~W., \& Reynolds, C.~S.\ 2006, \apj, 652, 1028

\bibitem[Brenneman et al.(2011)]{B11} 
Brenneman, L.~W., Reynolds, C.~S., Nowak, M.~A., et al.\ 2011, \apj, 736, 103

\bibitem[Brenneman et al.(2013)]{Brenneman2013} 
Brenneman, L.~W., Risaliti, G., Elvis, M., \& Nardini, E.\ 2013, \mnras, 429, 2662

\bibitem[Broderick et al.(2014)]{brod14} 
Broderick, A.~E., Johannsen, T., Loeb, A., \& Psaltis, D.\ 2014, \apj, 784, 7

\bibitem[Broderick et al.(2009)]{eea1} 
Broderick, A.~E., Loeb, A., \& Narayan, R.\ 2009, \apj, 701, 1357

\bibitem[Broderick \& Narayan(2007)]{grava-eh} 
Broderick, A.~E., \& Narayan, R.\ 2007, Classical and Quantum Gravity, 24, 659

\bibitem[Capozziello \& de Laurentis(2011)]{c1} 
Capozziello, S., \& de Laurentis, M.\ 2011, \physrep, 509, 167

\bibitem[Cardenas-Avendano et al.(2016a)]{v3-cjb1} 
Cardenas-Avendano, A., Jiang, J., \& Bambi, C.\ 2016a, \jcap, 4, 54

\bibitem[Cardenas-Avendano et al.(2016b)]{v3-cjb2} 
Cardenas-Avendano, A., Jiang, J., \& Bambi, C.\ 2016b, Physics Letters B, 760, 254

\bibitem[Cardoso et al.(2016)]{v3-ring} 
Cardoso, V., Franzin, E., \& Pani, P.\ 2016, Physical Review Letters, 116, 171101

\bibitem[Cardoso et al.(2014)]{cpr-m} 
Cardoso, V., Pani, P., \& Rico, J.\ 2014, \prd, 89, 064007

\bibitem[Carter(1968)]{carter68} 
Carter, B.\ 1968, Physical Review, 174, 1559

\bibitem[Carter(1971)]{nh1} 
Carter, B.\ 1971, Physical Review Letters, 26, 331

\bibitem[Casares \& Jonker(2014)]{v3-casares} 
Casares, J., \& Jonker, P.~G.\ 2014, \ssr, 183, 223

\bibitem[Chandrasekhar(1985)]{v3-chandra}
Chandrasekhar, S.\ 1985, {\it The Mathematical Theory of Black Holes} (Clarendon Press, Oxford, UK)

\bibitem[Chen et al.(2016)]{v2-newspin2} 
Chen, Z., Gou, L., McClintock, J.~E., et al.\ 2016, \apj, 825, 45

\bibitem[Chennamangalam \& Lorimer(2014)]{v3-pulsar-gc} 
Chennamangalam, J., \& Lorimer, D.~R.\ 2014, \mnras, 440, L86

\bibitem[Chiang et al.(2012)]{1652} 
Chiang, C.-Y., Reis, R.~C., Walton, D.~J., \& Fabian, A.~C.\ 2012, \mnras, 425, 2436

\bibitem[Chru{\'s}ciel et al.(2012)]{nh3} 
Chru{\'s}ciel, P.~T., Costa, J.~L., \& Heusler, M.\ 2012, Living Reviews in Relativity, 15, 7

\bibitem[Clowe et al.(2004)]{bullet} 
Clowe, D., Gonzalez, A., \& Markevitch, M.\ 2004, \apj, 604, 596

\bibitem[Coleman Miller \& Colbert(2004)]{r-imbh-cmc} 
Coleman Miller, M., \& Colbert, E.~J.~M.\ 2004, International Journal of Modern Physics D, 13, 1

\bibitem[Colpi et al.(1986)]{b-stars1} 
Colpi, M., Shapiro, S.~L., \& Wasserman, I.\ 1986, Physical Review Letters, 57, 2485

\bibitem[Daniel et al.(2010)]{test-gr-c} 
Daniel, S.~F., Linder, E.~V., Smith, T.~L., et al.\ 2010, \prd, 81, 123508

\bibitem[Czerny et al.(2011)]{4-czerny2011} 
Czerny, B., Hryniewicz, K., Niko{\l}ajuk, M., \& S{\c a}dowski, A.\ 2011, \mnras, 415, 2942

\bibitem[Dauser et al.(2013)]{dauser} 
Dauser, T., Garc\'ia, J., Wilms, J., et al.\ 2013, \mnras, 430, 1694

\bibitem[Davis et al.(2005)]{v2-davis1} 
Davis, S.~W., Blaes, O.~M., Hubeny, I., \& Turner, N.~J.\ 2005, \apj, 621, 372

\bibitem[Davis \& Hubeny(2006)]{v2-davis2} 
Davis, S.~W., \& Hubeny, I.\ 2006, \apjs, 164, 530

\bibitem[Davis \& Laor(2011)]{davis} 
Davis, S.~W., \& Laor, A.\ 2011, \apj, 728, 98

\bibitem[de La Calle P{\'e}rez et al.(2010)]{Calle-Perez2010} 
de La Calle P{\'e}rez, I., Longinotti, A.~L., Guainazzi, M., et al.\ 2010, \aap, 524, A50

\bibitem[De Villiers et al.(2003)]{hot-spot-s} 
De Villiers, J.-P., Hawley, J.~F., \& Krolik, J.~H.\ 2003, \apj, 599, 1238

\bibitem[Diener(2003)]{pat5} 
Diener, P.\ 2003, Classical and Quantum Gravity, 20, 4901 

\bibitem[Dodds-Eden et al.(2010)]{flares2} 
Dodds-Eden, K., Sharma, P., Quataert, E., et al.\ 2010, \apj, 725, 450

\bibitem[Doeleman et al.(2008)]{doeleman} 
Doeleman, S.~S., Weintroub, J., Rogers, A.~E.~E., et al.\ 2008, \nat, 455, 78

\bibitem[Done et al.(2013)]{4-done2013} 
Done, C., Jin, C., Middleton, M., \& Ward, M.\ 2013, \mnras, 434, 1955

\bibitem[Doneva et al.(2013)]{dandon} 
Doneva, D.~D., Yazadjiev, S.~S., Stergioulas, N., \& Kokkotas, K.~D.\ 2013, \prd, 88, 084060

\bibitem[Droz(1997)]{massinfl2} 
Droz, S.\ 1997, \prd, 55, 3575

\bibitem[Dvali \& Gomez(2013a)]{q1} 
Dvali, G., \& Gomez, C.\ 2013a, Fortschritte der Physik, 61, 742

\bibitem[Dvali \& Gomez(2013b)]{q2} 
Dvali, G., \& Gomez, C.\ 2013b, Physics Letters B, 719, 419

\bibitem[Dyson et al.(1920)]{eddington1919} 
Dyson, F.~W., Eddington, A.~S., \& Davidson, C.\ 1920, Royal Society of London Philosophical Transactions Series A, 220, 291

\bibitem[Eddington(1922)]{v3-eddington} 
Eddington, A.~S.\ 1922, {\it The Mathematical Theory of Relativity} (Cambridge University Press, London, UK)

\bibitem[Einstein(1916)]{einstein1916} 
Einstein, A.\ 1916, Annalen der Physik, 354, 769

\bibitem[Eisenhauer et al.(2008)]{gravity} 
Eisenhauer, F., Perrin, G., Brandner, W., et al.\ 2008, \procspie, 7013E, 69

\bibitem[Elvis et al.(2002)]{soltan1} 
Elvis, M., Risaliti, G., \& Zamorani, G.\ 2002, \apjl, 565, L75

\bibitem[Enolski et al.(2011)]{v2-hyper} 
Enolski, V.~Z., Hackmann, E., Kagramanova, V., Kunz, J., L\"ammerzahl, C.\ 2011, Journal of Geometry and Physics, 61, 899

\bibitem[Fabian et al.(1989)]{iron1} 
Fabian, A.~C., Rees, M.~J., Stella, L., \& White, N.~E.\ 1989, \mnras, 238, 729

\bibitem[Fabian et al.(2012)]{cyg3} 
Fabian, A.~C., Wilkins, D.~R., Miller, J.~M., et al.\ 2012, \mnras, 424, 217

\bibitem[Falcke \& Markoff(2013)]{falcke13} 
Falcke, H., \& Markoff, S.~B.\ 2013, Classical and Quantum Gravity, 30, 244003

\bibitem[Falcke et al.(2000)]{shadow-0} 
Falcke, H., Melia, F., \& Agol, E.\ 2000, \apjl, 528, L13

\bibitem[Fender et al.(2004)]{rev-jets} 
Fender, R.~P., Belloni, T.~M., \& Gallo, E.\ 2004, \mnras, 355, 1105

\bibitem[Fender et al.(2010)]{no-jets} 
Fender, R.~P., Gallo, E., \& Russell, D.\ 2010, \mnras, 406, 1425

\bibitem[Fish et al.(2013)]{fish13} 
Fish, V., Alef, W., Anderson, J., et al.\ 2013, arXiv:1309.3519

\bibitem[Fragos \& McClintock(2015)]{fragos15} 
Fragos, T., \& McClintock, J.~E.\ 2015, \apj, 800, 17

\bibitem[Fragos et al.(2010)]{fragos10k} 
Fragos, T., Tremmel, M., Rantsiou, E., \& Belczynski, K.\ 2010, \apjl, 719, L79

\bibitem[Frieman et al.(2008)]{darkenergy} 
Frieman, J.~A., Turner, M.~S., \& Huterer, D.\ 2008, \araa, 46, 385

\bibitem[Frolov \& Vilkovisky(1981)]{frolov81} 
Frolov, V.~P., \& Vilkovisky, G.~A.\ 1981, Physics Letters B, 106, 307

\bibitem[Gallo et al.(2005)]{Gallo2005} 
Gallo, L.~C., Fabian, A.~C., Boller, T., \& Pietsch, W.\ 2005, \mnras, 363, 64

\bibitem[Gallo et al.(2011)]{Gallo2011} 
Gallo, L.~C., Miniutti, G., Miller, J.~M., et al.\ 2011, \mnras, 411, 607

\bibitem[Garc\'ia et al.(2015)]{gx339b}
Garc\'ia, J.~A., Steiner, J.~F., McClintock, J.~E., et al.\ 2015, \apj, 813, 84

\bibitem[Genzel et al.(1997)]{v2-genzel2} 
Genzel, R., Eckart, A., Ott, T., \& Eisenhauer, F.\ 1997, \mnras, 291, 219

\bibitem[Genzel et al.(2003)]{flares1} 
Genzel, R., Sch{\"o}del, R., Ott, T., et al.\ 2003, \nat, 425, 934

\bibitem[Genzel et al.(1996)]{v2-genzel1} 
Genzel, R., Thatte, N., Krabbe, A., Kroker, H., \& Tacconi-Garman, L.~E.\ 1996, \apj, 472, 153

\bibitem[Geroch(1970)]{mmm1} 
Geroch, R.\ 1970, Journal of Mathematical Physics, 11, 2580

\bibitem[Ghasemi-Nodehi \& Bambi(2016)]{v3-masume} 
Ghasemi-Nodehi, M., \& Bambi, C.\ 2016, European Physical Journal C, 76, 290

\bibitem[Ghasemi-Nodehi et al.(2015)]{masoumeh} 
Ghasemi-Nodehi, M., Li, Z., \& Bambi, C.\ 2015, European Physical Journal C, 75, 315

\bibitem[Ghez et al.(2005)]{ghez} 
Ghez, A.~M., Salim, S., Hornstein, S.~D., et al.\ 2005, \apj, 620, 744

\bibitem[Giacomazzo et al.(2011)]{super3} 
Giacomazzo, B., Rezzolla, L., \& Stergioulas, N.\ 2011, \prd, 84, 024022

\bibitem[Giddings(2014)]{q3} 
Giddings, S.~B.\ 2014, \prd, 90, 124033

\bibitem[Gillessen et al.(2010)]{flares} 
Gillessen, S., Eisenhauer, F., Perrin, G., et al.\ 2010, \procspie, 7734, 77340Y

\bibitem[Gillessen et al.(2009)]{v2-genzel4} 
Gillessen, S., Eisenhauer, F., Trippe, S., et al.\ 2009, \apj, 692, 1075

\bibitem[Gimon \& Ho{\v r}ava(2009)]{v2-gh} 
Gimon, E.~G., \& Ho{\v r}ava, P.\ 2009, Physics Letters B, 672, 299

\bibitem[Glampedakis \& Babak(2006)]{tests-gw1} 
Glampedakis, K., \& Babak, S.\ 2006, Classical and Quantum Gravity, 23, 4167

\bibitem[Gossan et al.(2012)]{gossan} 
Gossan, S., Veitch, J., \& Sathyaprakash, B.~S.\ 2012, \prd, 85, 124056

\bibitem[Gou et al.(2009)]{lmcx1} 
Gou, L., McClintock, J.~E., Liu, J., et al.\ 2009, \apj, 701, 1076

\bibitem[Gou et al.(2011)]{cyg1}
 Gou, L., McClintock, J.~E., Reid, M.~J., et al.\ 2011, \apj, 742, 85
 
\bibitem[Gou et al.(2014)]{cyg2} 
 Gou, L., McClintock, J.~E., Remillard, R.~A., et al.\ 2014, \apj, 790, 29

\bibitem[Gou et al.(2010)]{62} 
Gou, L., McClintock, J.~E., Steiner, J.~F., et al.\ 2010, \apjl, 718, L122

\bibitem[Gould(2000)]{v3-micro1} 
Gould, A.\ 2000, \apj, 535, 928

\bibitem[Hamaus et al.(2009)]{hot-spot-m} 
Hamaus, N., Paumard, T., M{\"u}ller, T., et al.\ 2009, \apj, 692, 902

\bibitem[Hansen(1974)]{mmm2} 
Hansen, R.~O.\ 1974, Journal of Mathematical Physics, 15, 46

\bibitem[Hansen \& Yunes(2013)]{ny-nja} 
Hansen, D., \& Yunes, N.\ 2013, \prd, 88, 104020

\bibitem[Harko et al.(2009a)]{harko3} 
Harko, T., Kov{\'a}cs, Z., \& Lobo, F.~S.~N.\ 2009a, Classical and Quantum Gravity, 26, 215006

\bibitem[Harko et al.(2009b)]{harko4} 
Harko, T., Kov{\'a}cs, Z., \& Lobo, F.~S.~N.\ 2009b, \prd, 80, 044021

\bibitem[Harko et al.(2010)]{harko5} 
Harko, T., Kov{\'a}cs, Z., \& Lobo, F.~S.~N.\ 2010, Classical and Quantum Gravity, 27, 105010

\bibitem[Hawking(1972)]{haw-r} 
Hawking, S.~W.\ 1972, Communications in Mathematical Physics, 25, 152

\bibitem[Herdeiro \& Radu(2014)]{v2-hr0} 
Herdeiro, C.~A.~R., \& Radu, E.\ 2014, Physical Review Letters, 112, 221101 

\bibitem[Herdeiro \& Radu(2015)]{v2-hr} 
Herdeiro, C.~A.~R., \& Radu, E.\ 2015, International Journal of Modern Physics D, 24, 1542014

\bibitem[Hughes et al.(1994)]{teuk} 
Hughes, S.~A., Keeton, C.~R., II, Walker, P., et al.\ 1994, \prd, 49, 4004

\bibitem[Jacobson \& Venkataramani(1995)]{jacobson} 
Jacobson, T., \& Venkataramani, S.\ 1995, Classical and Quantum Gravity, 12, 1055

\bibitem[Jetzer(1992)]{b-stars2} 
Jetzer, P.\ 1992, \physrep, 220, 163

\bibitem[Jiang et al.(2015a)]{jjc1} 
Jiang, J., Bambi, C., \& Steiner, J.~F.\ 2015a, \jcap, 5, 025

\bibitem[Jiang et al.(2015b)]{jjc2} 
Jiang, J., Bambi, C., \& Steiner, J.~F.\ 2015b, \apj, 811, 130

\bibitem[Jiang et al.(2016)]{jjc3} 
Jiang, J., Bambi, C., \& Steiner, J.~F.\ 2016, \prd, 93, 123008

\bibitem[Johannsen(2012)]{sgra-tj} 
Johannsen, T.\ 2012, Advances in Astronomy, 2012, 486750

\bibitem[Johannsen(2013a)]{pat2} 
Johannsen, T.\ 2013a, \prd, 87, 124017

\bibitem[Johannsen(2013b)]{reg-j1} 
Johannsen, T.\ 2013b, \prd, 88, 044002

\bibitem[Johannsen(2014)]{reg-j2} 
Johannsen, T.\ 2014, \prd, 90, 064002

\bibitem[Johannsen(2015)]{v2-rev-tim} 
Johannsen, T.\ 2016, Classical and Quantum Gravity, 33, 113001

\bibitem[Johannsen \& Psaltis(2010)]{shadow-jp} 
Johannsen, T., \& Psaltis, D.\ 2010, \apj, 718, 446

\bibitem[Johannsen \& Psaltis(2011a)]{jp-m} 
Johannsen, T., \& Psaltis, D.\ 2011a, \prd, 83, 124015

\bibitem[Johannsen \& Psaltis(2011b)]{qpojp} 
Johannsen, T., \& Psaltis, D.\ 2011b, \apj, 726, 11 

\bibitem[Johannsen \& Psaltis(2013)]{iron-jp} 
Johannsen, T., \& Psaltis, D.\ 2013, \apj, 773, 57

\bibitem[Joshi \& Malafarina(2011)]{daniele} 
Joshi, P.~S., \& Malafarina, D.\ 2011, International Journal of Modern Physics D, 20, 2641

\bibitem[Kalogera \& Baym(1996)]{kb96} 
Kalogera, V., \& Baym, G.\ 1996, \apjl, 470, L61

\bibitem[Kerr(1963)]{v3-kerr} 
Kerr, R.~P.\ 1963, Physical Review Letters, 11, 237

\bibitem[Kesden et al.(2005)]{no-isco1} 
Kesden, M., Gair, J., \& Kamionkowski, M.\ 2005, \prd, 71, 044015

\bibitem[King \& Kolb(1999)]{kk99} 
King, A.~R., \& Kolb, U.\ 1999, \mnras, 305, 654

\bibitem[Kolehmainen \& Done(2010)]{gx339} 
Kolehmainen, M., \& Done, C.\ 2010, \mnras, 406, 2206

\bibitem[Kong et al.(2014)]{lingyao} 
Kong, L., Li, Z., \& Bambi, C.\ 2014, \apj, 797, 78

\bibitem[Konoplya et al.(2016)]{v3-krz} 
Konoplya, R., Rezzolla, L., \& Zhidenko, A.\ 2016, \prd, 93, 064015

\bibitem[Konoplya \& Zhidenko(2016)]{v3-kz} 
Konoplya, R., \& Zhidenko, A.\ 2016, Physics Letters B, 756, 350

\bibitem[Kormendy \& Richstone(1995)]{kormendy} 
Kormendy, J., \& Richstone, D.\ 1995, \araa, 33, 581

\bibitem[Krawczynski(2012)]{polar3} 
Krawczynski, H.\ 2012, \apj, 754, 133

\bibitem[Kumar \& Pringle(1985)]{kumar85} 
Kumar, S., \& Pringle, J.~E.\ 1985, \mnras, 213, 435

\bibitem[Lattimer(2012)]{v3-lattimer} 
Lattimer, J.~M.\ 2012, Annual Review of Nuclear and Particle Science, 62, 485

\bibitem[Lehto \& Valtonen(1996)]{valtonen3} 
Lehto, H.~J., \& Valtonen, M.~J.\ 1996, \apj, 460, 207

\bibitem[Li et al.(2009)]{polar1} 
Li, L.-X., Narayan, R., \& McClintock, J.~E.\ 2009, \apj, 691, 847

\bibitem[Li et al.(2005)]{cfm2} 
Li, L.-X., Zimmerman, E.~R., Narayan, R., \& McClintock, J.~E.\ 2005, \apjs, 157, 335

\bibitem[Li \& Bambi(2013a)]{z1} 
Li, Z., \& Bambi, C.\ 2013a, \jcap, 3, 031

\bibitem[Li \& Bambi(2013b)]{spin-cz} 
Li, Z., \& Bambi, C.\ 2013b, \prd, 87, 124022

\bibitem[Li \& Bambi(2014a)]{zilong041} 
Li, Z., \& Bambi, C.\ 2014a, \jcap, 1, 041

\bibitem[Li \& Bambi(2014b)]{hotspot-f3} 
Li, Z., \& Bambi, C.\ 2014b, \prd, 90, 024071

\bibitem[Li et al.(2014)]{hotspot-f1} 
Li, Z., Kong, L., \& Bambi, C.\ 2014, \apj, 787, 152

\bibitem[Lin et al.(2015)]{x-nan} 
Lin, N., Li, Z., Arthur, J., Asquith, R., \& Bambi, C.\ 2015, \jcap, 9, 038

\bibitem[Lin et al.(2015)]{v3-nan} 
Lin, N., Tsukamoto, N., Ghasemi-Nodehi, M., \& Bambi, C.\ 2015, European Physical Journal C, 75, 599

\bibitem[Liu et al.(2015)]{hotspot-f2} 
Liu, D., Li, Z., \& Bambi, C.\ 2015, \jcap, 1, 020

\bibitem[Liu et al.(2015)]{polar4} 
Liu, D., Li, Z., Cheng, Y., \& Bambi, C.\ 2015, European Physical Journal C, 75, 383

\bibitem[Liu et al.(2008)]{liu08} 
Liu, J., McClintock, J.~E., Narayan, R., Davis, S.~W., \& Orosz, J.~A.\ 2008, \apjl, 679, L37

\bibitem[Liu et al.(2010)]{liu10} 
Liu, J., McClintock, J.~E., Narayan, R., Davis, S.~W., \& Orosz, J.~A.\ 2010, \apjl, 719, L109

\bibitem[Liu et al.(2012)]{pulsar2} 
Liu, K., Wex, N., Kramer, M., Cordes, J.~M., \& Lazio, T.~J.~W.\ 2012, \apj, 747, 1

\bibitem[Lohfink et al.(2013)]{Lohfink2013} 
Lohfink, A.~M., Reynolds, C.~S., Jorstad, S.~G., et al.\ 2013, \apj, 772, 83

\bibitem[Lohfink et al.(2012)]{L12} 
Lohfink, A.~M., Reynolds, C.~S., Miller, J.~M., et al.\ 2012, \apj, 758, 67

\bibitem[Lorimer \& Kramer(2005)]{pulsar-r}
Lorimer, D.~R., \& Kramer, M.\ 2005, {\it Handbook of Pulsar Astronomy} (Cambridge University Press, Cambridge, UK)

\bibitem[Lu \& Torres(2003)]{iron-t} 
Lu, Y., \& Torres, D.~F.\ 2003, International Journal of Modern Physics D, 12, 63

\bibitem[Macedo et al.(2013)]{v2-mpcc} 
Macedo, C.~F.~B., Pani, P., Cardoso, V., \& Crispino, L.~C.~B.\ 2013, \apj, 774, 48

\bibitem[Manko et al.(2000)]{manko2} 
Manko, V.~S., Mielke, E.~W., \& Sanabria-G{\'o}mez, J.~D.\ 2000, \prd, 61, 081501

\bibitem[Manko \& Novikov(1992)]{manko1} 
Manko, V.~S., \& Novikov, I.~D.\ 1992, Classical and Quantum Gravity, 9, 2477

\bibitem[Maoz(1998)]{maoz} 
Maoz, E.\ 1998, \apjl, 494, L181

\bibitem[Markoff et al.(2001)]{nhs1} 
Markoff, S., Falcke, H., Yuan, F., \& Biermann, P.~L.\ 2001, \aap, 379, L13

\bibitem[Martocchia \& Matt(1996)]{matt2} 
Martocchia, A., \& Matt, G.\ 1996, \mnras, 282, L53

\bibitem[Maselli et al.(2015)]{qpo-rome} 
Maselli, A., Gualtieri, L., Pani, P., Stella, L., \& Ferrari, V.\ 2015, \apj, 801, 115

\bibitem[Matt et al.(1991)]{matt1} 
Matt, G., Perola, G.~C., \& Piro, L.\ 1991, \aap, 247, 25

\bibitem[McClintock et al.(2004)]{eeb1} 
McClintock, J.~E., Narayan, R., \& Rybicki, G.~B.\ 2004, \apj, 615, 402

\bibitem[McClintock et al.(2014)]{cfm-rev} 
McClintock, J.~E., Narayan, R., \& Steiner, J.~F.\ 2014, \ssr, 183, 295

\bibitem[McClintock et al.(2006)]{1915} 
McClintock, J.~E., Shafee, R., Narayan, R., et al.\ 2006, \apj, 652, 518

\bibitem[McKinney(2005)]{mckinney} 
McKinney, J.~C.\ 2005, \apjl, 630, L5

\bibitem[Merritt et al.(2010)]{will3} 
Merritt, D., Alexander, T., Mikkola, S., \& Will, C.~M.\ 2010, \prd, 81, 062002

\bibitem[Middleton et al.(2014)]{m31} 
Middleton, M.~J., Miller-Jones, J.~C.~A., \& Fender, R.~P.\ 2014, \mnras, 439, 1740

\bibitem[Mignemi \& Stewart(1993)]{v3-EdGB} 
Mignemi, S., \& Stewart, N.~R.\ 1993, \prd, 47, 5259

\bibitem[Miller et al.(2013)]{1915b} 
Miller, J.~M., Parker, M.~L., Fuerst, F., et al.\ 2013, \apjl, 775, L45

\bibitem[Miniutti et al.(2007)]{Miniutti2007} 
Miniutti, G., Fabian, A.~C., Anabuki, N., et al.\ 2007, \pasj, 59, S315

\bibitem[Miniutti et al.(2009)]{Miniutti2009a} 
Miniutti, G., Panessa, F., de Rosa, A., et al.\ 2009, \mnras, 398, 255

\bibitem[Misner et al.(1973)]{MTW73} 
Misner, C.~W., Thorne, K.~S., \& Wheeler, J.~A.\ 1973, {\it Gravitation} (W.H.~Freeman and Co., San Francisco, California)

\bibitem[Motta et al.(2014)]{qpo2} 
Motta, S.~E., Belloni, T.~M., Stella, L., Mu{\~n}oz-Darias, T., \& Fender, R.\ 2014, \mnras, 437, 2554

\bibitem[Narayan(2005)]{narayan1} 
Narayan, R.\ 2005, New Journal of Physics, 7, 199

\bibitem[Narayan \& Heyl(2002)]{eec1} 
Narayan, R., \& Heyl, J.~S.\ 2002, \apjl, 574, L139

\bibitem[Narayan \& McClintock(2008)]{eeb2} 
Narayan, R., \& McClintock, J.~E.\ 2008, \nar, 51, 733

\bibitem[Narayan \& McClintock(2012)]{si-jets} 
Narayan, R., \& McClintock, J.~E.\ 2012, \mnras, 419, L69

\bibitem[Narayan et al.(1995)]{adaf95} 
Narayan, R., Yi, I., \& Mahadevan, R.\ 1995, \nat, 374, 623

\bibitem[Nardini et al.(2011)]{Nardini2011} 
Nardini, E., Fabian, A.~C., Reis, R.~C., \& Walton, D.~J.\ 2011, \mnras, 410, 1251

\bibitem[Newman et al.(1965)]{v2-nj-alg2} 
Newman, E.~T., Couch, E., Chinnapared, K., et al.\ 1965, Journal of Mathematical Physics, 6, 918

\bibitem[Newman \& Janis(1965)]{v2-nj-alg} 
Newman, E.~T., \& Janis, A.~I.\ 1965, Journal of Mathematical Physics, 6, 915

\bibitem[Novikov \& Thorne(1973)]{nt1} 
Novikov, I.~D., \& Thorne, K.~S.\ 1973, in {\it Black Holes}, edited by C.~De~Witt and B.~De~Witt(Gordon and Breach, New York, US), 343

\bibitem[Oppenheimer \& Snyder(1939)]{op-sn} 
Oppenheimer, J.~R., \& Snyder, H.\ 1939, Physical Review, 56, 455

\bibitem[{\"O}zel et al.(2010)]{ozel} 
{\"O}zel, F., Psaltis, D., Narayan, R., \& McClintock, J.~E.\ 2010, \apj, 725, 1918

\bibitem[Page \& Thorne(1974)]{nt2} 
Page, D.~N., \& Thorne, K.~S.\ 1974, \apj, 191, 499

\bibitem[Pani et al.(2010)]{super1} 
Pani, P., Barausse, E., Berti, E., \& Cardoso, V.\ 2010, \prd, 82, 044009

\bibitem[Patrick et al.(2011a)]{P11} 
Patrick, A.~R., Reeves, J.~N., Lobban, A.~P., Porquet, D., \& Markowitz, A.~G.\ 2011a, \mnras, 416, 2725

\bibitem[Patrick et al.(2011b)]{P12} 
Patrick, A.~R., Reeves, J.~N., Porquet, D., et al.\ 2011b, \mnras, 411, 2353

\bibitem[Parker et al.(2015)]{v2-isco2} 
Parker, M.~L., Tomsick, J.~A., Miller, J.~M., et al.\ 2015, \apj, 808, 9

\bibitem[Pei et al.(2016)]{v4-jet} 
Pei, G., Nampalliwar, S., Bambi, C., \& Middleton, M.~J.\ 2016, European Physical Journal C, 76, 534

\bibitem[Penrose(1969)]{wccc} 
Penrose, R.\ 1969, Nuovo Cimento Rivista Serie, 1, 252

\bibitem[Perez et al.(1997)]{qpo3} 
Perez, C.~A., Silbergleit, A.~S., Wagoner, R.~V., \& Lehr, D.~E.\ 1997, \apj, 476, 589

\bibitem[Plant et al.(2015)]{v2-isco1} 
Plant, D.~S., Fender, R.~P., Ponti, G., Mu{\~n}oz-Darias, T., \& Coriat, M.\ 2015, \aap, 573, A120

\bibitem[Poisson(2004)]{poisson} 
Poisson, E.\ 2004, {\it A RelativistÕs Toolkit: The Mathematics of Black-Hole Mechanics} (Cambridge University Press, Cambridge, UK)

\bibitem[Poisson \& Israel(1990)]{massinfl1} 
Poisson, E., \& Israel, W.\ 1990, \prd, 41, 1796

\bibitem[Postman et al.(2012)]{a2261} 
Postman, M., Lauer, T.~R., Donahue, M., et al.\ 2012, \apj, 756, 159

\bibitem[Price(1972)]{price} 
Price, R.~H.\ 1972, \prd, 5, 2419

\bibitem[Psaltis et al.(2008)]{kpsaltis} 
Psaltis, D., Perrodin, D., Dienes, K.~R., \& Mocioiu, I.\ 2008, Physical Review Letters, 100, 119902

\bibitem[Pun et al.(2008a)]{harko1} 
Pun, C.~S.~J., Kov{\'a}cs, Z., \& Harko, T.\ 2008a, \prd, 78, 024043

\bibitem[Pun et al.(2008b)]{harko2} 
Pun, C.~S.~J., Kov{\'a}cs, Z., \& Harko, T.\ 2008b, \prd, 78, 084015

\bibitem[Reis et al.(2009)]{swift} 
Reis, R.~C., Fabian, A.~C., Ross, R.~R., \& Miller, J.~M.\ 2009, \mnras, 395, 1257

\bibitem[Reis et al.(2011)]{1752} 
Reis, R.~C., Miller, J.~M., Fabian, A.~C., et al.\ 2011, \mnras, 410, 2497

\bibitem[Reis et al.(2012)]{maxi} 
Reis, R.~C., Miller, J.~M., Reynolds, M.~T., Fabian, A.~C., \& Walton, D.~J.\ 2012, \apj, 751, 34

\bibitem[Remillard \& McClintock(2006)]{mcrem06} 
Remillard, R.~A., \& McClintock, J.~E.\ 2006, \araa, 44, 49

\bibitem[Reynolds(2014)]{iron2} 
Reynolds, C.~S.\ 2014, \ssr, 183, 277

\bibitem[Reynolds et al.(1999)]{reverb1} 
Reynolds, C.~S., Young, A.~J., Begelman, M.~C., \& Fabian, A.~C.\ 1999, \apj, 514, 164

\bibitem[Rezzolla et al.(2003)]{qpo5} 
Rezzolla, L., Yoshida, S., Maccarone, T.~J., \& Zanotti, O.\ 2003, \mnras, 344, L37

\bibitem[Rezzolla \& Zhidenko(2014)]{zhidenko} 
Rezzolla, L., \& Zhidenko, A.\ 2014, \prd, 90, 084009

\bibitem[Rhoades \& Ruffini(1974)]{rr74} 
Rhoades, C.~E., \& Ruffini, R.\ 1974, Physical Review Letters, 32, 324 

\bibitem[Risaliti et al.(2013)]{Risaliti2013} 
Risaliti, G., Harrison, F.~A., Madsen, K.~K., et al.\ 2013, \nat, 494, 449

\bibitem[Risaliti et al.(2009)]{Risaliti2009b} 
Risaliti, G., Miniutti, G., Elvis, M., et al.\ 2009, \apj, 696, 160

\bibitem[Risaliti et al.(2011)]{martin} 
Risaliti, G., Nardini, E., Elvis, M., Brenneman, L., \& Salvati, M.\ 2011, \mnras, 417, 178

\bibitem[Robinson(1975)]{nh2} 
Robinson, D.~C.\ 1975, Physical Review Letters, 34, 905
  
\bibitem[Russell et al.(2013)]{russell+} 
Russell, D.~M., Gallo, E., \& Fender, R.~P.\ 2013, \mnras, 431, 405  
  
\bibitem[Ryan(1995)]{tests-gw0} 
Ryan, F.~D.\ 1995, \prd, 52, 5707  
  
\bibitem[Savonije(1978)]{v3-sav} 
Savonije, G.~J.\ 1978, \aap, 62, 317  
  
\bibitem[Schee \& Stuchl{\'{\i}}k(2009)]{shadow-ss} 
Schee, J., \& Stuchl{\'{\i}}k, Z.\ 2009, International Journal of Modern Physics D, 18, 983  

\bibitem[Schmoll et al.(2009)]{Schmoll2009} 
Schmoll, S., Miller, J.~M., Volonteri, M., et al.\ 2009, \apj, 703, 2171

\bibitem[Schnittman \& Krolik(2009)]{polar2} 
Schnittman, J.~D., \& Krolik, J.~H.\ 2009, \apj, 701, 1175

\bibitem[Schnittman \& Krolik(2010)]{sandm} 
Schnittman, J.~D., \& Krolik, J.~H.\ 2010, \apj, 712, 908

\bibitem[Sch{\"o}del et al.(2002)]{v2-genzel3} 
Sch{\"o}del, R., Ott, T., Genzel, R., et al.\ 2002, \nat, 419, 694

\bibitem[Shafee et al.(2006)]{sh06} 
Shafee, R., McClintock, J.~E., Narayan, R., et al.\ 2006, \apjl, 636, L113

\bibitem[Soffitta et al.(2013)]{xipe} 
Soffitta, P., Barcons, X., Bellazzini, R., et al.\ 2013, Experimental Astronomy, 36, 523

\bibitem[Sperhake et al.(2009)]{v2-scp} 
Sperhake, U., Cardoso, V., Pretorius, F., et al.\ 2009, Physical Review Letters, 103, 131102

\bibitem[Steiner \& McClintock(2012)]{alignment2} 
Steiner, J.~F., \& McClintock, J.~E.\ 2012, \apj, 745, 136

\bibitem[Steiner et al.(2013)]{jack-jets} 
Steiner, J.~F., McClintock, J.~E., \& Narayan, R.\ 2013, \apj, 762, 104

\bibitem[Steiner et al.(2014)]{lmcx3}
Steiner, J.~F., McClintock, J.~E., Orosz, J.~A., et al.\ 2014, \apjl, 793, L29

\bibitem[Steiner et al.(2012a)]{h1743} 
Steiner, J.~F., McClintock, J.~E., \& Reid, M.~J.\ 2012a, \apjl, 745, L7

\bibitem[Steiner et al.(2010)]{r_in} 
Steiner, J.~F., McClintock, J.~E., Remillard, R.~A., et al.\ 2010, \apjl, 718, L117

\bibitem[Steiner et al.(2012b)]{lmcx1b} 
Steiner, J.~F., Reis, R.~C., Fabian, A.~C., et al.\ 2012b, \mnras, 427, 2552

\bibitem[Steiner et al.(2011)]{xte} 
Steiner, J.~F., Reis, R.~C., McClintock, J.~E., et al.\ 2011, \mnras, 416, 941
  
\bibitem[Steiner et al.(2016)]{v2-newspin1} 
Steiner, J.~F., Walton, D.~J., Garc\'ia, J.~A., et al.\ 2016, \apj, 817, 154
  
\bibitem[Stella \& Vietri(1999)]{qpo1} 
Stella, L., \& Vietri, M.\ 1999, Physical Review Letters, 82, 17  
  
\bibitem[Straub et al.(2012)]{x-1a} 
Straub, O., Vincent, F.~H., Abramowicz, M.~A., Gourgoulhon, E., \& Paumard, T.\ 2012, \aap, 543, A83  
  
\bibitem[Stuchl{\'{\i}}k \& Kotrlov{\'a}(2009)]{qpos} 
Stuchl{\'{\i}}k, Z., \& Kotrlov{\'a}, A.\ 2009, General Relativity and Gravitation, 41, 1305  
  
\bibitem[Svoboda et al.(2012)]{high-q-3} 
Svoboda, J., Dov{\v c}iak, M., Goosmann, R.~W., et al.\ 2012, \aap, 545, A106  
  
\bibitem[Tagger \& Melia(2006)]{nhs3} 
Tagger, M., \& Melia, F.\ 2006, \apjl, 636, L33  
  
\bibitem[Tan et al.(2012)]{Tan2012} 
Tan, Y., Wang, J.~X., Shu, X.~W., \& Zhou, Y.\ 2012, \apjl, 747, L11  
  
\bibitem[Tchekhovskoy et al.(2010)]{tchek1} 
Tchekhovskoy, A., Narayan, R., \& McKinney, J.~C.\ 2010, \apj, 711, 50

\bibitem[Tchekhovskoy et al.(2011)]{tchek2} 
Tchekhovskoy, A., Narayan, R., \& McKinney, J.~C.\ 2011, \mnras, 418, L79  
  
\bibitem[Thornburg(2007)]{pat3} 
Thornburg, J.\ 2007, Living Reviews in Relativity, 10, 3  
  
\bibitem[Thorne(1974)]{th74} 
Thorne, K.~S.\ 1974, \apj, 191, 507  
  
\bibitem[Timmes et al.(1996)]{v2-remnant} 
Timmes, F.~X., Woosley, S.~E., \& Weaver, T.~A.\ 1996, \apj, 457, 834  
  
\bibitem[T{\"o}r{\"o}k et al.(2005)]{qpo4b} 
T{\"o}r{\"o}k, G., Abramowicz, M.~A., Klu{\'z}niak, W., \& Stuchl{\'{\i}}k, Z.\ 2005, \aap, 436, 1  
  
\bibitem[Torres(2002)]{cfm-t} 
Torres, D.~F.\ 2002, Nuclear Physics B, 626, 377  
  
\bibitem[Tournear et al.(2003)]{tournear} 
Tournear, D., Raffauf, E., Bloom, E.~D., et al.\ 2003, \apj, 595, 1058  
  
\bibitem[Trap et al.(2011)]{flares3} 
Trap, G., Goldwurm, A., Dodds-Eden, K., et al.\ 2011, \aap, 528, A140  
  
\bibitem[Trippe et al.(2007)]{trippe07} 
Trippe, S., Paumard, T., Ott, T., et al.\ 2007, \mnras, 375, 764  
  
\bibitem[Tsukamoto et al.(2014)]{naoki} 
Tsukamoto, N., Li, Z., \& Bambi, C.\ 2014, \jcap, 6, 043  

\bibitem[Uttley et al.(2014)]{utt-14}
Uttley, P., Cackett, E.~M., Fabian, A.~C., Kara, E., \& Wilkins, D.~R.\ 2014, The Astronomy and Astrophysics Review, 22, 72

\bibitem[Valtonen et al.(2011)]{valtonen2} 
Valtonen, M.~J., Mikkola, S., Lehto, H.~J., et al.\ 2011, \apj, 742, 22

\bibitem[Valtonen et al.(2010)]{valtonen1} 
Valtonen, M.~J., Mikkola, S., Merritt, D., et al.\ 2010, \apj, 709, 725

\bibitem[Vigeland(2010)]{bumpy2} 
Vigeland, S.~J.\ 2010, \prd, 82, 104041 

\bibitem[Vigeland \& Hughes(2010)]{bumpy1} 
Vigeland, S.~J., \& Hughes, S.~A.\ 2010, \prd, 81, 024030

\bibitem[Vigeland et al.(2011)]{vigeland-y-s} 
Vigeland, S., Yunes, N., \& Stein, L.~C.\ 2011, \prd, 83, 104027

\bibitem[Vincent et al.(2015)]{x-1b} 
Vincent, F.~H., Yan, W., Straub, O., Zdziarski, A.~A., \& Abramowicz, M.~A.\ 2015, \aap, 574, A48

\bibitem[Volkov \& Gal'tsov(1989)]{volkov1} 
Volkov, M.~S., \& Gal'tsov, D.~V.\ 1989, Soviet Journal of Experimental and Theoretical Physics Letters, 50, 346

\bibitem[Volkov \& Gal'tsov(1999)]{volkov2} 
Volkov, M.~S., \& Gal'tsov, D.~V.\ 1999, \physrep, 319, 1 

\bibitem[Wald(1984)]{v3-wald} 
Wald, R.~M.\ 1984, {\it General Relativity} (University of Chicago Press, Chicago, Illinois)

\bibitem[Wald \& Iyer(1991)]{v3-wi} 
Wald, R.~M., \& Iyer, V.\ 1991, \prd, 44, R3719

\bibitem[Walton et al.(2013)]{W13} 
Walton, D.~J., Nardini, E., Fabian, A.~C., Gallo, L.~C., \& Reis, R.~C.\ 2013, \mnras, 428, 2901

\bibitem[Walton et al.(2012)]{1650} 
Walton, D.~J., Reis, R.~C., Cackett, E.~M., Fabian, A.~C., \& Miller, J.~M.\ 2012, \mnras, 422, 2510

\bibitem[Wang et al.(2006)]{soltan2} 
Wang, J.-M., Chen, Y.-M., Ho, L.~C., \& McLure, R.~J.\ 2006, \apjl, 642, L111

\bibitem[Wei \& Liu(2013)]{wei13} 
Wei, S.-W., \& Liu, Y.-X.\ 2013, \jcap, 11, 063

\bibitem[Weisskopf et al.(2008)]{ixpe} 
Weisskopf, M.~C., Bellazzini, R., Costa, E., et al.\ 2008, \procspie, 7011, 70111I

\bibitem[Wex \& Kopeikin(1999)]{pulsar1} 
Wex, N., \& Kopeikin, S.~M.\ 1999, \apj, 514, 388

\bibitem[Wilkins \& Fabian(2011)]{high-q-1} 
Wilkins, D.~R., \& Fabian, A.~C.\ 2011, \mnras, 414, 1269

\bibitem[Wilkins \& Fabian(2012)]{high-q-2} 
Wilkins, D.~R., \& Fabian, A.~C.\ 2012, \mnras, 424, 1284

\bibitem[Will(2008)]{will2} 
Will, C.~M.\ 2008, \apjl, 674, L25

\bibitem[Will(2014)]{will} 
Will, C.~M.\ 2014, Living Reviews in Relativity, 17, 4

\bibitem[Williams et al.(2004)]{v3-ppn2} 
Williams, J.~G., Turyshev, S.~G., \& Boggs, D.~H.\ 2004, Physical Review Letters, 93, 261101

\bibitem[Woosley \& Bloom(2006)]{wb06} 
Woosley, S.~E., \& Bloom, J.~S.\ 2006, \araa, 44, 507

\bibitem[Yoon et al.(2006)]{yoon06} 
Yoon, S.-C., Langer, N., \& Norman, C.\ 2006, \aap, 460, 199 

\bibitem[Yuan \& Narayan(2014)]{x-rev} 
Yuan, F., \& Narayan, R.\ 2014, \araa, 52, 529

\bibitem[Yuan et al.(2004)]{yifei} 
Yuan, Y.-F., Narayan, R., \& Rees, M.~J.\ 2004, \apj, 606, 1112

\bibitem[Yunes \& Hughes(2010)]{ny-sah-10} 
Yunes, N., \& Hughes, S.~A.\ 2010, \prd, 82, 082002

\bibitem[Yunes \& Pretorius(2009)]{chern-simons} 
Yunes, N., \& Pretorius, F.\ 2009, \prd, 79, 084043

\bibitem[Yunes \& Siemens(2013)]{rev-gw-13} 
Yunes, N., \& Siemens, X.\ 2013, Living Reviews in Relativity, 16, 9

\bibitem[Yunes et al.(2016)]{v3-yyp} 
Yunes, N., Yagi, K., \& Pretorius, F.\ 2016, arXiv:1603.08955

\bibitem[Yusef-Zadeh et al.(2006)]{nhs2} 
Yusef-Zadeh, F., Roberts, D., Wardle, M., Heinke, C.~O., \& Bower, G.~C.\ 2006, \apj, 650, 189

\bibitem[Zhang et al.(2015)]{z-l-y-15} 
Zhang, F., Lu, Y., \& Yu, Q.\ 2015, \apj, 809, 127

\bibitem[Zhang et al.(1997)]{cfm1} 
Zhang, S.~N., Cui, W., \& Chen, W.\ 1997, \apjl, 482, L155

\bibitem[Zhang et al.(2016)]{v4-extp} 
Zhang, S.~N., Feroci, M., Santangelo, A., et al.\ 2016, arXiv:1607.08823

\bibitem[Zhou et al.(2016)]{v3-menglei} 
Zhou, M., Cardenas-Avendano, A., Bambi, C., Kleihaus, B., \& Kunz, J.\ 2016, \prd, 94, 024036

\bibitem[Zoghbi et al.(2010)]{Zoghbi2010} 
Zoghbi, A., Fabian, A.~C., Uttley, P., et al.\ 2010, \mnras, 401, 2419
  
\end{thebibliography}
\end{document}